\documentclass[prx,letterpaper,nobalancelastpage,twocolumn,superscriptaddress,nofootinbib,longbibliography]{revtex4-2}

\usepackage{xr}
\usepackage{graphicx}
\usepackage{amsmath}
\usepackage{bbold}
\usepackage{amssymb}
\usepackage[english]{babel}
\usepackage{color}
\usepackage[version=4]{mhchem}
\usepackage[hidelinks]{hyperref}
\usepackage{dsfont}
\usepackage{placeins}
\usepackage{mathtools}
\usepackage[normalem]{ulem}
\usepackage{varwidth}
\usepackage{siunitx}
\setcitestyle{super}

\usepackage{soul}
\newcommand{\tex}{t_\mathrm{ex}}

\newcommand{\Caltech}{California Institute of Technology, Pasadena, CA 91125, USA}
\newcommand{\MIT}{Center for Theoretical Physics, Massachusetts Institute of Technology, Cambridge, MA 02139, USA}
\newcommand{\IAIFI}{The NSF AI Institute for Artificial Intelligence and Fundamental Interactions, Cambridge, MA 02139, USA}
\newcommand{\Stanford}{Department of Electrical Engineering, Stanford University, Stanford, CA, USA}

\usepackage{caption}

\DeclareCaptionLabelSeparator{bar}{ \textbf{\textbar}~}
\usepackage{enumitem}
\setlist{nolistsep}

\let\oldchi\chi
\renewcommand{\chi}{%
  \raisebox{0.44ex}{$\oldchi$}%
}

\makeatletter
\newcommand*{\addFileDependency}[1]{ 
  \typeout{(#1)}
  \@addtofilelist{#1}
  \IfFileExists{#1}{}{\typeout{No file #1.}}
}
\makeatother

\newcommand*{\myexternaldocument}[2]{%
  \externaldocument{#1/#2}%
    \addFileDependency{#2.tex}%
    \addFileDependency{#1/#2.aux}%
}
\myexternaldocument{build}{supplement}

\begin{document}

\title{Benchmarking highly entangled states on a 60-atom analog quantum simulator}
\author{Adam L. Shaw,$^{1,*,\dag}$ Zhuo Chen,$^{2,3,*}$ Joonhee Choi,$^{1,4,*}$ Daniel K. Mark,$^{2,*}$ Pascal Scholl,$^1$ \\Ran Finkelstein,$^1$  Andreas Elben,$^1$ Soonwon Choi,$^{2,\dag}$ and Manuel Endres$^{1,\dag}$}
\noaffiliation
\affiliation{\Caltech}
\affiliation{\MIT}
\affiliation{\IAIFI}
\affiliation{\Stanford}

\maketitle


\textbf{Quantum systems have entered a competitive regime where classical computers must make approximations to represent highly entangled quantum states~\cite{Preskill2018QuantumBeyond,Arute2019QuantumProcessor}. However, in this beyond-classically-exact regime, fidelity comparisons between quantum and classical systems have so far been limited to digital quantum devices~\cite{Arute2019QuantumProcessor,Wu2021StrongProcessor,Zhu2022QuantumSampling,Morvan2023PhaseSampling}, and it remains unsolved how to estimate the actual entanglement content of experiments~\cite{Vidal2002ComputableEntanglement}. Here we perform fidelity benchmarking and mixed-state entanglement estimation with a 60-atom analog Rydberg quantum simulator, reaching a high entanglement entropy regime where exact classical simulation becomes impractical. Our benchmarking protocol involves extrapolation from comparisons against an \textit{approximate} classical algorithm, introduced here, with varying entanglement limits. We then develop and demonstrate an estimator of the experimental mixed-state entanglement~\cite{Vidal2002ComputableEntanglement}, finding our experiment is competitive with state-of-the-art digital quantum devices performing random circuit evolution~\cite{Arute2019QuantumProcessor,Wu2021StrongProcessor,Zhu2022QuantumSampling,Morvan2023PhaseSampling}. Finally, we compare the experimental fidelity against that achieved by various approximate classical algorithms, and find that only the algorithm we introduce is able to keep pace with the experiment on the classical hardware we employ. Our results enable a new paradigm for evaluating the ability of both analog and digital quantum devices to generate entanglement in the beyond-classically-exact regime, and highlight the evolving divide between quantum and classical systems.
}\footnote{\label{a}$^*$ These authors contributed equally}
\footnote{\label{b}$^{\dag}$ ashaw@caltech.edu, soonwon@mit.edu, mendres@caltech.edu}

\begin{figure}[t!]
        \centering
	\includegraphics[width=89mm]{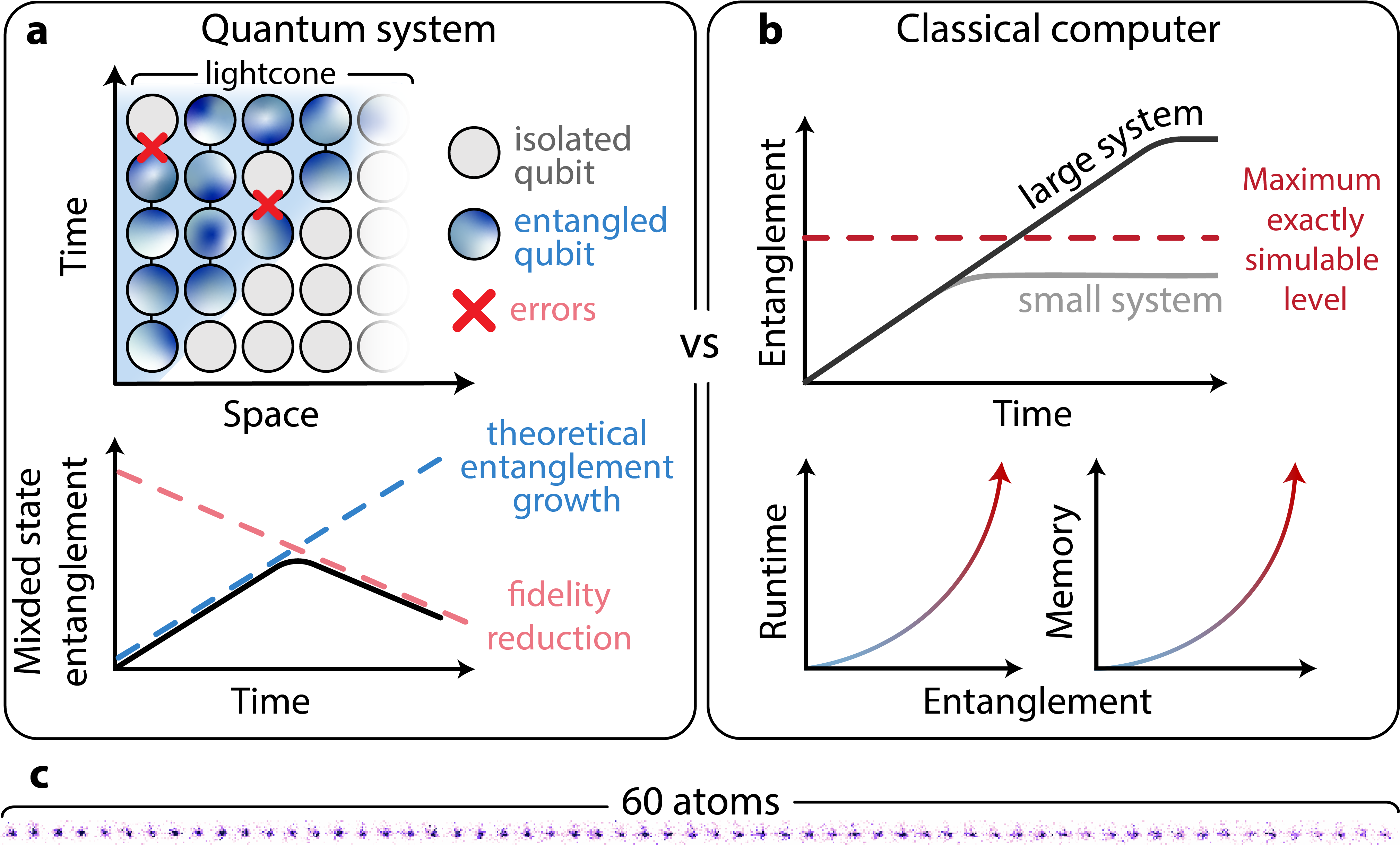}
	\caption{\textbf{Entanglement in quantum and classical systems. a.} In quantum systems, entanglement spreads between neighboring particles before saturating at an extensive level. However, entanglement growth is hampered by experimental errors which reduce the fidelity, limiting entanglement build-up. \textbf{b.} On the other hand, classical computers employ approximate simulation algorithms which often can only capture a limited degree of entanglement to avoid an exponential increase in cost, meaning they cannot exactly simulate dynamics at large system sizes and long evolution times. \textbf{c.} Here we compare quantum devices and classical algorithms in their ability to prepare highly entangled states using a Rydberg quantum simulator with up to 60 atoms in a one-dimensional array (shown as a fluorescence image).
	} 
	\vspace{-0.5cm}
	\label{Fig1}
\end{figure}

\begin{figure*}[t!]
	\centering
	\includegraphics[width=181mm]{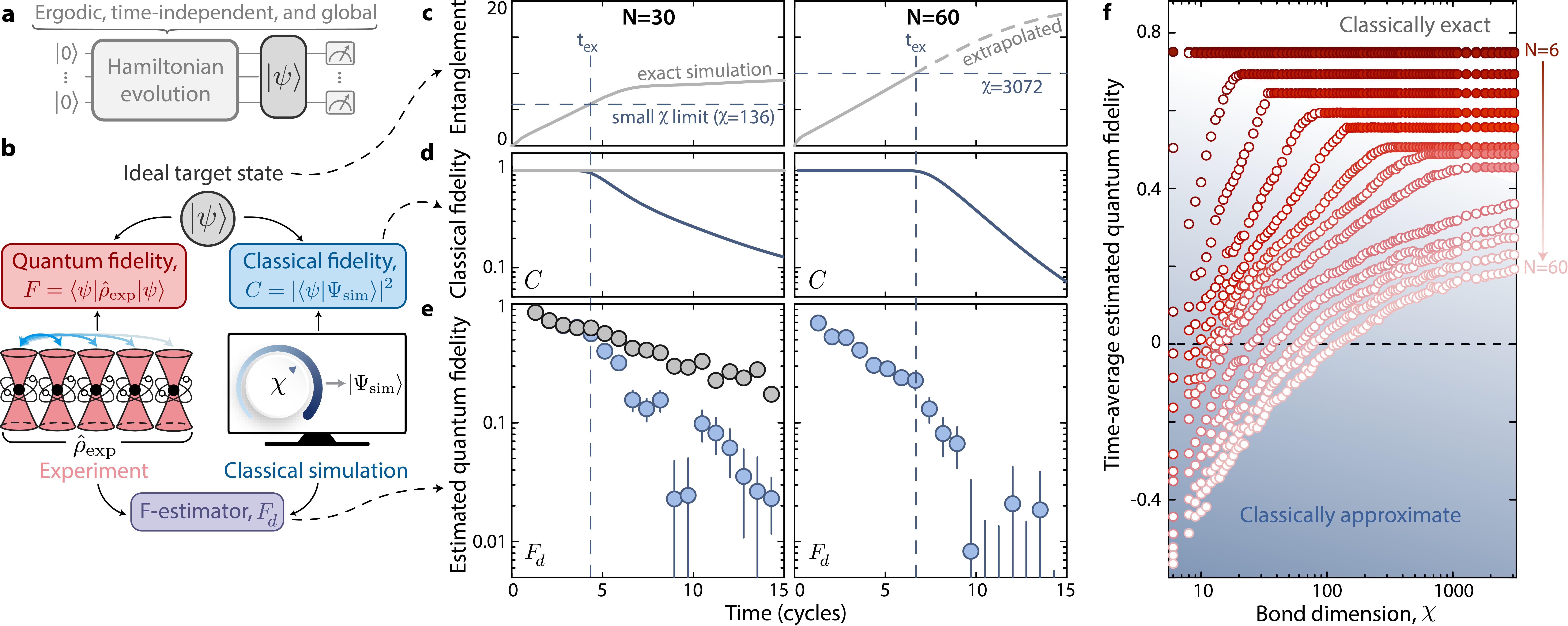}
	\caption{\textbf{Failure of fidelity estimation with an approximate classical algorithm. a.} We use a Rydberg quantum simulator and a classical computer to simulate a time-independent, high-temperature quench starting from the all-zero state, targeting an ideal pure state, $|\psi\rangle$. \textbf{b.} The classical algorithm is characterized by a bond dimension, $\chi$, which limits the maximum simulable entanglement, resulting in smaller-than-unity classical simulation fidelity, $C$. We estimate the quantum fidelity, $F$, with a cross-correlation between measurement outcomes of the classical and quantum systems, termed~\cite{Mark2023BenchmarkingLett} $F_d$. \textbf{c, d, e.} (Top) Half-cut von Neumann entanglement entropy of $|\psi\rangle$, (Middle) classical simulation fidelity, (Bottom) estimated experimental quantum fidelity. We study benchmarking against an exact simulation (gray) or an approximate simulation with limited bond dimension (blue). \textbf{c,} For a system size of $N=30$ (left panels), using too small of a bond dimension sets a cap in the simulation entanglement. \textbf{d.} This causes the classical fidelity to fall at a time, $\tex$, when the entanglement of the target state becomes too large. \textbf{e.} At approximately $\tex$, the estimated experimental quantum fidelity also drops. For the largest system size, $N=60$ (right panels), $\tex$ is well before when the entanglement saturates, even for the largest bond dimension we employ. The time-axis is normalized by the Rabi frequency (Methods). \textbf{f.} The estimated fidelity (averaged over all times in \textbf{e}) increases with bond dimension (open markers), before saturating (closed markers) at a bond dimension capturing the necessary entanglement. For the largest system sizes, saturation is not achieved using the available classical resources.
        } 
	\vspace{-0.5cm}
	\label{Fig2}
\end{figure*}

Classical computers generally struggle to exactly represent highly entangled states~\cite{Preskill2012QuantumFrontier,Ghosh2023ComplexityEntanglement,Kechedzhi2023EffectiveExperiments}, in the sense of entanglement entropy. This has raised interest in the potential of quantum devices to efficiently solve certain classically-hard problems~\cite{Shor1994AlgorithmsFactoring,Mosca1999TheComputer}, but modern noisy-intermediate-scale-quantum~\cite{Preskill2018QuantumBeyond,Bharti2022NoisyAlgorithms} (NISQ) devices are limited by experimental errors (Fig.~\ref{Fig1}a). This makes it a key goal to benchmark NISQ devices in the highly entangled regime where exact classical simulation becomes infeasible (Fig.~\ref{Fig1}b); for example, state-of-the-art classical simulation of Hamiltonian time evolution generating highly entangled states with exact global fidelity is currently limited to 38 qubits~\cite{Hauru2021SimulationEvolution}.

One such approach is to study the fidelity of preparing a highly entangled target state of interest~\cite{Arute2019QuantumProcessor}, with several efficient fidelity estimators~\cite{Neill2018AQubits,Cross2019ValidatingCircuits,Choi2023PreparingChaos,Mark2023BenchmarkingLett,Proctor2022MeasuringComputers} having been introduced in recent years. However, in the beyond-classically-exact regime, these protocols have only been applied to digital quantum devices, with no such demonstrations on analog quantum simulators~\cite{Cirac2012GoalsSimulation}, i.e. quantum devices tailored to efficiently encode select problems of interest~\cite{Daley2022PracticalSimulation,Browaeys2020Many-bodyAtoms,Monroe2021ProgrammableIons,Flamini2019PhotonicReview,Gross2017QuantumLattices,Hartmann2016QuantumPhotons,Altman2021QuantumOpportunities,Houck2012On-chipCircuits}.

In this work, we perform fidelity estimation with an analog quantum simulator targeting highly entangled states which are impractical to represent exactly on a classical computer. Our Rydberg quantum simulator~\cite{Browaeys2020Many-bodyAtoms,Choi2023PreparingChaos} has recently demonstrated~\cite{Scholl2023ErasureSimulator} two-qubit entanglement fidelities of ${\sim}0.999$, spurring this study with up to 60 atoms~\cite{Shaw2023Dark-StateArray} in a one-dimensional array (Fig.~\ref{Fig1}c). We stress that we target high entanglement entropy states that require an exponential number of coefficients to represent classically, as distinct from Greenberger–Horne–Zeilinger (GHZ), cluster, or stabilizer states, which are efficiently representable on a classical computer at all system sizes~\cite{Bridgeman2017Hand-wavingNetworks} (Ext. Data Fig.~\ref{EFig:hilbert_map}).

Our fidelity estimation is based on extrapolation from benchmarking against many \textit{approximate} classical simulations, namely, matrix product state (MPS) algorithms which cap the maximum simulation entanglement to avoid the aforementioned exponential increase in classical cost~\cite{Vidal2003EfficientComputations,Zhou2020WhatComputers,Bridgeman2017Hand-wavingNetworks} (Fig.~\ref{Fig1}b). In one-dimension, early-time entanglement growth is system-size independent, so at short times the MPS representation is exact for essentially arbitrarily large systems. When entanglement growth surpasses the entanglement cap, the MPS is no longer a faithful reference, but we can extrapolate the fidelity through a combination of varying the system size, evolution time and simulation entanglement limit.

Using the fidelity, we derive and demonstrate a simple proxy of the experimental mixed state entanglement~\cite{Vidal2002ComputableEntanglement}, which to date has been notoriously difficult to measure in large systems. Our proxy serves as a universal quality-factor requiring only the fidelity with, and the entanglement of, the ideal target pure state. This enables comparisons between our experiment and state-of-the-art digital quantum devices~\cite{Arute2019QuantumProcessor,Wu2021StrongProcessor,Zhu2022QuantumSampling,Morvan2023PhaseSampling,Moses2023AProcessor}, with which we are competitive. 

Ultimately, we compare the fidelity of our experiment against that achieved by a variety of approximate classical algorithms, including several not based on MPS. Using a single node of the Caltech central computing cluster, none of the tested algorithms is able to match the experimental fidelity in the high-entanglement regime, except for an improved algorithm we introduce, termed Lightcone-MPS. Even with this new algorithm, classical costs reach a regime requiring high-performance computing to match the experiment's performance.

\vspace{0.25cm}
\noindent\textbf{Fidelity estimation with approximate algorithms}\newline
A key quantity when studying quantum systems is the \textit{fidelity}~\cite{Nielsen2010QuantumInformation}, $F=\langle\psi|\hat{\rho}_\mathrm{exp}|\psi\rangle$, where $|\psi\rangle$ is a pure state of interest, and $\hat{\rho}_\mathrm{exp}$ is the experimental mixed state. For digital devices studying deep circuits, the fidelity can be estimated via the linear-cross-entropy~\cite{Neill2018AQubits,Arute2019QuantumProcessor}, a cross-correlation between measurement outcomes of an experiment and an exact classical simulation. A modified cross-entropy, termed~\cite{Mark2023BenchmarkingLett} $F_d$, was proposed for both analog and digital systems, and demonstrated on Rydberg~\cite{Choi2023PreparingChaos} and superconducting~\cite{Zhang2023AMetamaterial} analog quantum simulators (Methods). $F_d$ is efficiently sampled (Ext. Data Fig.~\ref{EFig:sample_complexity}) as 
\begin{align}
    F_d = 2 \frac{\frac{1}{M} \sum_{m=1}^M p(z_m) / p_\text{avg}(z_m)}{\sum_z p(z)^2 / p_\text{avg}(z)} - 1,
\end{align}
where $M$ is the number of measurements, $z_m$ is the experimentally measured bitstring, $p(z)$ is the probability of measuring $z$ with no errors following quench evolution, and $p_\text{avg}(z)$ is the time-averaged probability of measuring $z$. Importantly, $F_d\approx F$ for a wide class of physical systems, as long as the rescaled probabilities $p(z)/p_\text{avg}(z)$ follow the so-called Porter-Thomas distribution~\cite{Mark2023BenchmarkingLett}. Still, a stringent requirement remains: access to an exact classical simulation to obtain $p(z)$, precluding direct fidelity estimation at large system sizes. We circumvent this constraint by introducing a method to estimate the fidelity by benchmarking against \textit{approximate} classical simulations.

We consider a comparison (Fig.~\ref{Fig2}b) between an ideal high-entanglement target pure state, $|\psi\rangle$, the experimental mixed state, $\hat{\rho}_\mathrm{exp}$, and a pure state from classical MPS simulation, $|\Psi_\text{sim}\rangle$. We introduce an improved MPS time-evolution algorithm using an optimal decomposition of Hamiltonian dynamics into quantum circuits~\cite{Haah0QuantumHamiltonians,Tran2019LocalityInteractions}, which we term Lightcone-MPS (Methods). The MPS is parameterized by a \textit{bond dimension}, $\chi$, that defines the maximum simulable entanglement, which scales as $\log(\chi)$. Starting from an all-zero state, we program a time-independent, global quench under the one-dimensional Ising-like Rydberg Hamiltonian (Fig.~\ref{Fig2}a, for Hamiltonian details see Ext. Data Fig.~\ref{EFig:experiment}, Methods). Hamiltonian parameters lead to high-temperature thermalization~\cite{Kaufman2016QuantumSystem,Abanin2019ColloquiumEntanglement}, such that describing $|\psi\rangle$ at late times requires an exponential number of classical coefficients~\cite{Choi2023PreparingChaos}.

For a system size of $N{=}30$ (Fig.~\ref{Fig2}c-e, left), we can exactly classically simulate these dynamics (Fig.~\ref{Fig2}d, grey); by exact, we mean the classical fidelity, $C=|\langle\Psi_\mathrm{sim}|\psi\rangle|^2$, stays near unity for all times. We numerically observe the entanglement of the target state increases linearly at early times, before eventual near-saturation (Fig.~\ref{Fig2}c). Moreover, the estimated experimental quantum fidelity, $F_d$, shows apparent exponential decay due to experimental errors~\cite{Choi2023PreparingChaos} (Fig.~\ref{Fig2}e, grey).

\begin{figure}[t!]
	\centering
	\includegraphics[width=89mm]{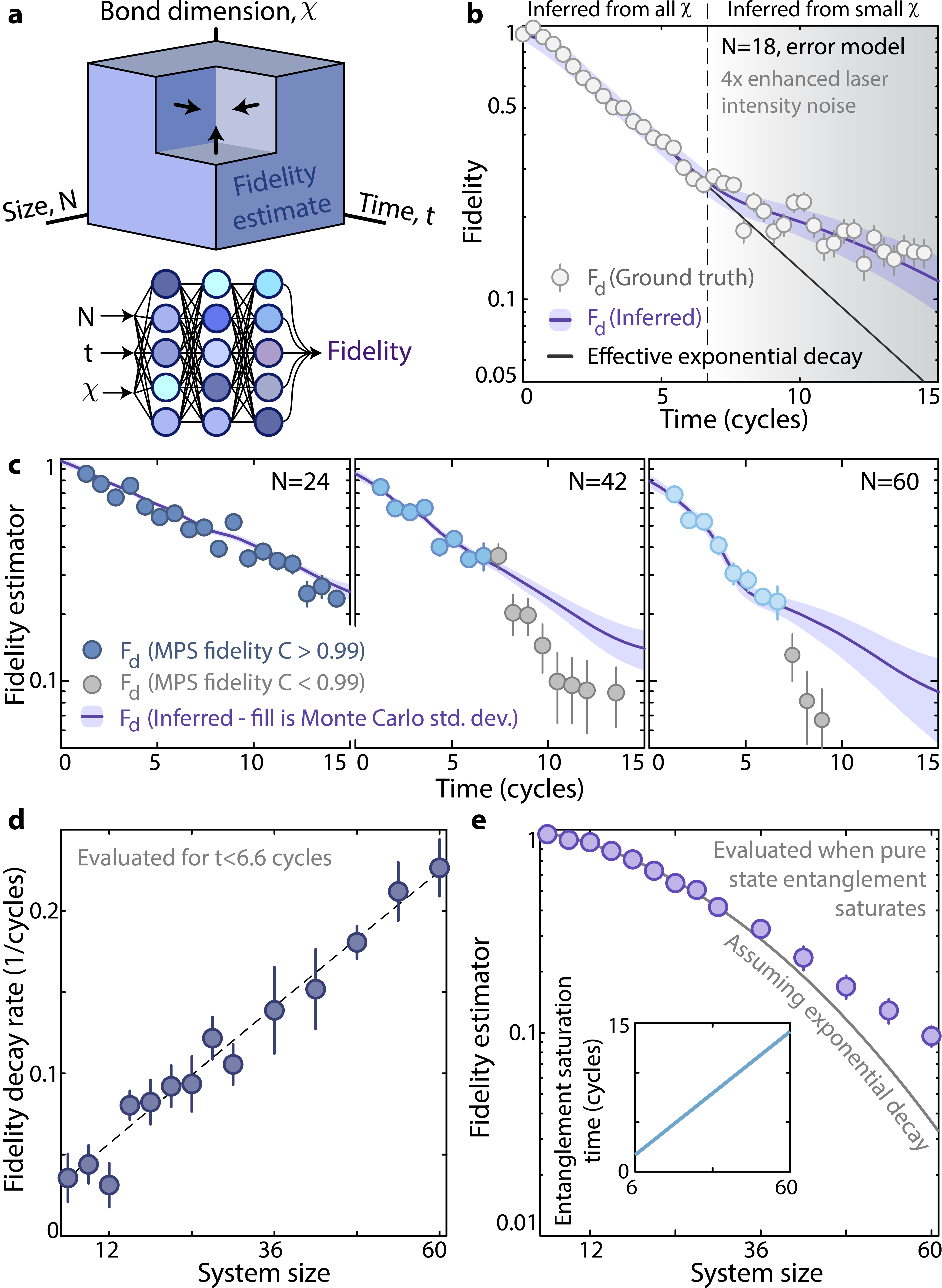}
	\caption{\textbf{Fidelity-benchmarking a 60-atom system. a.} We employ a Monte Carlo inference approach to extrapolate the fidelity at large system sizes and long evolution times. Specifically, we train 1500 neural networks, each instantiated with randomized (hyper)parameters, to predict $F_d$ as a function of size, time, and bond dimension, and take the ensemble average as the predicted value. \textbf{b.} We test this procedure using error model simulations from $N=8$ to 18 with increased laser intensity noise to emulate the fidelity expected for the experimental $N=60$ dataset. For $t>6.6$ cycles and $N>15$, we only train on bond dimensions below the level necessary for exact simulation in order to mimic constraints at large system sizes. We observe two behaviors: 1) the ensemble prediction is consistent with the ground truth, and 2) the fidelity appears to follow a non-exponential form. See Methods for further crosschecks, as well as analytic evidence for the origin of the non-exponential decay due to non-Markovian noise. \textbf{c.} Experimental fidelities for $N$ up to 60; markers are grayscale where the classical fidelity (with $\chi=3072$) is less than 0.99. \textbf{d.} Early-time fidelity decay rate as a function of system size, consistent with linear system size scaling. \textbf{e.} Fidelity at the time (inset) where the pure state entanglement saturates, with $F_d=0.095(11)$ at $N=60$; the error bar is the standard error over Monte Carlo inferences added in quadrature with the underlying sampling error.
        } 
	\vspace{-0.5cm}
	\label{Fig3}
\end{figure}

However, the situation changes when using an \textit{approximate} classical simulation. Now, the classical fidelity begins to decay (Fig.~\ref{Fig2}d, blue) after the time, $t_\mathrm{ex}$, when the ideal entanglement exceeds the limit set by the bond dimension (Fig.~\ref{Fig2}c, blue), meaning the classical simulation is no longer a faithful reference of the ideal dynamics. Most importantly, we find that after $t_\mathrm{ex}$ the experimental benchmarked fidelity \textit{also deviates downwards} (Fig.~\ref{Fig2}e, blue), indicating that $F_d$ no longer accurately estimates the fidelity to the ideal state. For the largest system sizes (for instance, $N=60$ in Fig.~\ref{Fig2}c-e, right), $t_\mathrm{ex}$ occurs well before the entanglement is predicted to saturate, even for the largest bond dimension we can realistically use. We estimate the classical fidelity in this case using the product of MPS truncation errors~\cite{Zhou2020WhatComputers}, which we find is accurate in the regime in which we operate (Ext. Data Fig.~\ref{EFig:f_trun}).

Essentially, $F_d$ appears to be an amalgam of both classical and quantum fidelities, only estimating the quantum fidelity to the ideal state in the limit of the classical simulation being perfect. To test this behavior for all system sizes, we study the benchmarked value of $F_d$ averaged over all experimental times (Fig.~\ref{Fig2}f). Consistently we see for too small of a bond dimension (open markers), $F_d$ is reduced. In some cases the requirement that $p(z)/p_\text{avg}(z)$ follow a Porter-Thomas distribution can be violated, resulting in $F_d$ even becoming unphysically negative. As bond dimension increases, $F_d$ rises, before reaching a \textit{saturation bond dimension}, $\chi_0(N,t)$, which depends on system size and time (closed markers). For the largest system sizes and times, however, the saturation bond dimension is beyond the capabilities of any current classical hardware~\cite{Hauru2021SimulationEvolution}.

\begin{figure*}[t!]
	\centering
	\includegraphics[width=181mm]{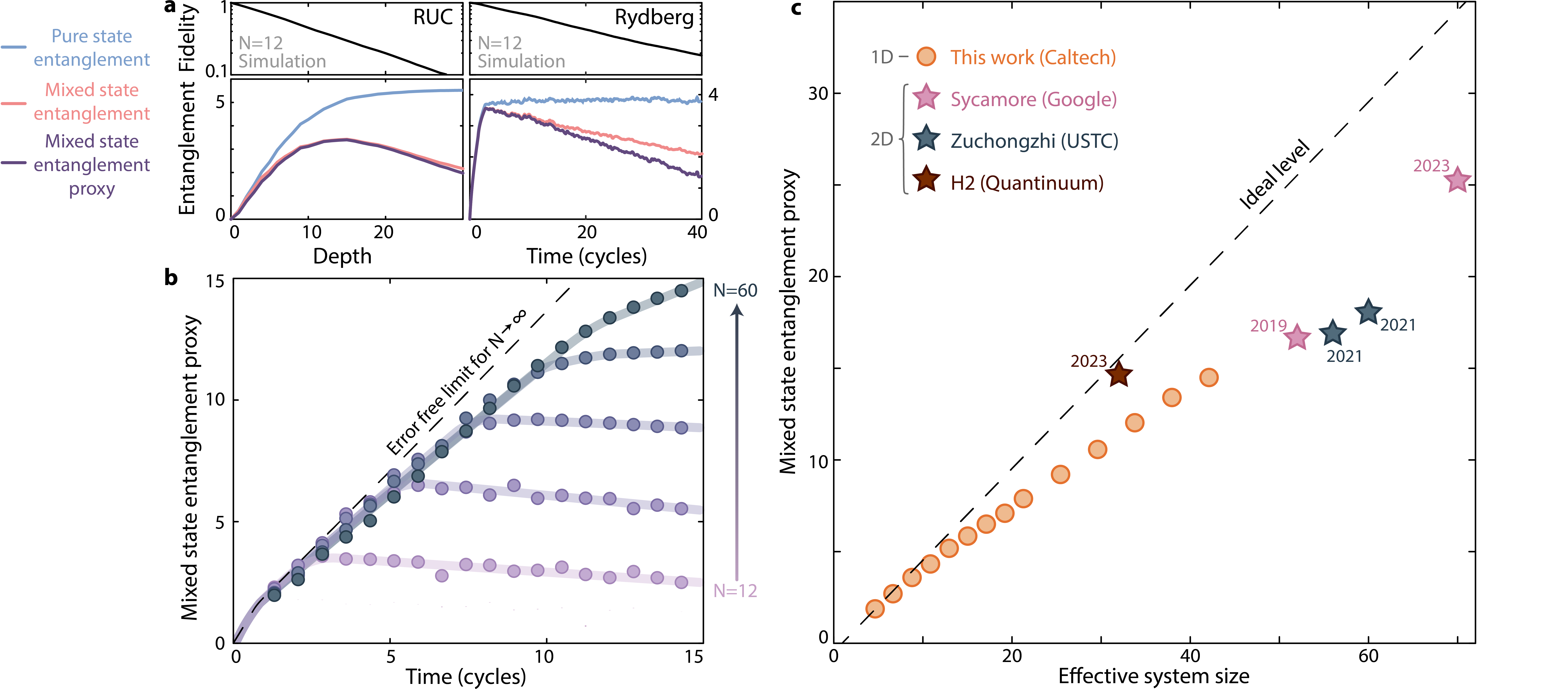}
	\caption{\textbf{Experimental mixed state entanglement. a.} We develop an experimentally measurable proxy that lower-bounds the log negativity, which is a measure of mixed state entanglement. Here we demonstrate this proxy with error model simulations of random unitary circuit and Rydberg evolution. \textbf{b.} The experimental mixed state entanglement-proxy; solid lines are guides to the eye. \textbf{c.} The maximum entanglement-proxy for our experiment can be compared against that of literature examples performing global fidelity estimation with digital quantum processors: Sycamore~\cite{Arute2019QuantumProcessor,Morvan2023PhaseSampling}, Zuchongzhi~\cite{Wu2021StrongProcessor,Zhu2022QuantumSampling}, and H2~\cite{Moses2023AProcessor} (text indicates release year). For literature examples, the x-axis is the number of qubits, while for our experiment the effective system size is defined as the number of qubits with the same Hilbert space dimension as our experiment under the Rydberg blockade constraint (Methods), and is for instance ${\sim}42$ at $N=60$.
 }
	\vspace{-0.5cm}
	\label{Fig4}
\end{figure*}

If the noise affecting the system was purely Markovian then the fidelity would decay exponentially~\cite{Dalzell2021RandomNoise} and it would be possible to measure the fidelity at early times before $t_\mathrm{ex}$ to learn the exponential decay rate, and then extrapolate in time to estimate the late-time fidelity. Indeed, we note this is an advantage of the $F_d$ metric we employ here, because it accurately estimates the fidelity earlier than other estimators like the standard linear cross-entropy~\cite{Choi2023PreparingChaos,Mark2023BenchmarkingLett}. However, extrapolating to late times is non-trivial in our case due to non-Markovian noise sources often affecting analog quantum systems. In particular, with analytic and numerical analysis we show that shot-to-shot Hamiltonian parameter fluctuations (e.g. laser intensity variations) induce \textit{sub-exponential fidelity decay} at low fidelities (Methods, Theorem~\ref{thm:gaussian_fidelity}, Ext. Data Fig.~\ref{EFig:nonexp_decay}).

Instead, we use a model-agnostic extrapolation by leveraging a large amount of data with three independent parameters: evolution time, system size, and bond dimension normalized by its saturation value (Fig.~\ref{Fig3}a, Methods). We can calculate $F_d$ in seven of the octants of this parameter space - the only outlier is the high-entanglement regime of interest. We thus use a Monte Carlo inference approach by training an ensemble~\cite{Ganaie2022EnsembleReview} of initially randomized neural networks to predict $F_d$ given an input $N$, $\chi$, and $t$; $F_d$ at large system sizes and long evolution times is then estimated as the ensemble average when $\chi\rightarrow\chi_0$ (Ext. Data Fig.~\ref{EFig:neural_net}). We emphasize that essentially we are simply performing curve fitting of the smoothly varying function $F_d(N,\chi,t)$, for which we can directly simulate many ground truth data. 

We check that this protocol consistently reproduces fidelities at small system sizes, does not appear to overfit the experiment (Ext. Data Fig.~\ref{EFig:validation}), is insensitive to hyperparameters such as the neural net topology and size, and that predictions are converged as a function of the bond dimension (Ext. Data Fig.~\ref{EFig:learnable}). We further reaffirm that our method extrapolates correctly by replicating our entire procedure in a smaller scale wherein the quantum device is replaced by numerical error model simulations up to $N=18$ atoms (Methods). For $t>6.6$ cycles and $N>15$, the training data only consists of low bond dimensions to emulate the limitations of the large-$N$ experimental data. Even still, the extrapolated fidelity is in excellent agreement with the ground truth data (Fig.~\ref{Fig3}b, Ext. Data Fig.~\ref{EFig:validation}), and reproduces the sub-exponential fidelity decay predicted analytically (Theorem~\ref{thm:gaussian_fidelity}).

Ultimately, we apply Monte Carlo inference to the full experimental dataset for system sizes up to $N=60$ atoms (Fig.~\ref{Fig3}c; see Ext. Data Figs.~\ref{EFig:alldata} and~\ref{EFig:alldata_vschi} for all data). At high fidelities (${\gtrsim}0.2$), we observe nearly exponential decay, with a rate scaling linearly with system size (Fig.~\ref{Fig3}d). At low fidelity, however, the Monte Carlo prediction again reproduces the expected sub-exponential response.  We estimate the fidelity to produce the target state when the entanglement is expected to saturate (Fig.~\ref{Fig3}e), yielding $F_d=0.095(11)$ at $N=60$.

To our knowledge, this is the first global fidelity estimation in the classically-inexact regime with an analog quantum simulator, and it furthermore showcases benchmarking a quantum device by extrapolating from approximate classical simulations. We expect this approach to be scalable; by studying the convergence of predicted fidelities as a function of bond dimension, our approach appears feasible for up to an order-of-magnitude more atoms than we employ here (Ext. Data Fig.~\ref{EFig:montecarloscale}).

\vspace{0.25cm}
\noindent\textbf{Experimental mixed state entanglement}\newline
Having benchmarked the fidelity of our Rydberg quantum simulator, we now turn to investigate the actual half-chain bipartite entanglement content of the experiment. In the past, several studies have investigated entanglement properties of (nearly) pure states by estimating the second R\'enyi entropy in (sub-)systems up to 10 particles~\cite{Islam2015MeasuringSystem,Kaufman2016QuantumSystem,Linke2018MeasuringComputer,Brydges2019ProbingMeasurements}. However, the actual output of an experiment can be a highly mixed state, with markedly different entanglement content than the target pure state. For this reason, it is desirable to directly quantify \textit{mixed state entanglement} measures. Unfortunately, extensions of most pure state entanglement measures to the case of mixed states are defined variationally, and as such are incalculable for even moderately sized systems~\cite{Plenio2005LogarithmicConvex}.

\begin{figure}[t!]
	\centering
	\includegraphics[width=86mm]{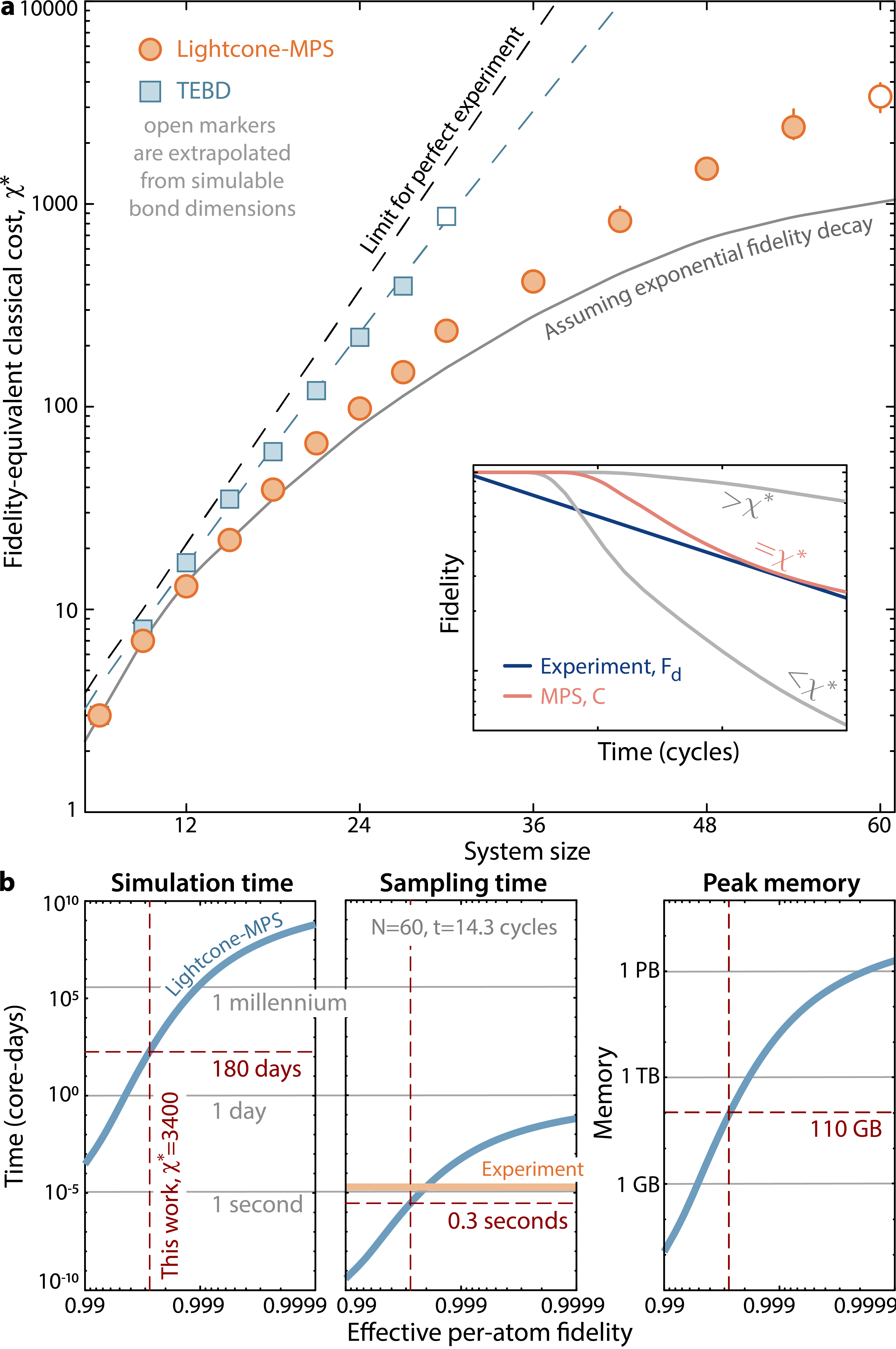}
	\caption{\textbf{Classical cost to simulate the experiment. a.} The equivalent classical cost of the experiment, as quantified by the minimum bond dimension, $\chi^*$, for the classical simulation to maintain a higher fidelity than the experiment across all times, i.e. for $C>F_d$ (inset). We consider several classical algorithms (for example, time-evolving block decimation, TEBD~\cite{Vidal2004EfficientSystems}), all of which become impractical at moderate system sizes. This necessitates the introduction of our Lightcone-MPS algorithm (Methods), which reaches a maximum value of $\chi^*=3400$ for $N=60$. \textbf{b.} Predicted MPS costs (simulation time, sampling time, and peak memory usage) to operate at $\chi^*$ as a function of the experimental per-atom fidelity (see text). Times are representative of a single 16-core node on the Caltech cluster (Methods).
 } 
	\vspace{-0.5cm}
	\label{Fig5}
\end{figure}

An alternative, computable measure of mixed state entanglement is the log negativity~\cite{Vidal2002ComputableEntanglement}, $\mathcal{E}_N$, which is an upper bound to the distillable entanglement of the system~\cite{Plenio2005LogarithmicConvex}. However, measuring the value of the negativity na\"ively requires tomography of the full system density matrix, which is infeasible even for intermediate scale quantum systems~\cite{ODonnell2016EfficientTomography,Haah2017Sample-OptimalStates}. In the past, experiments have been limited to demonstrating necessary conditions for a non-vanishing negativity, which can only reveal the binary presence of mixed state entanglement~\cite{Elben2020Mixed-StateMeasurements,Mooney2021Whole-DeviceComputer}.

Here we derive and demonstrate an entanglement-proxy, $\mathcal{E}_P$, which can lower-bound the extensive mixed state entanglement (quantified by log negativity). For a mixed state, $\hat{\rho}$, with fidelity, $F$, to a target pure state, $|\psi\rangle$, with known entanglement, $\mathcal{E}_N(|\psi\rangle)$, our mixed state entanglement-proxy is
\begin{align}
\mathcal{E}_P(\hat{\rho})\equiv \mathcal{E}_N(|\psi\rangle) + \log_2(F).
\label{eqn:ln_lower_bound}
\end{align}

Intuitively, $\mathcal{E}_P$ is a proxy evaluating the competition between the growth of the error-free entanglement, $\mathcal{E}_N(|\psi\rangle)$, versus the error-sensitive fidelity, as $F<1$ reduces the mixed state entanglement. When $\hat{\rho}$ is an isotropic state (an admixture of a maximally entangled state and a maximally mixed state), it has been shown~\cite{Vidal2002ComputableEntanglement,Lee2003Convex-roofSystems} that $\mathcal{E}_N(\hat{\rho})=\mathrm{max}(\mathcal{E}_P(\hat{\rho}),0)$ at large system sizes. Further, we show the same holds for a Haar-random state admixed with a maximally mixed state -- the expected output~\cite{Dalzell2021RandomNoise} of deep noisy random unitary circuits (RUCs) -- as long as the fidelity is large compared to the inverse of the half-chain Hilbert space dimension (Methods). 

More generally, we prove $\mathcal{E}_P$ is a lower bound for $\mathcal{E}_N$ for \textit{any} mixed state assuming $|\psi\rangle$ is the highest fidelity state to $\hat{\rho}$, and becomes tighter as the system size increases (Ext. Data Fig.~\ref{EFig:negativity_bounds}). Importantly, violations of this assumption can only lead to small violations of our bound in the worst case for physically realistic conditions with local or quasi-static errors, as we show with both analytic (Theorems~\ref{thm:commutator_negativity_bound} and~\ref{thm:fidelity_negativity_bound}) and numeric (Ext. Data Figs.~\ref{Efig:errors_of_negativity_rydberg} and~\ref{Efig:errors_of_negativity}) support in the Methods. 

We demonstrate the efficacy of $\mathcal{E}_P$ on both noisy RUC evolution and error model simulation of our Rydberg dynamics (Fig.~\ref{Fig4}a, Methods). In both cases, the target pure state log negativity increases and saturates, while the exactly calculated mixed state log negativity reaches a maximum before decaying at late times, behavior which the entanglement-proxy $\mathcal{E}_P$ replicates as a lower bound.

We then plot the experimental entanglement-proxy (Fig.~\ref{Fig4}b), where $\mathcal{E}_N(|\psi\rangle)$ is extrapolated from small system sizes (Ext. Data Fig.~\ref{EFig:entanglement_growth}), and $F$ is found from Monte Carlo inference. We observe the entanglement proxy peaks before falling at late times; this peak value increases (Fig.~\ref{Fig4}c) as a function of effective system size defined as the number of qubits with the same Hilbert space dimension as our experiment under the Rydberg blockade constraint (${\sim}42$ for $N{=}60$).

Importantly, with Eq.~(\ref{eqn:ln_lower_bound}) we can directly compare the results of our present study against RUC evolution in state-of-the-art digital quantum devices~\cite{Arute2019QuantumProcessor,Wu2021StrongProcessor,Zhu2022QuantumSampling,Morvan2023PhaseSampling,Moses2023AProcessor} (Fig.~\ref{Fig4}c). We find we are within ${\sim}2$ \textit{ebits} of early tests of quantum advantage~\cite{Arute2019QuantumProcessor} (an ebit is the entanglement of a two-qubit Bell state). For literature examples, we assume targeted states are Haar-random~\cite{Bhosale2012EntanglementStatistics,Datta2010NegativityStates}, while for our experiment we conservatively use the extrapolated log negativity, which is ${\sim}2$ ebits below the expectation for Haar-random states at the largest system sizes (Ext. Data Fig.~\ref{EFig:entanglement_growth}). 

The mixed-state entanglement-proxy $\mathcal{E}_P$ can serve as a useful quality-factor of the ability for different experiments to produce highly entangled states, including preparation methods besides quench evolution such as quasi-adiabatic ground state preparation (Ext. Data Figs.~\ref{Efig:full_process} and~\ref{EFig:sweep_entanglement}), and could be a more widely applicable alternative to other measures, such as quantum volume~\cite{Cross2019ValidatingCircuits}, for directing efforts to improve NISQ-era quantum systems.

\vspace{0.25cm}
\noindent\textbf{The classical cost of quantum simulation}\newline
We finally ask: which device, quantum or classical, has a higher fidelity of reproducing a high-entanglement pure target state of interest? Equivalently, in terms of fidelity, what are the minimum classical resources required for a classical computer to outperform the quantum device?

To answer this, we compare the fidelity of the experiment against that of the MPS with varying bond dimension. We define the critical bond dimension for a given system size, $\chi^*$, as the minimum bond dimension for which the classical fidelity always exceeds the estimated experimental fidelity. Importantly, this controls the costs of classical simulation -- for instance, MPS simulation time scales as $\mathcal{O}(N\chi^3)$. We find $\chi^*$ continually increases as a function of system size (Fig.~\ref{Fig5}a), reaching a maximum value of $\chi^*=3400$ for $N=60$ (Ext. Data Fig.~\ref{EFig:chi_extrap}), and apparently continuing to increase beyond that point. 

In performing this study, we used our new Lightcone-MPS algorithm, but considered several alternative approximate classical algorithms, including path integral~\cite{Arute2019QuantumProcessor}, matrix product operator~\cite{Noh2020EfficientDimension}, time-dependent variational principle~\cite{Haegeman2011Time-DependentLattices}, Schrieffer-Wolff transformation~\cite{Bluvstein2021ControllingArrays}, and neural net~\cite{Gutierrez2022RealStates} approaches (Methods); however, we found the equivalent classical cost of these methods quickly became infeasible, typically well before $N=60$. As an example, we show $\chi^*$ for a more conventional MPS approach using time-evolving block decimation~\cite{Vidal2004EfficientSystems} (Fig.~\ref{Fig5}a).

All calculations used a single 16-core node of the Caltech central computing cluster (Methods). On this machine, we estimate that running the Lightcone-MPS simulation for $N=60$ and $\chi^*=3400$ would entail a peak memory usage of ${\sim}110$ GB (scaling as $\mathcal{O}(N\chi^2)$), and would take ${\sim}$11.3 days, or $11.3{\times}16\approx180$ core-days; sampling from the resultant MPS would take ${\sim}0.3$ core-seconds per sample (scaling as $\mathcal{O}(N\chi^2)$). For comparison, the experimental cycle time is ${\sim}1.7$ s, limited by array loading and imaging; the actual quantum simulation time is only ${\sim}1\, \mu$s per shot. While the classical computer can utilize multiple cores, so too can the experiment be parallelized over multiple atom-array chains simultaneously, which in fact we do already at small system sizes.

We predict these classical costs are highly-sensitive to the effective per-atom fidelity, $\mathcal{F}$, defined by $\mathcal{F}^{N t}\equiv F(N,t)$ (Fig.~\ref{Fig5}b, Methods). For instance, the simulation time scales as ${\sim}(1{-}{\mathcal{F})^{-10}}$ around the experimental $\mathcal{F}$. While specialized classical hardware~\cite{Hauru2021SimulationEvolution,Ganahl2023DensityUnits,Haner20170.5Circuit} may more readily perform the present approximate classical simulations, we thus expect small improvements in the quantum fidelity may soon make the experiment out of reach of even these more advanced classical systems.

\vspace{0.25cm}
\noindent\textbf{Outlook}\newline
As quantum systems tackle tasks of rising complexity, it is increasingly important to understand their ability to produce states in the highly entangled, beyond-classically-exact regime. Here we have studied this regime directly by measuring the global fidelity of an analog quantum simulator with up to 60 atoms, the first such demonstration to our knowledge.

A careful analysis (Ext. Data Fig.~\ref{EFig:montecarloscale}) indicates that with reasonable classical resources, our Monte Carlo inference protocol is scalable to an order-of-magnitude larger system sizes than were studied here, potentially enabling fidelity estimation for system sizes with $N\sim500$. It is also applicable for digital devices~\cite{Arute2019QuantumProcessor,Wu2021StrongProcessor,Zhu2022QuantumSampling,Morvan2023PhaseSampling,Moses2023AProcessor} that are affected by non-Markovian noises like control errors~\cite{Arute2019QuantumProcessor}, which could then lead to non-exponential scaling of global fidelities in certain parameter regimes. Additionally, it could be applied to analog quantum simulators for itinerant particles~\cite{Mark2023BenchmarkingLett,Daley2022PracticalSimulation,Gross2017QuantumLattices,Young2023AnSampler}. Further, one may imagine applying the same basic technique to cross-platform comparisons~\cite{Elben2020Cross-PlatformDevices,Zhu2022Cross-platformStates,Carrasco2021TheoreticalVerification} between erroneous quantum devices by varying the decoherence of each, a form of zero-noise extrapolation~\cite{Li2017EfficientMinimization,Temme2017ErrorCircuits}.

In addition, we have addressed a longstanding problem by introducing a simple proxy of the experimental mixed state entanglement. This entanglement-proxy can serve as a universal quality-factor comparable amongst analog and digital quantum devices as a guide for improving future systems, and may act as a probe for detecting topological order~\cite{Lee2013EntanglementOrder,Lu2020DetectingNegativity} and measurement-induced criticality~\cite{Sang2021EntanglementCriticality}.

Finally, we have studied the equivalent classical cost of our experiment on the level of global fidelity, which we note could be greatly increased through the use of erasure conversion~\cite{Wu2022ErasureArrays,Scholl2023ErasureSimulator,Ma2023High-fidelityQubit}. Similar techniques could be applied to quantify the classical cost of measuring physical observables~\cite{Trivedi2022QuantumSimulators,Kechedzhi2023EffectiveExperiments}, and to benchmark the performance of approximate classical algorithms themselves through comparison to high fidelity quantum data. While here we have focused on one-dimensional systems to exploit the power of MPS representations, using higher-dimensional systems~\cite{Scholl2021QuantumAtoms,Ebadi2021QuantumSimulator}, while maintaining high fidelities, may prove even more difficult for classical algorithms. We emphasize that in contrast to many previous experiments~\cite{Arute2019QuantumProcessor,Wu2021StrongProcessor,Zhu2022QuantumSampling,Morvan2023PhaseSampling} which explicitly targeted spatiotemporally complex quantum evolution when exploring the limits of classical simulation, here the dynamics we have studied are one-dimensional and both space- and time-independent, yet still begin to reach a regime of classical intractability. Ultimately, our results showcase the present and potential computational power of analog quantum simulators, encouraging an auspicious future for these platforms~\cite{Daley2022PracticalSimulation}.

\vspace{0.25cm}
\noindent\textbf{Acknowledgements}\newline
We thank John Preskill, Garnet Chan, Hsin-Yuan Huang, Miles Stoudenmire, Alexander Baumgärtner, Gyohei Nomura, Elie Bataille, Kon Leung, and Richard Tsai for their feedback on this work. We acknowledge support from the NSF QLCI program (2016245), the DOE (DE-SC0021951), the Institute for Quantum Information and Matter, an NSF Physics Frontiers Center (NSF Grant PHY-1733907), the DARPA ONISQ program (W911NF2010021), the AFOSR YIP (FA9550-19-1-0044), Army Research Office MURI program (W911NF2010136), NSF CAREER award 2237244 and NSF CAREER award 1753386. Support is also acknowledged from the U.S. Department of Energy, Office of Science, National Quantum Information Science Research Centers, Quantum Systems Accelerator. ZC acknowledges the DARPA 134371-5113608 award and NSF 10434. PS acknowledges support from the IQIM postdoctoral fellowship. RF acknowledges support from the Troesh postdoctoral fellowship. AE\ acknowledges funding by the German National Academy of Sciences Leopoldina under the grant number LPDS 2021-02 and by the Walter Burke Institute for Theoretical Physics at Caltech.

\vspace{0.25cm}
\noindent\textbf{Author contributions}\newline
A.L.S., P.S., and R.F. performed the experiments. A.L.S., Z.C., and J.C. performed the data analysis. A.L.S., Z.C., J.C., D.K.M., A.E. contributed to underlying theory and associated numerics. A.L.S. and M.E. wrote the manuscript with contributions and input from all authors. S.C, and M.E. supervised this project.

\vspace{0.25cm}
\noindent\textbf{Data and code availability}\newline
The data and codes supporting this study are available from the corresponding author upon reasonable request.
\\
\\

\FloatBarrier
\bibliography{references.bib}
\bibliographystyle{adamref}

\FloatBarrier

\end{document}


\title{Methods: Benchmarking highly entangled states on a 60-atom analog quantum simulator}
\author{Adam L. Shaw,$^{1,*,\dag}$ Zhuo Chen,$^{2,3,*}$ Joonhee Choi,$^{1,4,*}$ Daniel K. Mark,$^{2,*}$ Pascal Scholl,$^1$ \\Ran Finkelstein,$^1$  Andreas Elben,$^1$ Soonwon Choi,$^{2,\dag}$ and Manuel Endres$^{1,\dag}$}
\noaffiliation
\affiliation{\Caltech}
\affiliation{\MIT}
\affiliation{\IAIFI}
\affiliation{\Stanford}

\maketitle

\setcounter{figure}{0}
\captionsetup[figure]{labelfont={bf},name={Ext. Data Fig.},labelsep=bar,justification=raggedright,font=small}

\captionsetup[table]{labelfont={bf},name={Ext. Data Table},labelsep=bar,justification=raggedright,font=small}
\FloatBarrier
\vspace{-5mm}
\tableofcontents

\footnote{\label{a}$^*$ These authors contributed equally}
\footnote{\label{b}$^{\dag}$ ashaw@caltech.edu, soonwon@mit.edu, mendres@caltech.edu}

\clearpage
\begin{figure}[t!]
	\centering
	\includegraphics[width=83mm]{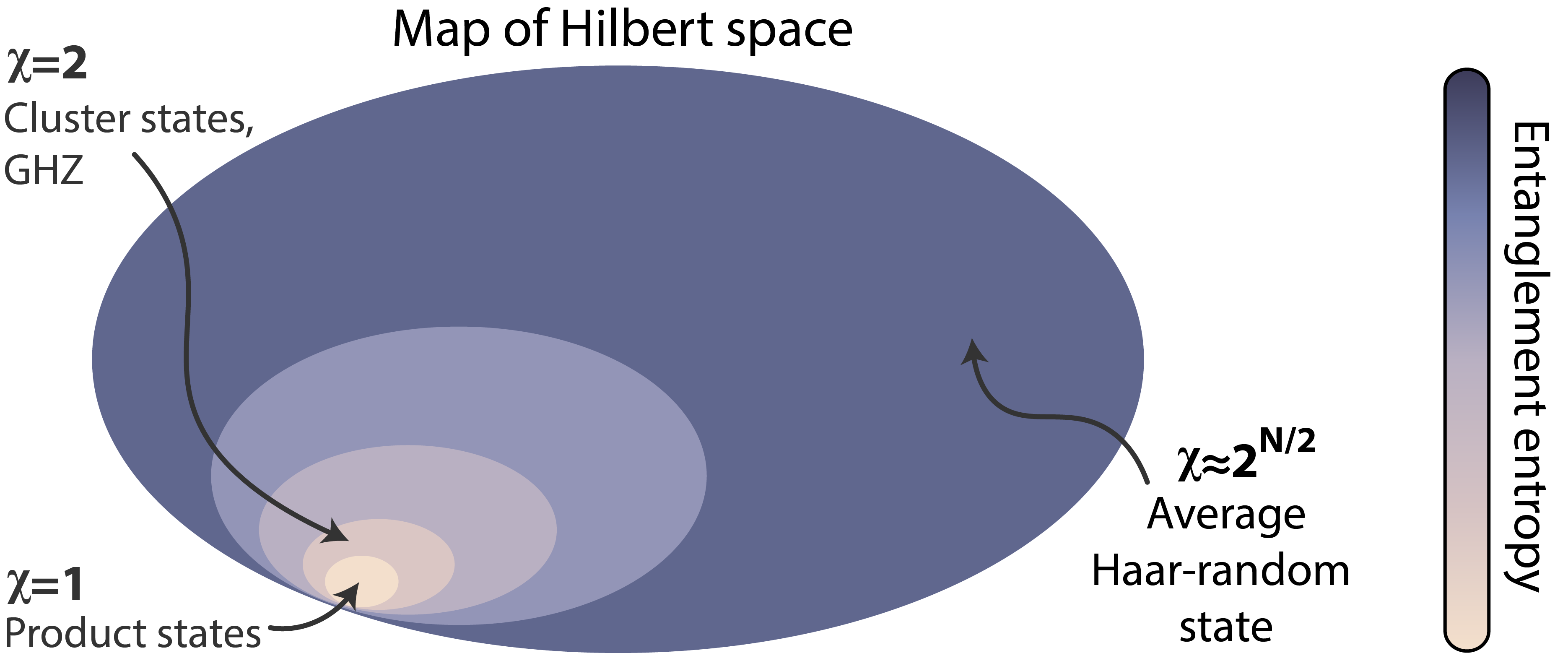}
	\caption{\textbf{Map of the relevant regions of Hilbert space.} For any intermediate or large system size, $N$, an average Haar random state in the Hilbert space will have nearly maximal entanglement entropy (up to the Page correction~\cite{Page1993AverageSubsystem}), and correspondingly is described by an MPS with approximately maximal bond dimension. By contrast, states such as GHZ state and cluster states -- which are colloquially referred to as maximally-entangled -- actually have very little extensive entanglement entropy, and can be written in an efficient MPS representation with only $\chi=2$.
        } 
	\vspace{-0.0cm}
	\label{EFig:hilbert_map}
\end{figure}

\begin{figure*}[ht!]
	\centering
	\includegraphics[width=181mm]{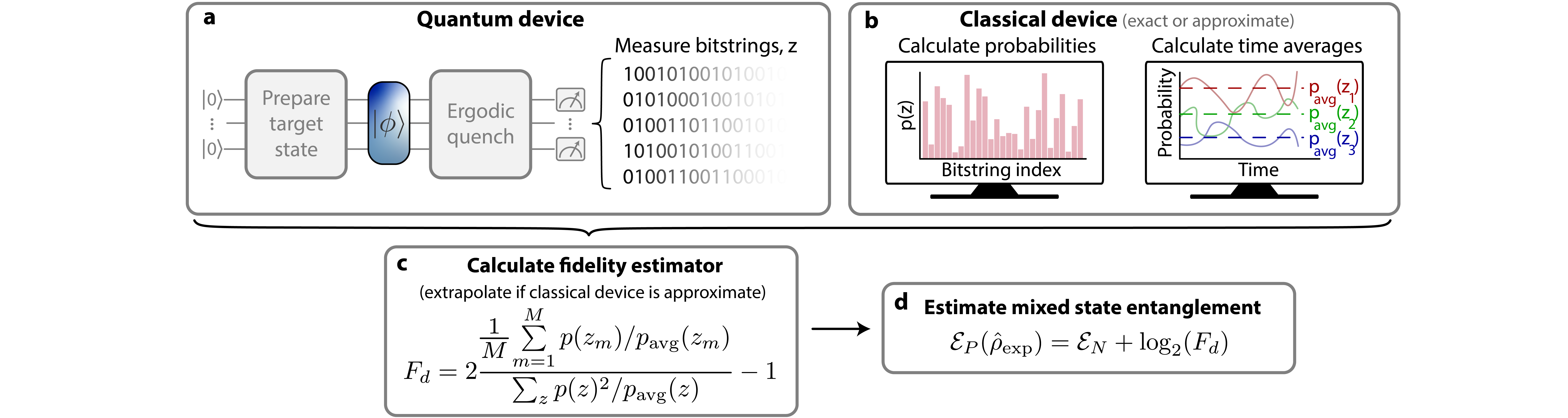}
	\caption{\textbf{Schematic for mixed state entanglement estimation of a target quantum state. a.} The quantum device prepares a target state~\cite{Choi2023PreparingChaos} of interest, $|\psi\rangle$, before performing an ergodic quench, and measuring a set of resulting bitstrings. The experimental fidelity drops both during state preparation and the quench due to errors (not pictured). \textbf{b.} A classical computer then performs noise-free simulations of the entire dynamics (either exactly or approximately), calculates both the bitstring probabilities and the time-averaged bitstring probabilities, $p_\text{avg}$, and further estimates the second moment of the bitstring probability distribution. \textbf{c.} The fidelity is then estimated via a cross-entropy type quantity~\cite{Choi2023PreparingChaos,Mark2023BenchmarkingLett,Zhang2023AMetamaterial}, $F_d$. If the classical simulation were approximate, $F_d$ is predicted either via Monte Carlo inference or direct extrapolation. \textbf{d.} The negativity of the target state is estimated or assumed, and the experimental mixed state entanglement-proxy is calculated.
        } 
	\vspace{-0.0cm}
	\label{Efig:full_process}
\end{figure*}
\newpage
\FloatBarrier
\subsection{Description of the Experiment}
\subsubsection{State preparation and readout}
\begin{figure*}[h!]
	\centering
        \vspace{-0.5cm}
	\includegraphics[width=181mm]{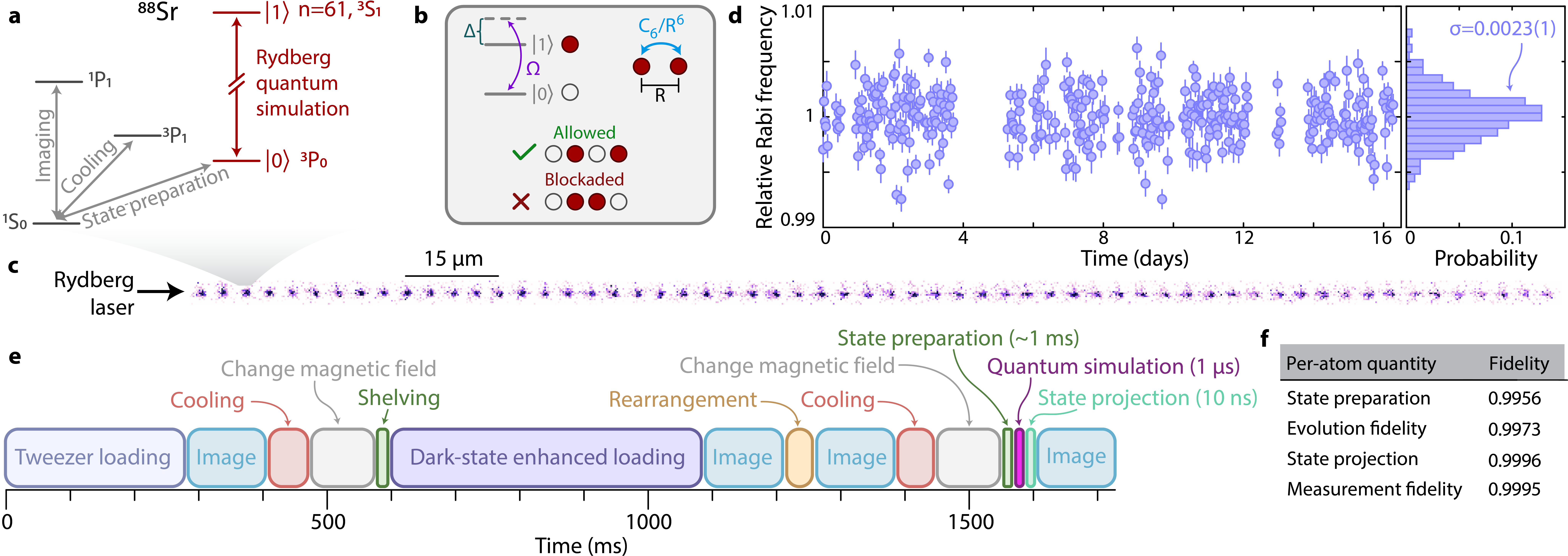}
	\caption{\textbf{Description of the experiment. a.} We use a Rydberg quantum simulator, based on trapping an array of single atoms of bosonic strontium-88. We treat the metastable $5s5p$ $^{3}P_{0}$ state as an effective ground state, and consider dynamics in the Rydberg manifold. State projection is performed by autoionizing the Rydberg state~\cite{Madjarov2020High-fidelityAtoms} (not shown). \textbf{b.} The Rydberg Hamiltonian is Ising-like, with single particle terms characterized by a detuning, $\Delta$, and Rabi frequency, $\Omega$, between the two qubit levels. Rydberg atoms are strongly interacting, characterized by the $C_6$ coefficient, and with a strength falling off approximately as $1/R^6$, where $R$ is the interatom distance. When two atoms are close enough together, simultaneous excitation to the Rydberg state is strongly suppressed, causing the so-called Rydberg blockade effect. We operate in a regime in which the nearest neighbor interaction strength is ${\sim}13\times$ the Rabi frequency, which is strongly blockaded (see also Ext. Data Fig.~\ref{EFig:blockaded}d). \textbf{c.} We study one-dimensional chains of atoms (fluorescence image), with the Rydberg excitation laser aligned along the longitudinal array axis to minimize inhomogeneity. \textbf{d.} A classical control system~\cite{Choi2023PreparingChaos} continuously interleaves data-taking with automated atom-based calibration of Hamiltonian parameters, resulting in highly stable operation over the course of multiple weeks. For example, the Rabi frequency is held constant with a relative standard deviation of only $0.0023(1)$. \textbf{e.} A single run of the experiment takes ${\sim}1.7$ s, almost entirely limited by array loading and imaging. The actual time required for quantum evolution is ${\sim}1\ \mu$s. We use dark-state enhanced loading~\cite{Shaw2023Dark-StateArray} to reach the largest array sizes used in this work. \textbf{f.} Representative per-atom fidelities; the reported per-atom evolution fidelity is defined as in Fig.~\ref{Fig5}b of the main text.
        } 
	\vspace{-0.0cm}
	\label{EFig:experiment}
\end{figure*}

Our experiment is a Rydberg atom array quantum simulator~\cite{Choi2023PreparingChaos,Browaeys2020Many-bodyAtoms} trapping individual strontium-88 atoms in optical tweezers~\cite{Cooper2018Alkaline-EarthTweezers,Norcia2018MicroscopicArrays}; see also Ext. Data Fig.~\ref{EFig:experiment}. Up-to-date details of our apparatus may be found in  previous works~\cite{Scholl2023ErasureSimulator,Choi2023PreparingChaos}. In brief, we use dark-state enhanced loading~\cite{Shaw2023Dark-StateArray} to load a 68-tweezer array with ${\sim}61$ atoms, which we then rearrange~\cite{Endres2016Atom-by-atomArrays,Barredo2016AnArrays} into defect free arrays of various sizes, with the array spacing calibrated with a laser-based ruler~\cite{Shaw2023Multi-ensembleMovements}. Atoms are initially in the $5s^2$ $^{1}S_{0}$ state, and are cooled on the narrow-line $5s^2$ $^{1}S_{0}$ $\leftrightarrow$ $5s5p$ $^{3}P_{1}$ transition close to their motional ground state. Atoms are prepared into the long-lived $5s5p$ $^{3}P_{0}$ \textit{clock state} with a preparation fidelity of 0.9956(1) per atom, which we then treat as a metastable ground state, $|0\rangle$. Atoms are then driven globally to the $5s61s$ $^{3}S_{1},m_{J}{=}0$ \textit{Rydberg state}, $|1\rangle$ while the traps are briefly blinked off.

Following Hamiltonian evolution, state projection is performed by autoionizing~\cite{Madjarov2020High-fidelityAtoms} the Rydberg atoms, leaving them dark to our fluorescent imaging, with a fidelity of ${\sim}0.9996$ per atom. Atoms in the clock state are pumped into the imaging cycle, from which we map atomic fluorescence to qubit state~\cite{Covey2019A,Madjarov2020High-fidelityAtoms} with a detection fidelity ${\gtrsim}0.9995$ per atom~\cite{Covey2019A,Scholl2023ErasureSimulator}. For an initially defect-free array, this results in a series of bitstrings associated with the measured qubit states in the array; note that we discard any experimental bitstrings for which initial rearrangement failed. As we load many more atoms than are needed when benchmarking small system sizes, we simultaneously excite and detect multiple non-interacting subensembles when possible, and accrue several thousand bitstrings per system size and time (Ext. Data Fig.~\ref{EFig:sample_complexity}a, inset).

\subsubsection{The Rydberg Hamiltonian}
The Hamiltonian of our system is well approximated by
\begin{align}
\hat{H}/h=\Omega\sum_i \hat{S}_i^x -\Delta\sum_i \hat{n}_i + \frac{C_6}{a^6} \sum_{i>j} \frac{\hat{n}_i \hat{n}_j}{|i-j|^6} \label{eq:RydbergHam}
\end{align}
describing a set of interacting two-level systems, labeled by site indices $i$ and $j$, driven by a laser with Rabi frequency $\Omega$ and detuning $\Delta$. The interaction strength is determined by the $C_6$ coefficient and the lattice spacing $a$. Operators are $\hat{S}^x_i =(\ket{1}_i\bra{0}_i + \ket{0}_i \bra{1}_i)/2$ and $\hat{n}_i = \ket{1}_i\bra{1}_i$, where $\ket{0}_i$ and $\ket{1}_i$ denote the electronic ground and Rydberg states at site $i$, respectively. Hamiltonian parameters are summarized in Ext. Data Fig.~\ref{EFig:ham_pams}b, and are chosen to lead to high-temperature thermalization~\cite{Kaufman2016QuantumSystem,Abanin2019ColloquiumEntanglement}, and chaotic behavior consistent with previous studies~\cite{Choi2023PreparingChaos}. 

An important feature of our system arises from the Rydberg blockade~\cite{Browaeys2020Many-bodyAtoms} effect (Fig.~\ref{Fig2}b of the main text). Notably, the nearest neighbor interaction strength is $13\times$ larger than the next largest energy scale (the Rabi frequency), greatly reducing the probability to have simultaneous excitation of neighboring atoms to Rydberg states. To first order, this reduces the Hilbert space dimension from $2^N$ to $\text{Fib}(N+2)\approx1.62^N$, where $\text{Fib}$ is the Fibonacci function. This means that for a system size of $N$ atoms, the effective Hilbert space size is only that of ${\sim}0.7 N$ qubits, restricting the maximum entanglement built-up. 

We account for this effect in our benchmarking protocol (discussed below), and, more concretely, in Fig.~\ref{Fig4} of the main text we plot the experimental mixed state entanglement-proxy against the effective system size, $\text{Fib}(N+2)$.

\subsubsection{Maximum likelihood estimation of Hamiltonian parameters}

\begin{figure*}[h!]
	\centering
        \vspace{0.5cm}
	\includegraphics[width=181mm]{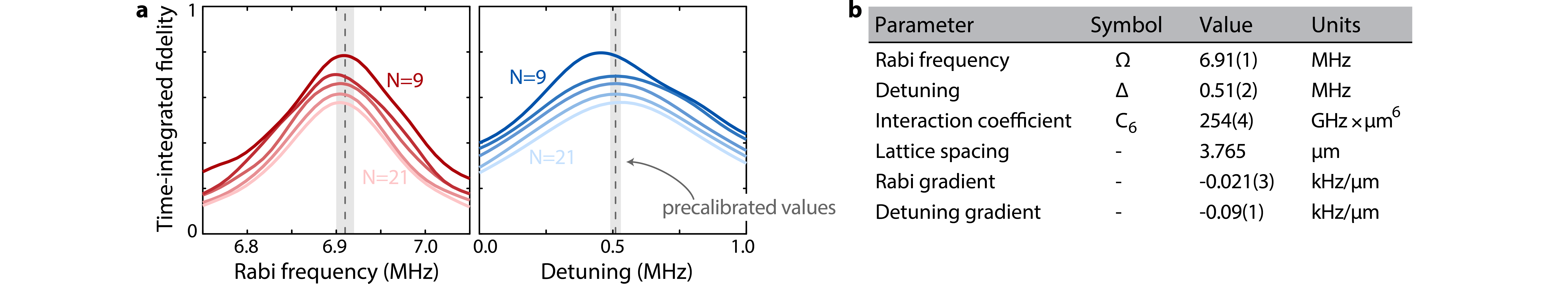}
	\caption{\textbf{Coherent Hamiltonian parameters. a.} We calibrate Hamiltonian parameters using a maximum-likelihood-estimation (MLE) technique based on varying parameters used during classical simulation, and finding the point of highest fidelity overlap with the experimental data; for further details, see Ref.~\cite{Choi2023PreparingChaos}. Fidelities are time-integrated over all experimental measurements. MLE calibrated values are consistent for system sizes up to $N=21$, as is expected as data taking for all system sizes (and times) is done in a randomized, interleaved fashion. Several smaller arrays are benchmarked simultaneously over the entire spatial extent of the largest 60-atom array to perform MLE on parameter gradients. \textbf{b.} Resultant MLE-calibrated Hamiltonian parameters (and gradients) used for all $N$ in the classical simulation.
        } 
	\vspace{-0.0cm}
	\label{EFig:ham_pams}
\end{figure*}

Hamiltonian parameters are precalibrated via \textit{ab initio} measurements, and then refined using maximum likelihood estimation (MLE) from fidelity measurements at small system sizes ($N\leq21$), as described in Ref.~\cite{Choi2023PreparingChaos}. MLE estimated parameters are insensitive to system size, so we believe they are accurate for the larger system sizes we test here. MLE results are consistent with pre-calibrated values, and tightly optimize the target Hamiltonian used in classical simulation (Ext. Data Fig.~\ref{EFig:ham_pams}a). Note that the real-space positions of atoms are taken into account when performing MLE on parameter gradients. For instance, several small islands of atoms are benchmarked over the same spatial extent as the largest 60-atom array.

The main experimental dataset was taken continuously over the course of 17 days. Parameters such as Rabi frequency, detuning, beam alignment, state preparation, and others were automatically calibrated via our home-built control architecture~\cite{Choi2023PreparingChaos}, resulting in high stability even over such a long data-taking period. The overall duty cycle of our experiment (the ratio of time spent on `science shots' versus the total wall time) was roughly 36\% over that period.

\subsubsection{Defining the experimental time unit}
Throughout this manuscript, we define the time unit in terms of `cycles', given by $t_\mathrm{cycle}=1/\Omega$, and thus concretely $t_\mathrm{cycle}\approx145$ ns. This timescale was chosen as $\Omega$ is the dominant energy-scale in the system (after the blockading nearest neighbor interaction energy). However, it is instructive to compare cycles versus a more natural timescale, namely the `unit-entanglement-time', $t_\mathrm{ebit}$, required to generate 1 ebit of entanglement in the early time linear growth regime (see Fig.~\ref{Fig2}c). For our system parameters, we numerically find this timescale is given by $t_\mathrm{ebit}\approx0.83\times t_\mathrm{cycle}\approx121$ ns. This definition is convenient, because it then allows us to directly compare our analog evolution \textit{times} against the equivalent \textit{depth} for digital circuit evolution. 

We numerically fit the early time linear entanglement growth for one-dimensional random unitary circuit (RUC) evolution, and compare against our experiment. For the gate-set of Ref.~\cite{Arute2019QuantumProcessor}, we find that $t_\mathrm{ebit}\approx1.1$ layers, meaning 1 ebit of entanglement is built up in 1.1 layers. While this value is gate-set dependent, the exact numerical values are not of too much importance, but serve to demonstrate that $t_\mathrm{cycle}$ used throughout this manuscript can be treated roughly as 1 circuit depth. For further details on comparing the evolution of analog and digital devices, see Ref.~\cite{Choi2023PreparingChaos}.

\newpage
\subsection{Fidelity estimation with a modified cross-entropy benchmark}
We employ a recently developed fidelity estimator~\cite{Mark2023BenchmarkingLett}, $F_d$, to evaluate the many-body fidelity of ergodic quench dynamics~\cite{Cotler2023EmergentFunctions} producing large entanglement. Importantly, $F_d$ can be efficiently sampled using only a small number of measurements as
\begin{align}
    F_d = 2 \frac{\frac{1}{M} \sum_{m=1}^M p(z_m) / p_\text{avg}(z_m)}{\sum_z p(z)^2 / p_\text{avg}(z)} - 1,
\end{align}
where $M$ is the number of measurements, $z_m$ is the experimentally measured bitstring at the $m^\mathrm{th}$ repetition, $p(z)$ is the theoretically calculated probability of measuring $z$ by a classical algorithm, and $p_\text{avg}(z)$ is the time-averaged probability of measuring $z$ over a certain time window. The denominator can be efficiently calculated by importance-sampling from the MPS~\cite{Ferris2012PerfectNetworks} as $ \mathbb{E}_{z\sim p(z)} p(z) / p_\text{avg}(z)$. For small system sizes where we accrue bitstrings from multiple non-interacting subarrays, we calculate $F_d$ for each individual `island' and then average their values, weighting by the number of measured bitstrings from each. This procedure is summarized schematically in Ext. Data Fig.~\ref{Efig:full_process}.

\begin{figure}[ht!]
    \vspace{5mm}
	\centering
	\includegraphics[width=86mm]{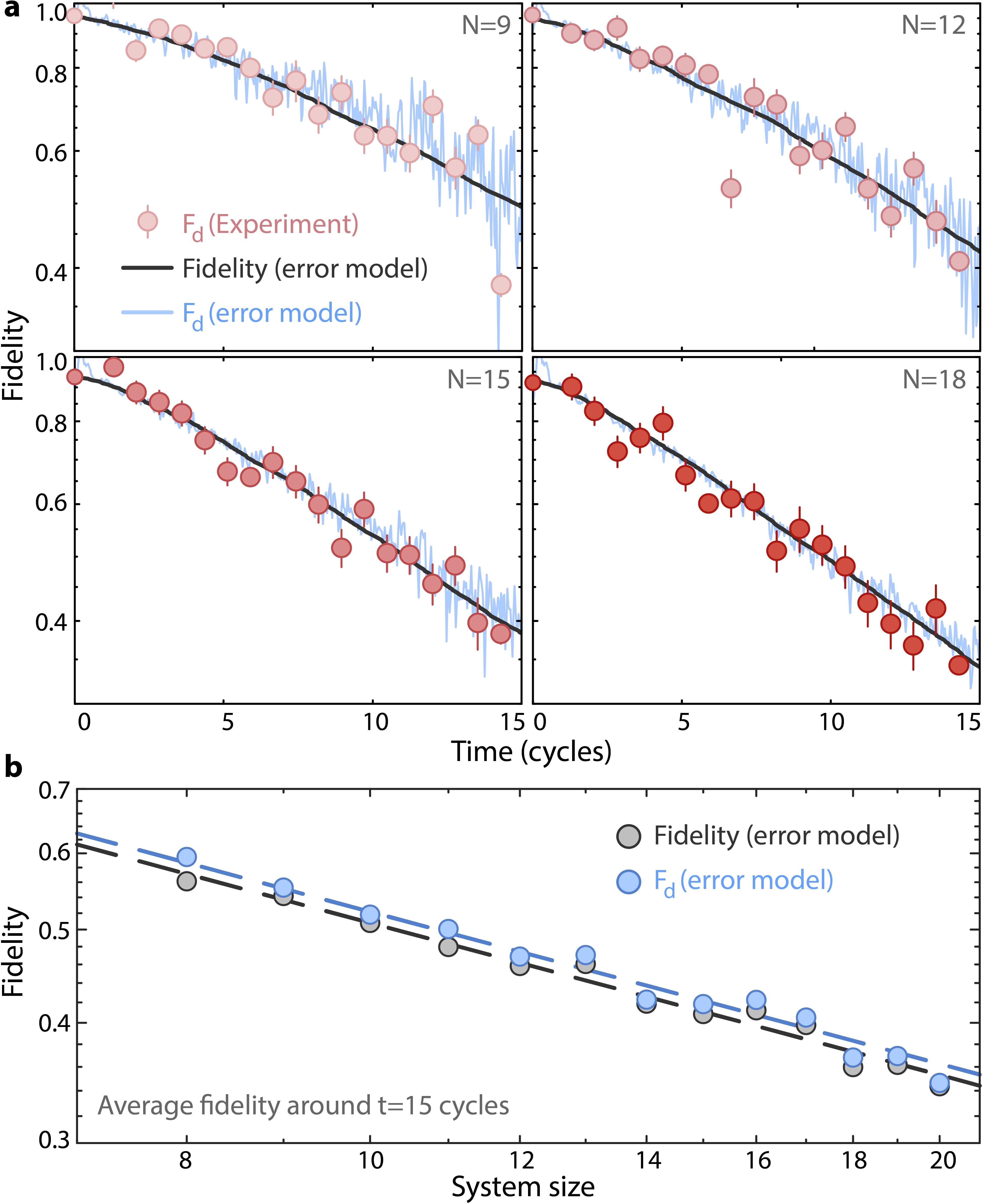}
	\caption{\textbf{Accuracy of the fidelity estimator. a.} Comparing fidelities estimated from experiment against both the true fidelity and the estimated fidelity for our error model, showing good agreement both between experiment and theory, and between the estimator $F_d$ and the true fidelity. $F_d$ shows systematic typicality fluctuations around the true fidelity, but these decrease in amplitude with increasing system size, see Ext. Data Fig.~\ref{EFig:sample_complexity}b. \textbf{b.} $F_d$ does exhibit a systematic multiplicative offset from the true fidelity in the time regime we study due to a time-delay needed for errors to scramble and become visible~\cite{Choi2023PreparingChaos,Mark2023BenchmarkingLett}. We find that $F_d\approx1.02 F$, regardless of system size around $t=15$ cycles.
        } 
	\vspace{-0.0cm}
	\label{EFig:fd_vs_f}
\end{figure}

We compute $p_\text{avg}(z)$ via time-averaging the calculated bitstring probabilities. Specifically, for each time point $t_i$ of the experiment, we estimate $p_\text{avg}(z)$ as a discrete average of the probability distributions within a window of $t_i\pm\SI{1.4}{\mu s}$, with an step size of approximately $\SI{28}{ns}$. This means around 100 points are averaged, which limits statistical sampling fluctuations. Points included in the averaging are weighted to emphasize data with similar values of classical fidelity, namely using a weight factor of $\min\{\frac{F_\text{svd}(t_i)}{ F_\text{svd}(t_j)}, \frac{F_\text{svd}(t_j)}{F_\text{svd}(t_i)}\}$. Here $F_\text{svd}$ is the product of MPS truncation errors~\cite{Bridgeman2017Hand-wavingNetworks,Zhou2020WhatComputers}, which we use to approximate the classical accuracy, and defined as
\begin{equation}
    F_\text{svd} = \prod_i \sum_{\alpha=1}^\oldchi s_{i, \alpha}^2,
\end{equation}
where $i$ runs over all MPS steps involving Schmidt value truncations, and $s_{i, \alpha}$ are the Schmidt values at truncation step $i$. Here, we assume the wave function is normalized where $\sum_{\alpha=1}^\infty s_{i, \alpha}^2=1$. While this estimation is only approximate, it can be extremely accurate when successive truncations are independent~\cite{Zhou2020WhatComputers}. We describe this approximation in more detail in Section~\ref{seq:estimating_fsvd} and Ext. Data Fig.~\ref{EFig:f_trun}. We weight in this way under the hypothesis that averaging classical simulations of very different accuracies (as typified by $F_\text{svd}$) will lead to a poorer estimation of $p_\text{avg}(z)$.

To account for the Rydberg blockade mechanism, we have the option to slightly modify~\cite{Choi2023PreparingChaos} the $F_d$ formula. Indexing the different blockade sectors by the number of blockade violations in that sector, $s$, we can write an approximator for $F_d$ to get an estimator for the blockade sector, $F_{d, s=0}$, only:
\begin{align}
    F_{d, s=0} = B_\text{thy}B_\text{exp}\Big(2 \frac{\frac{1}{M'} \sum_{m=1}^{M'} p'(z'_m) / p'_\text{avg}(z'_m)}{\sum_z p'(z)^2 / p'_\text{avg}(z)} - 1\Big).
\end{align}
Here $B_\text{exp}$ ($B_\text{thy}$) is the total probability for an experimental (simulation) bitstring to be in the blockaded Hilbert space, $s=0$, which further redefines the normalized probabilities, $p'(z) = p(z)/B_\text{thy}$. Note here $M'$ is the number of bitstrings, $z'_m$, measured in the blockaded Hilbert space.

\begin{figure*}[ht!]
	\centering
	\includegraphics[width=183mm]{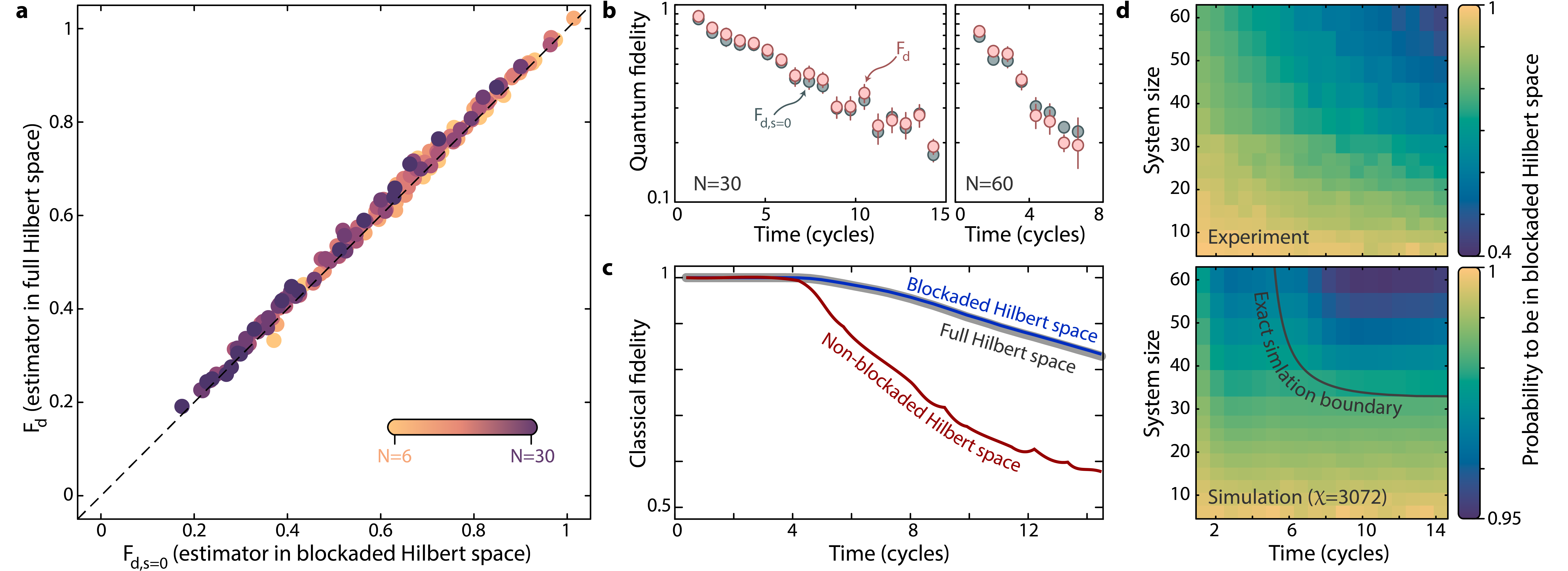}
	\caption{\textbf{Benchmarking in the blockaded Hilbert space. a.} $F_d$ versus $F_{d,s=0}$ (the fidelity benchmarked in only the blockaded Hilbert space, see text) for system sizes up to N=30, consistent with correlation with unity slope, but with a slight ${\sim}0.01$ offset. \textbf{b.} For exactly simulatable systems, $F_d$ typically always exceeds $F_{d,s=0}$ (left), but for larger systems (right), $F_d$ systematically begins underestimating $F_{d,s=0}$, even before the nominal exact simulation time. \textbf{c.} We understand this behavior as arising from the MPS preferentially truncating higher blockade subspaces, leading to the non-blockaded classical fidelity dropping below unity before it is visible in the full Hilbert space fidelity. This implies $F_{d,s=0}$ stays a more accurate estimator slightly longer than $F_d$. \textbf{d.} Probability to be in the blockaded subspace for both experiment and simulation. Note the different color-scale axes for the simulation and experiment, as many experimental errors can lead to apparently non-blockaded measurements.
    } 
	\vspace{-0.0cm}
	\label{EFig:blockaded}
\end{figure*}

Importantly, we find $F_{d,s=0}$ is more robust against the failure of the classical simulation algorithm, as compared to $F_{d,s>0}$. This is because MPS truncations more strongly affect the $s>0$ sectors in ways which are not necessarily visible only by looking at the MPS truncation fidelity (Ext. Data Fig.~\ref{EFig:blockaded}bc). 

For this reason, when estimating the quantum fidelity, we employ a two-pronged approach. Where $F_{\mathrm{svd}}>0.99$ we approximate $F_d$ as $F_{d,s=0}$, which from measurements at small system sizes we find is always conservative (Ext. Data Fig.~\ref{EFig:blockaded}a). Utilizing $F_{d,s=0}$ allows us to more confidently directly measure the fidelity out to intermediate times at the largest system sizes, and measure the according effective decay rate (Fig.~\ref{Fig3}c of the main text). However, where $F_{\mathrm{svd}}<0.99$, we estimate $F_d$ from our Monte Carlo inference procedure (described below), which is trained directly on the values of $F_d$. 

\begin{figure*}[ht!]
	\centering
	\includegraphics[width=183mm]{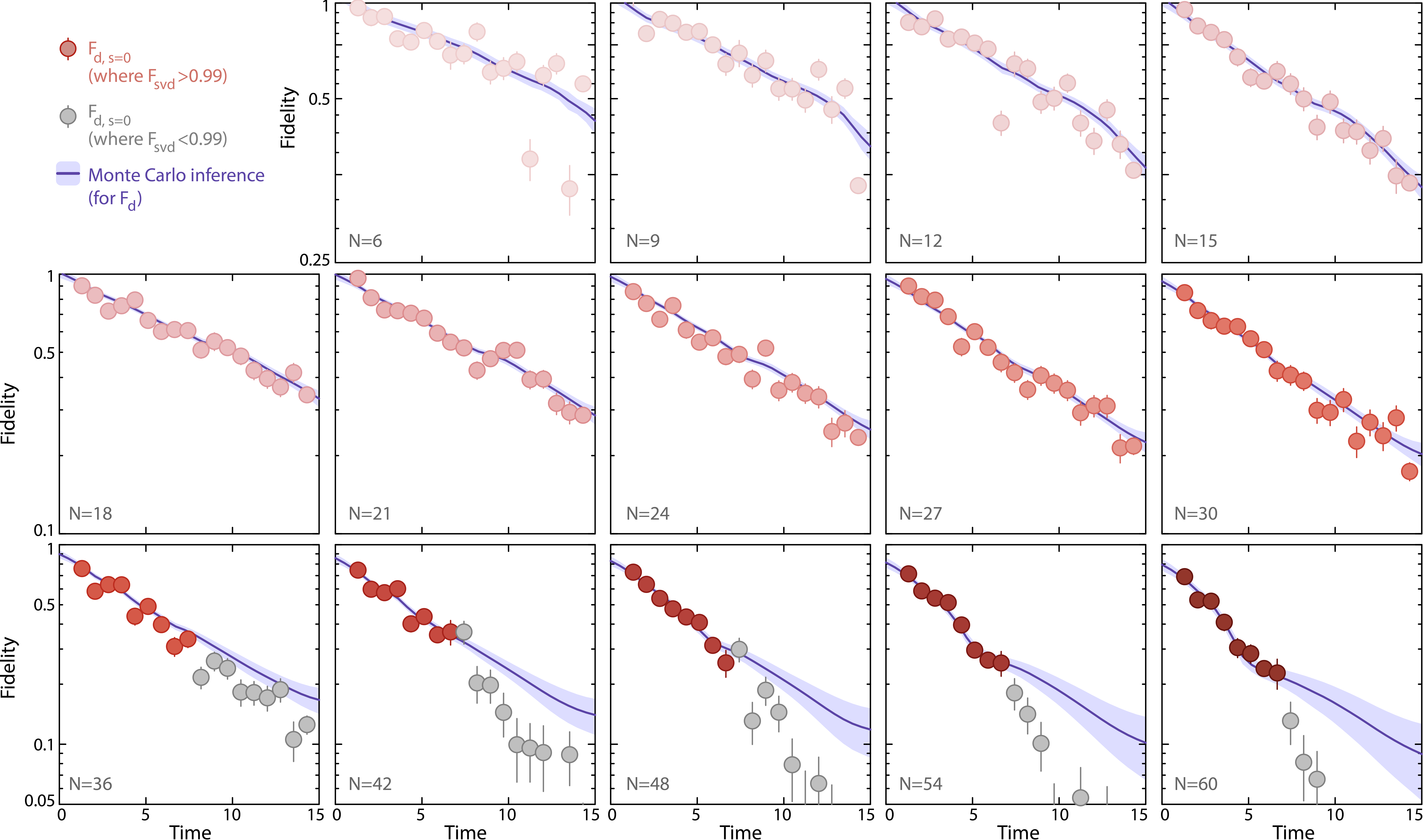}
	\caption{\textbf{Fidelity benchmarking for all system sizes and times. } Fidelity (quantified by $F_{d,s=0}$ with $\chi=3072$) as well as Monte Carlo inference for all system sizes. As in the main text, grayscale markers are those for which the classical fidelity is less than 0.99.
    } 
	\vspace{-0.0cm}
	\label{EFig:alldata}
\end{figure*}

To be concrete, in the main text, all data shown as markers in Figs.~\ref{Fig2} and~\ref{Fig3} is $F_{d,s=0}$ for $\chi=3072$. We summarize all of the data, both $F_{d,s=0}$ and the Monte Carlo fidelities, in Ext. Data Fig.~\ref{EFig:alldata}.

\vspace{5mm}

\newpage
\clearpage
\subsection{Non-Markovian origins of non-exponential fidelity decay}
\label{app:non-exp_decay}
In Fig.~\ref{Fig3} of the main text, we experimentally observe that the many-body fidelity appears to decrease exponentially at intermediate times, before bending sub-exponentially at late times, a behavior which seems to grow stronger with increasing system size (also visible in Ext. Data Fig.~\ref{EFig:alldata}). We believe this behavior originates from globally-correlated, non-Markovian Hamiltonian parameter fluctuations, which we note are often found in analog quantum simulators. In particular, a dominant noise source in our experiment is shot-to-shot variation of the Rabi frequency, for instance arising from fluctuations of the driving laser intensity, which we measure to be around ${\sim}0.5\%$, corresponding to a shot-to-shot variation of $\sigma{\sim}\SI{40}{kHz}$ (Ext. Data Fig.~\ref{EFig:nonexp_decay}a). Here, we present analytical and numerical support that such an error source can induce the observed fidelity decay behavior.

\subsubsection{Fidelity response to coherent Hamiltonian errors}
We discover a surprising dependence of the fidelity between states subject to coherent Hamiltonian errors.

\begin{theorem}[Gaussian fidelity response to coherent errors]
\label{thm:gaussian_fidelity}
Let $\ket{\Psi(t, \theta)} \equiv \exp(-i(\hat{H}_0+\theta \hat{V})t)\ket{\Psi_0}$ denote the initial state $\ket{\Psi_0}$ quench-evolved by a translationally-invariant local Hamiltonian $\hat{H}_0+\theta \hat{V}$ for a time $t$, where the perturbation $\hat{V} = \sum_x \hat{V}^x$ is a sum of local terms and $\theta$ quantifies the perturbation strength. We assume the Eigenstate Thermalization Hypothesis (ETH), i.e.~that the expectation values $\langle E | \hat{V}^x | E\rangle$ of each term $\hat{V}^x$ with respect to the eigenstates $|E\rangle$ of $\hat{H}_0$ are independent random samples from a fixed distribution. Then in the limit of large system size $N$ and long time $t$, the fidelity $F$ between two such states $|\Psi(t,\theta_1)\rangle$ and $|\Psi(t,\theta_2)\rangle$ is a Gaussian function of $(\theta_1-\theta_2)t$:
\begin{equation}
    F(\theta_1,\theta_2) \equiv  \abs{\langle\Psi(t,\theta_1)|\Psi(t,\theta_2)\rangle}^2 = \exp(- N \lambda (\theta_1-\theta_2)^2 t^2/2)~,\label{eq:Fidelity_Gaussian}
\end{equation}
for some constant $\lambda$ defined in Eq.~\eqref{eq:gaussian_fidelity_lambda}. 
\end{theorem}
This result is illustrated in Ext. Data Fig.~\ref{EFig:nonexp_decay}b and will be instrumental to showing our observed non-exponential fidelity decay (Corollary~\ref{cor:nonexp_decay}). This fidelity response is related to the Loschmidt echo, and our result is consistent with existing numerical observations in e.g.~Refs.~\cite{Gorin2006DynamicsDecay,Zurek2007GaussianEnvironments,Zangara2017LoschmidtIrreversibility}.

\vspace{5mm}
\begin{proof}
Without loss of generality, it suffices to let $\theta_1=0$ and $\theta_2 = \theta$. We compute the Taylor series of $F(\theta) \equiv F(0,\theta)$ and show that for large $N$, it agrees to all orders with the Gaussian series
\begin{align}
\exp\left(-\frac{\theta^2 t^2}{2\sigma^2}\right) = 1 - \frac{1}{2}\left(\frac{\theta t}{\sigma}\right)^2 + \frac{1}{8} \left(\frac{\theta t}{\sigma}\right)^4 + \cdots \label{eq:Gaussian_Taylor}
\end{align}

It is instructive to compute the first non-trivial term: the second derivative \begin{align}
\partial_\theta^2 F(\theta)|_{\theta=0} 
= \Big[&\ \ \ (\partial_\theta^2\langle{\Psi(t)}|{\Psi(t,\theta)}\rangle)\langle{\Psi(t,\theta)}|{\Psi(t)}\rangle\\
&+2(\partial_\theta \langle{\Psi(t)}|{\Psi(t,\theta)}\rangle)(\partial_\theta\langle{\Psi(t,\theta)}|{\Psi(t)}\rangle)\\
&+ \langle{\Psi(t)}|{\Psi(t,\theta)}\rangle(\partial_\theta^2\langle{\Psi(t,\theta)}|{\Psi(t)}\rangle) \Big]\Big\vert_{\theta=0}\nonumber
\end{align}

In order to evaluate these derivatives, we use the relation~\cite{Najfeld1995DerivativesComputation}:
\begin{align}
\partial_\theta \exp(i(\hat{H}_0+\theta \hat{V}) t)|_{\theta=0} \label{eq:Ham_derivative}
&= \int_0^t d\tau \exp(i \hat{H}_0 \tau) (i\hat{V}) \exp(i\hat{H}_0 (t-\tau))\\
&=it \sum_{E} V_{EE} \ketbra{E} e^{i E t}+i \sum_{E\neq E'} V_{EE'} \ket{E}\bra{E'} \frac{e^{i E' t} - e^{i E t}}{i(E'-E)} ~,\nonumber
\end{align}
where $\ket{E}$ are the eigenstates of $\hat{H}_0$ with eigenvalues $E$, and we have defined $V_{EE'} \equiv \bra{E'}\hat{V}\ket{E}$. For our purposes, Eq.~\eqref{eq:Ham_derivative} is valid even when $\hat{H}_0$ has degeneracies: the initial state $\ket{\Psi_0}$ projects any degenerate subspace onto a single eigenstate $\ket{E}$. 

The first term in Eq.~\eqref{eq:Ham_derivative} grows linearly with time $t$, while the summands in the second term oscillate and are sub-dominant. We subsequently neglect such terms. We obtain a similar expression for the second derivative:
\begin{align}
\partial_\theta^2 \exp(i(\hat{H}_0+\theta \hat{V}) t) \approx -t^2 \sum_{E} V_{EE}^2 \ketbra{E} e^{i E t}~.
\end{align}
This gives:
\begin{align}
\partial_\theta^2 F(\theta) \vert_{\theta = 0} &\approx - 2t^2 \Big[\sum_{E} \abs{c_E}^2 V_{EE}^2 - \big(\sum_{E} \abs{c_E}^2 V_{EE}\big)^2\Big]\nonumber\\
&\equiv -2t^2(\langle \hat{V}^2\rangle - \langle \hat{V}\rangle^2)\nonumber\\
&\equiv -2t^2 \kappa_2(\hat{V})~,
\end{align}
where $\abs{c_E} \equiv \abs{\langle E |\Psi_0 \rangle}^2$ are the populations of $\ket{\Psi_0}$ in the energy eigenbasis. We interpret the quantities $\sum_E\abs{c_E}^2 V_{EE}$ and $\sum_E \abs{c_E}^2 V_{EE}^2$ as averages of the vector $V_{EE}$ and $V^2_{EE}$ with respect to the distribution $\abs{c_E}^2$: we denote them $\langle \hat{V} \rangle$ and $\langle \hat{V}^2 \rangle$ respectively; the expression $\langle \hat{V}^2 \rangle-\langle \hat{V} \rangle^2$ is the variance, which we denote $\kappa_2$ for reasons made clear below.

Within the approximation made in Eq.~\eqref{eq:Ham_derivative}, $\partial^n_\theta F(\theta)|_{\theta=0}=0$ for all odd $n$. The next term can be expressed in terms of the moments $\langle \hat{V}^k\rangle$ as
\begin{align}
\partial_\theta^4 F(\theta) \vert_{\theta = 0} &\approx 2t^4(\langle \hat{V}^4\rangle - 4\langle \hat{V}^3\rangle\langle \hat{V}\rangle + 3 \langle \hat{V}^2\rangle^2 ) 
\nonumber\\
&= 2t^4\big[\kappa_4(\hat{V}) + 6\kappa_2(\hat{V})^2\big]~, \label{eq:fourth_derivative}
\end{align}
where $\kappa_4(\hat{V}) \equiv \langle (\hat{V} - \langle \hat{V} \rangle)^4 \rangle - 3 \langle (\hat{V} - \langle \hat{V} \rangle )^2\rangle^2$ is the \textit{fourth cumulant} of $\hat{V}$. Crucially, $\hat{V}=\sum_x \hat{V}^x$ is the sum of identical local terms: the Eigenstate Thermalization Hypothesis (ETH) predicts that their eigenstate overlaps $V_{EE}^x$ fluctuate about a smooth function of energy~\cite{DAlessio2016FromThermodynamics}. Furthermore, for sufficiently high-temperature states, correlations $\langle \hat{V}^x \hat{V}^y \rangle$ are short-ranged, and we can treat each $\hat{V}^x$ as an effectively independent random variable.

We utilize a further property of cumulants: they are additive, obeying $\kappa_n(\sum_x \hat{V}^x) = \sum_x \kappa_n(\hat{V}^x)$. Therefore, all cumulants scale linearly with $N$. When $N$ is large,  $\kappa_4(\hat{V}) \ll \kappa_2(\hat{V})^2$ and we can neglect the $\kappa_4$ term in Eq.~\eqref{eq:fourth_derivative}. This gives the expression
\begin{align}
F(\theta) \approx 1 - \frac{1}{2}2 \kappa_2(\hat{V}) (\theta t)^2 + \frac{1}{8} (2 \kappa_2 (\hat{V}))^2 (\theta t)^4 + \cdots,\label{eq:Fidelity_Taylor}
\end{align}
which agrees with Eq.~\eqref{eq:Gaussian_Taylor} when we identify
\begin{equation}
 \sigma^{-1} = 2\kappa_2(\hat{V}) = N \kappa_2(\hat{V}^x) \equiv N \lambda
 \label{eq:gaussian_fidelity_lambda}
\end{equation}
In fact, all higher cumulants are subleading to the $\kappa_2$ contribtion. As long as they can be neglected, the series Eq.~\eqref{eq:Gaussian_Taylor} and Eq.~\eqref{eq:Fidelity_Taylor} agree \textit{to all orders}. For example, the $(\theta t)^6$ term of \eqref{eq:Fidelity_Taylor} is $-2(\kappa_6(\hat{V})+30\kappa_4(\hat{V})\kappa_2(\hat{V})+60\kappa_2(\hat{V}))/6! \approx -(2\kappa_2(\hat{V}))^3/48$, in agreement with the sixth-order term of Eq.~\eqref{eq:Gaussian_Taylor}.

We prove this equivalence to all orders using properties of \textit{Bell polynomials}, $B_k$ (where $k$ is a polynomial index)~\cite{Bell1934ExponentialPolynomials}. The $n$-th derivative of the fidelity can be expressed as
\begin{align}
\partial^n_\theta F|_{\theta=0} = (-1)^{n/2}(\theta t)^n \sum_{k=0}^n (-1)^k {n \choose k} \langle \hat{V}^{n-k}\rangle  \langle \hat{V}^k \rangle~.\label{eq:Fidelity_nth_derivative}
\end{align}
We can rewrite this in terms of the cumulants $\kappa_1,\dots,\kappa_k$ with the cumulant-moment relation~\cite{Withers2009MomentsVersa}
\begin{align}
\langle \hat{V}^k\rangle  = B_k(\kappa_1,\kappa_2, \cdots, \kappa_k)~.
\end{align}
While the full expression is challenging, we are only interested in the coefficient of the $\kappa_2(\hat{V})^{n/2}$ term in Eq.~\eqref{eq:Fidelity_nth_derivative}. We first note that $\langle \hat{V}^k\rangle$ contains a $\kappa_2(\hat{V})^{k/2}$ term only for even $k$. To determine its coefficient, we use the recursion relation~\cite{Bell1934ExponentialPolynomials} $\partial_{\kappa_2} B_k = {k \choose 2} B_{k-2}$ to conclude that $\langle \hat{V}^k\rangle = \cdots + (k!/k!!) \kappa_2(\hat{V})^{k/2} + \cdots$, where $k!!=k(k-2)\cdots 2$ is the \textit{double factorial} of $k$. Returning to Eq.~\eqref{eq:Fidelity_nth_derivative}, we conclude that $\partial^n_\theta F|_{\theta=0} = \cdots + a_n \kappa_2(\hat{V})^{n/2} (\theta t)^n+ \cdots$, where
\begin{align}
a_n &= (-1)^{n/2}\sum_{k~\text{even}} {n \choose k} \frac{k!(n-k)!}{k!! (n-k)!!} \kappa_2^{n/2}\nonumber\\
&= (-2)^{n/2}\frac{n!}{n!!} \kappa_2^{n/2}~,
\end{align}
which is equal to the $n$-th derivative $\partial_\theta^n \exp(-\theta^2/(2\sigma^2))|_{\theta=0} = (-1)^{n/2}\frac{n!}{n!!} \sigma^{-n}$, thus concluding our derivation of the Gaussian fidelity response [Eq.~\eqref{eq:Fidelity_Gaussian}].
\end{proof}

\subsubsection{Non-exponential decay for Rydberg quantum simulators}
The non-exponential fidelity decay immediately follows from Theorem~\ref{thm:gaussian_fidelity}.
\begin{corollary}
\label{cor:nonexp_decay}
When the parameter value $\theta$ follows a Gaussian probability distribution $P(\theta)$ with mean $\theta'$ and standard deviation $\sigma$ 
\begin{align}
P(\theta) = \frac{1}{\sqrt{2\pi}\sigma}\exp\Big(-\frac{(\theta-\theta')^2}{2 \sigma^2}\Big)~,
\end{align}
Theorem~\ref{thm:gaussian_fidelity} and a straightforward integration give a universal form for the fidelity of the target state $\ket{\Psi(t,\theta_0)}$ to the mixed state $\rho(t) \equiv \int d\theta P(\theta) |\Psi(t,\theta)\rangle\langle\Psi(t,\theta)|$.
\begin{align}
F= \int d\theta P(\theta) F(\theta_0,\theta) &=\frac{1}{\sqrt{1+\Lambda}}\exp\Big(-\frac{\delta^2}{2}\frac{\Lambda}{1+\Lambda} \Big)~,
\label{eq:nonexp_decay}
\end{align}
where $\Lambda \equiv N t^2\lambda\sigma^2$ and $\delta=(\theta'-\theta_0)/\sigma$ is the overall relative parameter miscalibration.
\end{corollary}

\begin{figure*}[h!]
	\centering
    \vspace{5mm}
	\includegraphics[width=183mm]{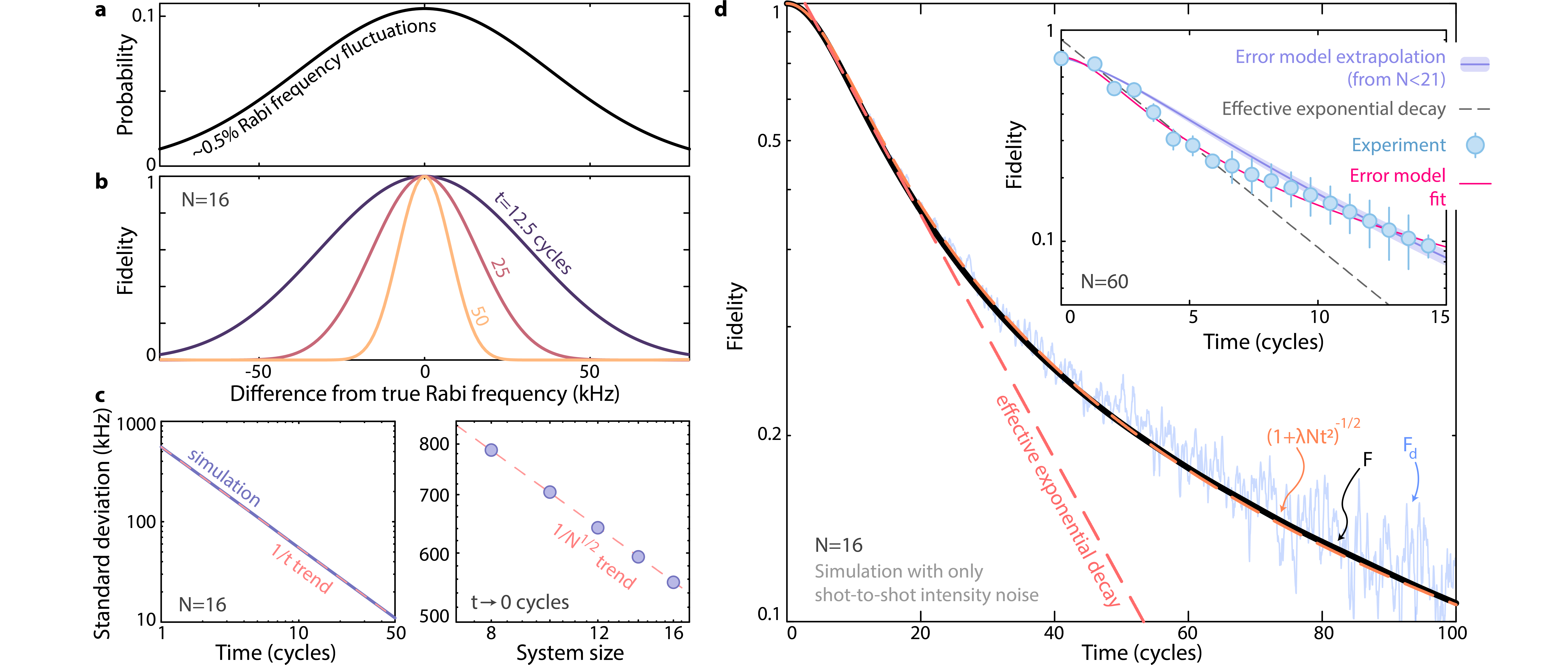}
	\caption{\textbf{Origins of the non-exponential fidelity decay. \textbf{a.}} A major noise source for our platform is Rabi frequency fluctuations (for instance arising from variation in the driving laser intensity), which have a shot-to-shot standard deviation of ${\sim}0.5\%$ (${\sim}\SI{40}{kHz}$). \textbf{b.} For global, correlated intensity noise, the fidelity is Gaussian with respect to error in the Rabi frequency. \textbf{c.} The standard deviation of this Gaussian dependence decreases as $1/t\sqrt{N}$. \textbf{d.} This behavior leads to an analytic solution for the fidelity decay of $F{\sim}(1+\lambda N t^2)^{-1/2}$, for free parameter $\lambda$. This decay is characterized by super-exponential curvature at early times, and sub-exponential curvature at late times. In the inset, we use our error model with all error sources enabled, and extrapolate to the expected fidelity behavior at $N=60$, seeing generally good agreement with the experiment, which is improved by directly fitting and allowing unity-level changes in the extrapolated error model parameters.
        } 
	\vspace{-0.0cm}
	\label{EFig:nonexp_decay}
\end{figure*}

This functional form leads to markedly different behavior as compared to simple exponential decay, as it becomes essentially like a power-law at late times (Ext. Data Fig.~\ref{EFig:nonexp_decay}d). In simulation, we benchmark an $N=16$ atom system experiencing only global intensity fluctuations -- but otherwise no noise -- and see an excellent agreement between the simulated fidelity, and a fit to Eq.~(\ref{eq:nonexp_decay}) with free $\lambda$. Notably, we see Eq.~(\ref{eq:nonexp_decay}) exhibits three distinct regimes: super-exponential behavior at very early time, effectively simple-exponential behavior at intermediate times, and finally sub-exponential behavior at late times. Notably, by studying multiple system sizes, we find the divergence between Eq.~(\ref{eq:nonexp_decay}) and an effective exponential decay model is mainly a function of fidelity, not time, and becomes visible when the fidelity reaches $F{\sim}0.2$, which in our dataset is only the case for the highest $N$ (Fig.~\ref{Fig3} of the main text).

As a quantitative comparison, we define an effective model of the fidelity as 
\begin{align}
F\approx F_0\exp(-\gamma t)\frac{1}{\sqrt{1+\lambda' t^2}}\exp\Big(-\frac{\delta^2}{2}\frac{\lambda' t^2}{1+\lambda' t^2} \Big),
\label{Eq:effective_decay}
\end{align}
that is, Eq.~(\ref{eq:nonexp_decay}) multiplied by an exponential decay factor to account for the other noise sources, where we have redefined the free parameter $\lambda'\equiv\lambda N \sigma^2$. We perform error model simulations from $7<N<21$ for two cases: 1) for only global, shot-to-shot Rabi and detuning noise, 2) for all other noise sources. Then, assuming multiplicativity of the error sources (Ext. Data Fig.~\ref{EFig:error_model}a),
from case (1) we fit the expected behavior of $\lambda'$, and from case (2) we fit the expected behavior of $\gamma$. We extrapolate the observed parameters all the way to $N=60$, and compare against the experiment (inset of Ext. Data Fig.~\ref{EFig:nonexp_decay}d). Even assuming perfect calibration ($\delta=0$) we see general agreement between the fidelity decay curve of the error model prediction and the experiment. 

However, the relative proportion of exponential decay versus non-exponential decay is difficult to calibrate versus experimental data at small system sizes, as this requires evolving out to very late times when other error sources (such as finite recapture probability of atoms in their traps) become noticeable. In addition, we expect that Hamiltonian parameter calibration is likely imperfect, at the very least up to our uncertainty in the Hamiltonian parameters. As a general sanity check then, we fit with $\gamma$, $\lambda'$, and $\delta$ free. Fitting $N=60$ (with $F_0$ fixed by the separately calibrated single-atom preparation fidelity), we find $\delta\approx0.2$, $\gamma\approx1/4\times\gamma_0$ and $\lambda'\approx4\times\lambda'_0$, where $\gamma_0$ and $\lambda'_0$ are the extrapolated predictions from error model. We note $\delta\approx0.2$ corresponds to a potential error of ${\approx}10$ kHz in the calibrated Rabi frequency, which is within our measured error bar (Ext. Data Fig.~\ref{EFig:ham_pams}b).

We also note that other error sources, in particular atomic position fluctuations, also likely have some contribution to the non-exponential decay behavior, which may improve the agreement above. However, the analytical expression for their effect becomes quite involved, and so we disregard them here for simplicity, and leave a more in-depth analytic and numerical study of the various error sources to future work.

\newpage
\subsection{Fidelity extrapolation through Monte Carlo inference}
To extrapolate fidelities into the high-entanglement (i.e. late-time, large system-size) regime, we use a Monte Carlo inference approach. More specifically, we train an ensemble~\cite{Ganaie2022EnsembleReview} of fully-connected regression neural networks, implemented in \textsc{MATLAB}, with three inputs (system size, evolution time, and bond dimension), and one output (estimated fidelity). The neural networks are instantiated with randomized hyperparameters (including the initial seed for the weights and biases). We emphasize that this approach is not trying to learn directly from experimental observations~\cite{Torlai2018Neural-networkTomography,Carrasquilla2019ReconstructingModels,Zhang2021DirectLearning,Huang2023ProvablyProblems}, and instead we are simply extrapolating a smoothly varying 3-to-1 function. Fundamentally, the goal of this approach is to utilize all aspects of the data simultaneously, rather than extrapolating along only a single axis at a time, for instance either via time-extrapolation or system-size scaling alone. 

\begin{figure*}[b!]
	\centering
    \vspace{5mm}
	\includegraphics[width=181mm]{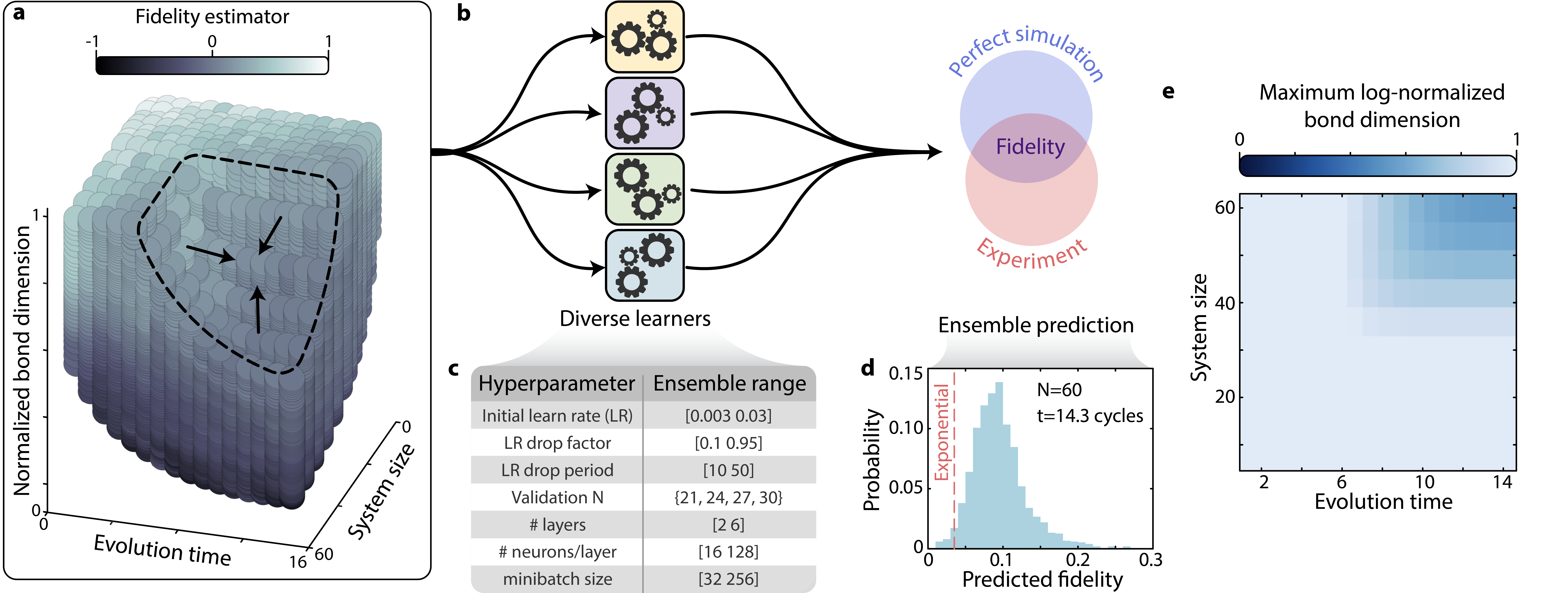}
	\caption{\textbf{Monte Carlo  fidelity inference. \textbf{a.}} All the data in this study, mapping the three inputs (system size, $N$, evolution time, $t$, and log-normalized bond dimension, $\tilde{\chi}=\log(\chi)/\log(\chi_0)$) to an estimated fidelity. Here $\chi_0$ is the time- and size-dependent saturation bond dimension (Ext. Data Fig.~\ref{EFig:chi_exact}e). \textbf{b.} To improve generalization into the unknown regime, we use a Monte Carlo inference approach. We train an ensemble of 1500 fully-connected neural networks, each instantiated and trained with random choices of parameters within reasonably specified ranges (defined in \textbf{c}); the Monte Carlo predicted fidelity is then taken as the ensemble average for a given input. \textbf{d.} Monte Carlo prediction for the fidelity of the $N=60$ system at the entanglement saturation time. Over ${\sim}98\%$ of the individual Monte Carlo instance predictions exceed the expectation from the early time exponential decay (see Fig.~\ref{Fig3} of the main text), resulting in an average of 0.095(11), where the error bar represents the standard error on the mean, added in quadrature with the intrinsic sampling error of the underlying data. \textbf{e.} Maximum log-normalized bond dimension available for training as a function of system size and time, for $\chi=3072$.
        } 
	\vspace{-0.0cm}
	\label{EFig:neural_net}
\end{figure*}

\subsubsection{Method description}
For training data, we use the fidelities from all combinations of the 14 system sizes, 18 evolution times, and 91 simulated bond dimensions (from $\chi=2$ to 3072), resulting in a total dataset of ${\sim}$23000 points (Ext. Data Fig.~\ref{EFig:neural_net}a). The network is then trained to learn the functional dependence of $F(N,\chi,t)$. 

To improve generalization and avoid fine-tuning, we employ a Monte Carlo approach through ensemble learning~\cite{Ganaie2022EnsembleReview}, where the training is repeated 1500 times. Each iteration is instantiated with a randomized seed, and every three iterations we randomly select from a range of hyperparameters, the domain of which are specified in Ext. Data Fig.~\ref{EFig:neural_net}c. The fidelity for a given set of inputs is then taken to be the unweighted ensemble average, while the error bar is taken as the ensemble standard error of the mean added in quadrature with the sampling error at a given time.

\begin{figure*}[t!]
	\centering
	\includegraphics[width=183mm]{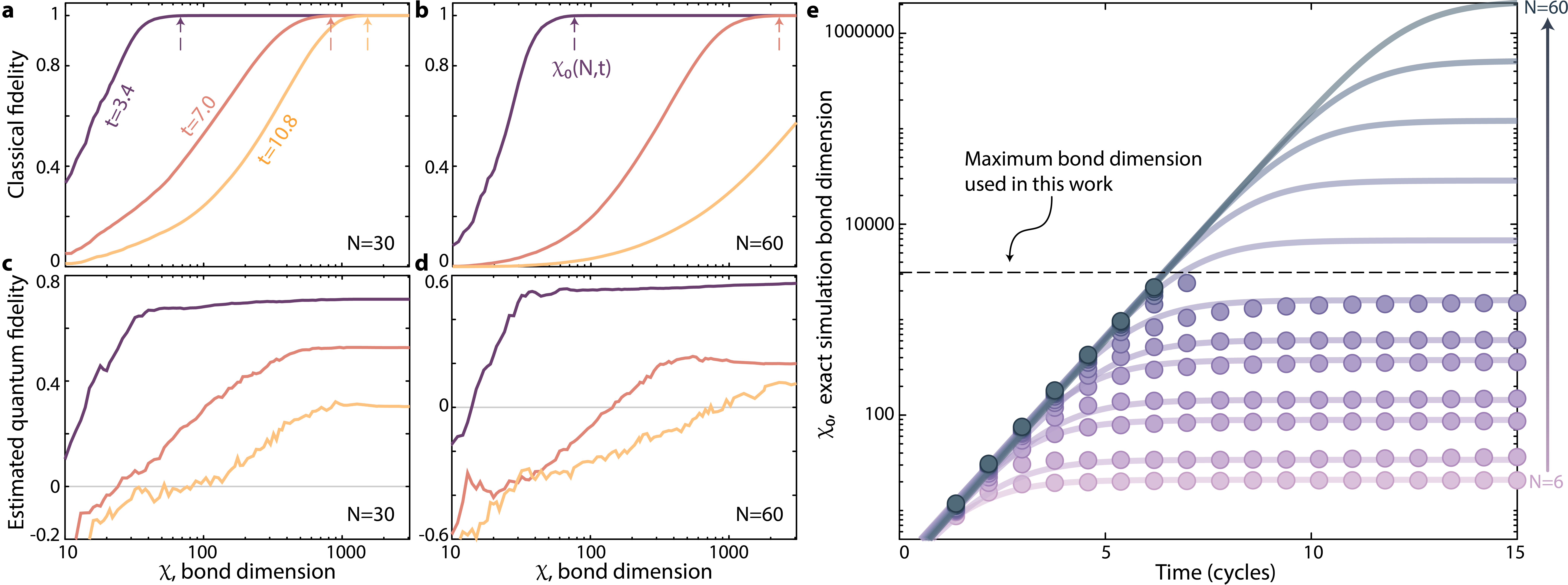}
	\caption{\textbf{Determining the log-normalized bond dimension. a, b.} As a function of bond dimension, the classical simulation fidelity rises before saturating to 1; we define $\chi_0$ (the saturation bond dimension), as the minimum value for which the classical fidelity is greater than 0.999. \textbf{c, d.} By $\chi_0$, the experimentally benchmarked fidelities have also saturated. \textbf{e.} We fit the dependence of $\chi_0$ as a function of system size and time (solid lines) in order to extrapolate beyond the range of $\chi$ values which we directly simulate in this work.
        } 
	\vspace{0.0cm}
	\label{EFig:chi_exact}
\end{figure*}

To preprocess the data, we log-normalize the bond dimensions as $\tilde{\chi}=\log(\chi)/\log(\chi_0)$, where $\chi_0$ is the size- and time-dependent saturation bond dimension required for the simulation to be exact (Ext. Data Fig.~\ref{EFig:chi_exact}e). This makes the learned function have a more consistent domain for different system sizes and evolution times. We also test if we train using simply $\log(\chi)$. We stress that in doing so, the largest trained $\log(\chi)\approx8$, but for instance for $N=60$ we predict the fidelity at $\log(\chi)\approx14.5$, with no training data in the interim regime. We find consistent mean predictions between the two methods, but the standard deviation of Monte Carlo instances is reduced by training on $\tilde{\chi}$.

We determine $\chi_0$ by finding the minimum $\chi$, at a given system size and evolution time, where the MPS fidelity is greater than 0.999. We extrapolate the behavior of $\chi_0$ into the high-entanglement regime of large system sizes and late evolution times (Ext. Data Fig.~\ref{EFig:chi_exact}e). We further normalize the input (subtracting off the mean and dividing by the standard deviation) to improve learnability.

To avoid overfitting, we randomly select a few $N$ from 21 to 30, for which we excise from the training any data with $\tilde{\chi}>2/3$ and $t>6.6$ cycles. We then use the data for these system sizes and times with $\tilde{\chi}\sim1$ as the validation dataset, and stop the training when the validation loss -- defined by the prediction root mean square error (RMSE) -- does not decrease for 50 epochs in a row (see for instance Ext. Data Fig.~\ref{EFig:validation}b).

\begin{figure*}[ht!]
	\centering \includegraphics[width=183mm]{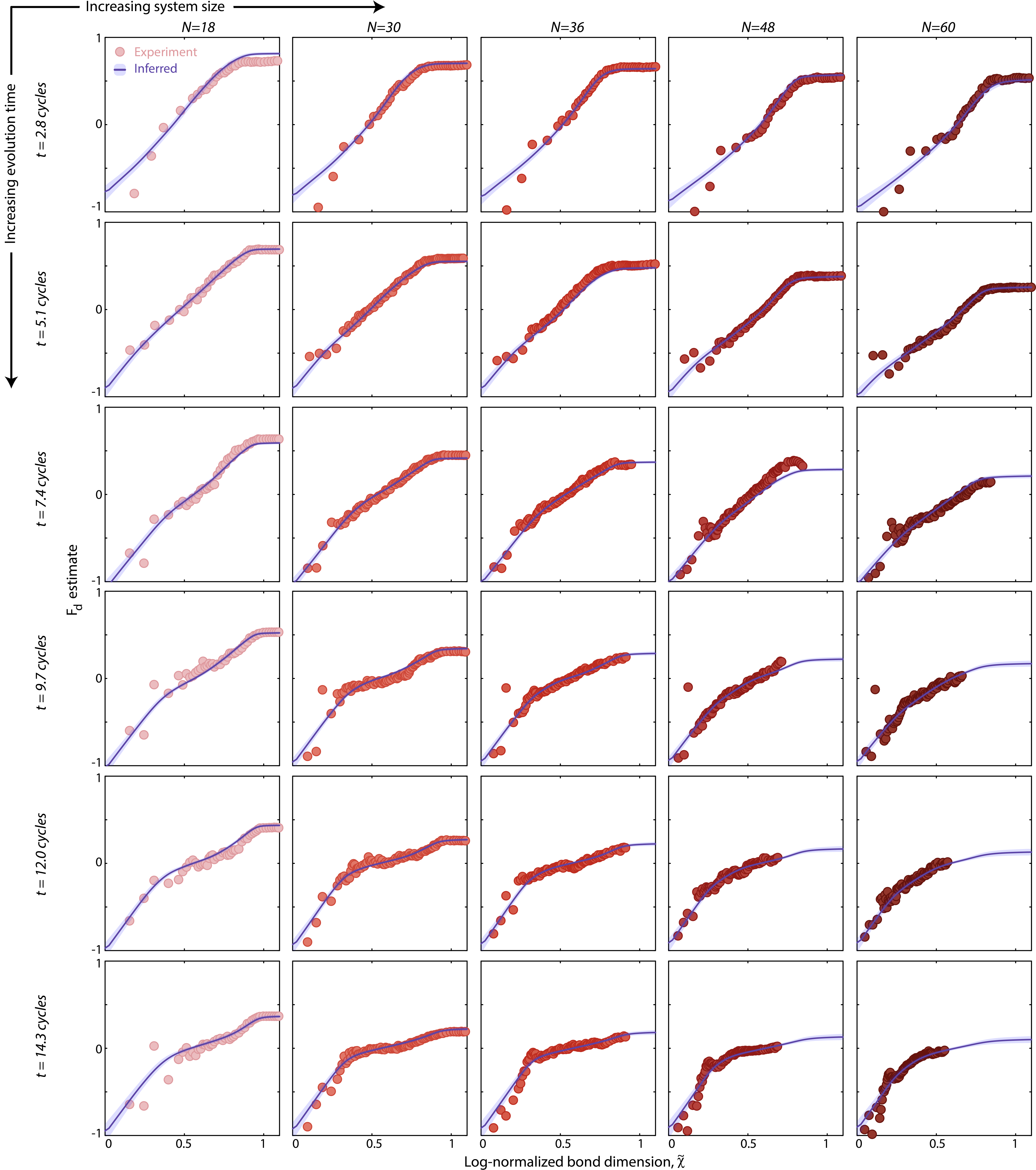}
	\caption{\textbf{Monte Carlo inference for all data. a.} Fidelity-estimator, $F_d$, for various system sizes and evolution times as a function of the log-normalized bond dimension, $\tilde{\chi}=\log(\chi)/\log(\chi_0)$. The fidelity is estimated at $\tilde{\chi}=1$. A key feature to note is that as system size or evolution time increases (i.e. as $\chi_0$ increases), the data points shift left: the classical simulation breakdown regime is when no data is available at $\tilde{\chi}=1$.
    } 
	\vspace{0.0cm}
	\label{EFig:alldata_vschi}
\end{figure*}

We summarize a partial subset of the data showing the Monte Carlo inference as a function of $N$, $t$, and $\chi$ in Ext. Data Fig.~\ref{EFig:alldata_vschi}.

\subsubsection{Consistency checks}
In order to test that this approach is accurate in estimating the fidelity, we perform several consistency checks, ranging from varying the number of learnable parameters, testing the result as a function of the amount of experimental training data, checking against small system sizes in experiment and error model where we have exact ground truth data, and more. Several of these tests are described below.
\\

\paragraph*{Small system sizes, experimental---}
For the largest system sizes we do not have ground truth data for $F_d$ in the late-time regime, and so cannot directly verify our model's efficacy. However, we can emulate the effect in this regime at smaller system sizes where we do have ground truth data in all regimes.

\begin{figure*}[b!]
	\centering
	\includegraphics[width=181mm]{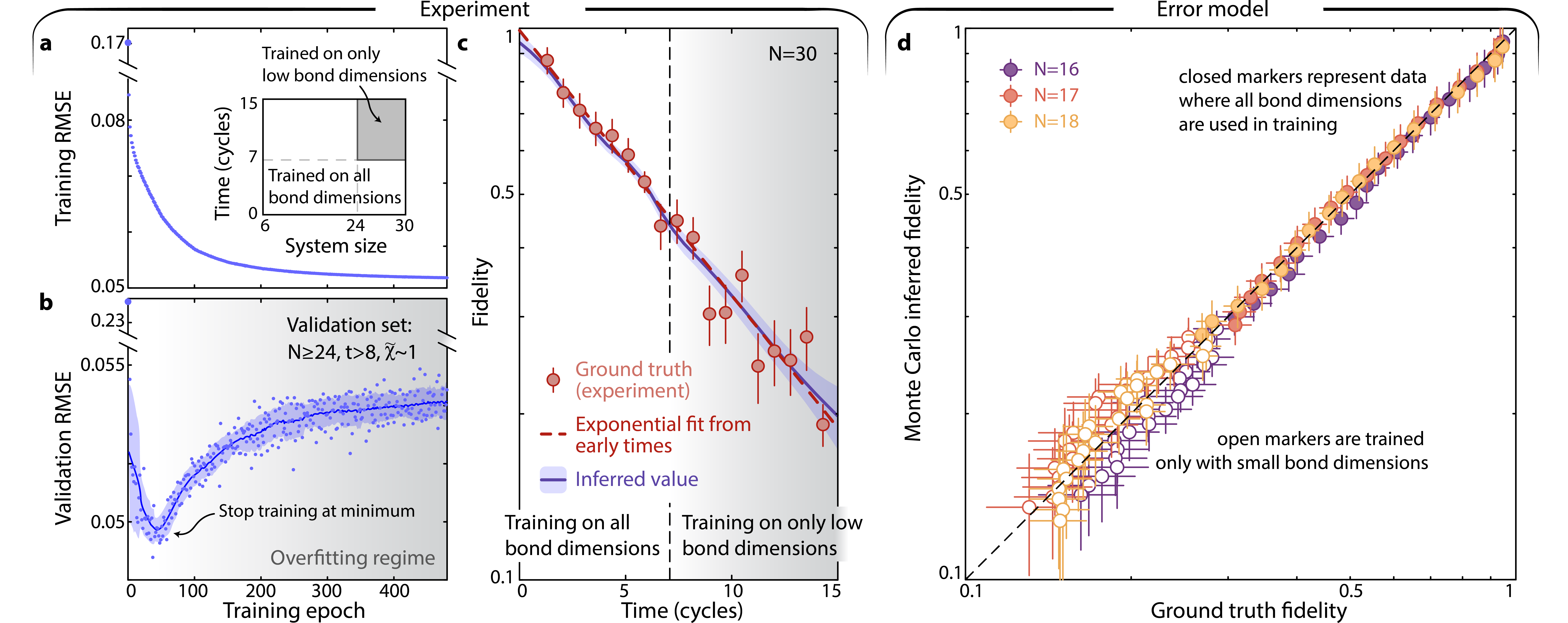}
	\caption{\textbf{Evaluating the Monte Carlo inference at small system sizes. a, b.} As a cross-check of our approach, we use the same training methodology employed on the full dataset, but truncate to only include $N\leq30$ where we have the full ground truth. We bipartition the data (inset), and for the large system sizes and late evolution times we only directly train with $\tilde{\chi}<2/3$. Averaging over 1500 instantiations, we find the training loss decreases consistently, but validation loss at  $\tilde{\chi}{\sim}1$ reaches a minimum after ${\sim}50$ epochs, and as expected proceeds into an over-fitting regime after. \textbf{c.} Model predictions show the Monte Carlo inference accurately predicts fidelities at late times, despite not having access to the underlying ground truth data at large bond dimensions in this regime. \textbf{d.} We perform a similar test using error model simulations of our experiment, as for instance shown in Fig.~\ref{Fig3}b of the main text. We train with $N=8$ to 18, excising  $\tilde{\chi}>2/3$ for $N>15$ and $t>6.6$ cycles. The resultant Monte Carlo inference predictions are well correlated with the ground truth, across the whole range of fidelities. X-error bars represent estimated typicality fluctuations, y-error bars represent the Monte Carlo inference standard deviation.
        } 
	\vspace{-0.0cm}
	\label{EFig:validation}
\end{figure*}

We employ the same training methodology as on the full dataset, but only consider $N\leq30$ where we have the full ground truth. For $t>6.6$ cycles and $N\geq24$, we excise from the training data all bond dimensions for which $\tilde{\chi}>2/3$. This value is chosen as it is approximately the maximum normalized bond dimension available at $N=60$ (Ext. Data Fig.~\ref{EFig:neural_net}e). Concretely, this approach is essentially pretending as if $N\geq24$ is beyond the exactly simulable regime. In reality, of course, we have the ground truth data to compare against, and can for instance use it to study the behavior of the loss functions (Ext. Data Fig.~\ref{EFig:validation}ab). 

Averaging over all Monte Carlo instantiations, we find the training loss decreases consistently as a function of training epoch, but the validation loss (where the ground truth $N\geq$24 results are used as the validation data set) reaches a minimum after ${\sim}50$ epochs, after which it proceeds into an overfitting regime. As with the full dataset, we thus stop the training for each instantiation at the point where the validation loss is non-decreasing for ${\sim}50$ epochs in a row, and then take the result from the best epoch. We note that both losses drop most dramatically after only a single epoch of training from initially randomized configurations, meaning the gradient descent quickly finds a nearly optimal regime, implying that the loss function is both smooth and non-flat.

We then compare directly between the Monte Carlo predicted $F_d$ and the ground truth value for $N=30$ across all times. We see good agreement even in the emulated breakdown regime at long times (Ext. Data Fig.~\ref{EFig:validation}c), with no indication that the learned fidelities are misestimating the true values, despite not having access to the underlying ground truth data at large bond dimensions. 
\\

\paragraph*{Small system sizes, error model---}
As described, we can check our model's efficacy at small system sizes directly on experimental data, but the non-exponential fidelity decay behavior is not yet strongly visible. In the main text, we reach this regime using Monte Carlo fidelity inference on error model simulations (Fig.~\ref{Fig3} of the main text). Here we further describe this analysis.

We perform error model simulations for $N=8-18$ up to the maximum experimental time, but artificially increase the shot-to-shot Rabi frequency variation by a factor of 4 in order to roughly emulate the predicted experimental fidelities at large $N$. We choose to do this, rather than simulate for much longer timescales where the behavior would become naturally evident (Ext. Data Fig.~\ref{EFig:nonexp_decay}d), because we want to replicate as much as possible the fact that for the experimental dataset the entanglement is still growing for most times for the largest system sizes. Performing error model simulations out to ${\sim}100$ cycles would dominate the training by a regime where the entanglement had already saturated for these small system sizes.

We perform MPS simulations for all bond dimensions up to $\chi_0(N=18,t\rightarrow\infty)=89$, and benchmark the error model with each. Though we run the error model with a finer timestep of ${\sim}10$ ns, we only select data at the experimentally measured times for training. We then emulate the experimental dataset by treating all $N<16$ as ``small N'' where we have full ground truth data. We excise from the training any data for which $\tilde{\chi}>2/3$, $N>15$, and $t>6.6$ cycles, to emulate the available data for the full experimental dataset. We use $N=13-15$ as the validation dataset to prevent overfitting. We then train in the same way as for the experiment, and compare Monte Carlo predictions versus ground truth output (Ext. Data Fig.~\ref{EFig:validation}d). We find the two are consistent, both in terms of the values obtained, and the non-exponential fidelity decay, indicating the Monte Carlo inference is able to learn this behavior.
\\

\begin{figure*}[h!]
	\centering
    \vspace{5mm}
	\includegraphics[width=181mm]{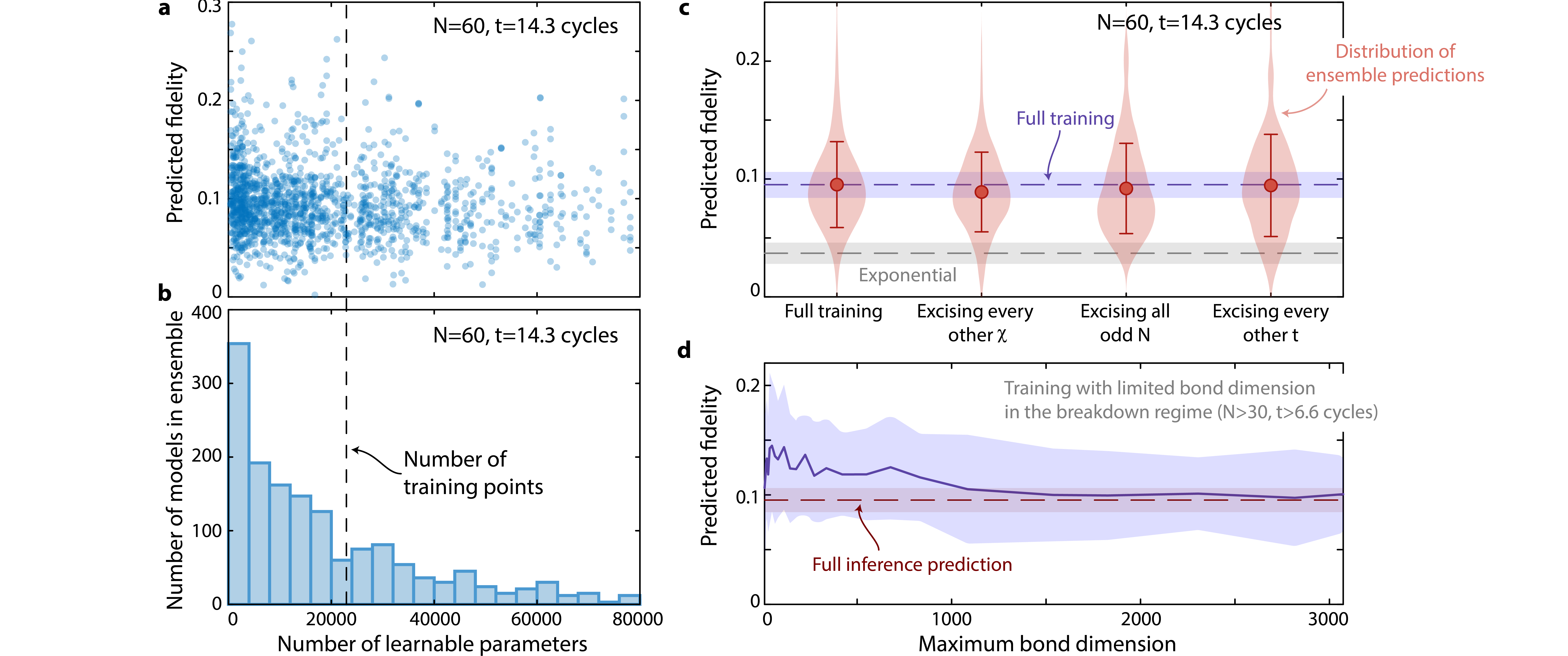}
	\caption{\textbf{Robustness of Monte Carlo inference. a, b.} For the majority (${\sim}70\%$) of hyperparameter combinations sampled during the Monte Carlo inference (Ext. Data Fig.~\ref{EFig:neural_net}c), the number of learnable parameters is smaller than the number of training points (${\sim}23000$). Further, we see no correlation between number of learnable parameters (nor any individual specific choice of hyperparameter) with the predicted fidelity. \textbf{c.} The Monte Carlo mean prediction is insensitive to the choice of data: removing half of the data along any axis leads to the same predicted fidelity. Note that while the ``Full training'' is composed of 1500 models, the ``excised'' ensembles only feature 192 trained models. Markers show mean and standard deviation for each distribution. \textbf{d.} To check the convergence of our protocol, we repeat the Monte Carlo inference procedure with varying maximum bond dimension allowed in the breakdown regime of large system sizes and long evolution times. We find the fidelity (at $N=60$ and the latest experimental time) is generally consistent over the entire range, with convergence achieved around a bond dimension of 1500.
        } 
	\vspace{-0.0cm}
	\label{EFig:learnable}
\end{figure*}

\paragraph*{Number of learnable parameters---}
For the full dataset, a total of ${\sim}23000$ training data points are generated, while depending on the choice of hyperparameters, the number of learnable parameters can vary from ${\sim}350-83000$. For ${\sim}70\%$ of the Monte Carlo instances, the number of learnable parameters is lower than the number of training points (Ext. Data Fig.~\ref{EFig:learnable}ab). Further we do not see any strong correlation between the predicted fidelity and the number of learnable parameters, or more generally between the fidelity and either network depth or width.
\\

\paragraph*{Reducing the experimental dataset---}
We perform the same training procedure on several variants of reduced experimental datasets, where we excise from the training half of the bond dimensions, system sizes, or evolution times used. We find the resulting Monte Carlo mean is always consistent with the prediction from training on the full dataset (Ext. Data Fig.~\ref{EFig:learnable}c).
\\

\paragraph*{Maximum bond dimension used---}
The maximum bond dimension used in the training is constrained by our computational resources, but we can still study if the Monte Carlo prediction appears to be converged as a function of the bond dimension. In Ext. Data Fig.~\ref{EFig:learnable}d we systematically scale the training bond dimension used in the breakdown regime of large system sizes and long evolution times, and find that the results appear converged with a bond dimension of ${\sim}1500$, about a factor of 2 lower than the maximum bond dimension we consider in this work. However, we note that by supplementing the experimental data with numerical data obtained from our error model simulations, this required bond dimension is dramatically reduced (see Section G, Ext. Data Fig.~\ref{EFig:montecarloscale}).

\clearpage
\newpage
\subsection{Scalability of fidelity estimation}
\subsubsection{Fidelity estimation with approximate algorithms for Markovian noise}
A major advantage of the fidelity estimator which we use, $F_d$, is its accuracy at early times~\cite{Mark2023BenchmarkingLett,Choi2023PreparingChaos}. This is in contrast to the original linear cross-entropy which only becomes accurate for deep circuits~\cite{Arute2019QuantumProcessor}. As we establish in the main text, using tensor network simulation methods such as MPS, we are able to compute exact classical simulation references for essentially arbitrarily large systems up to constant time. Assuming the noise affecting the system is purely Markovian (i.e. the fidelity decays exponentially~\cite{Dalzell2021RandomNoise}), then to estimate the fidelity in the late-time, high-entanglement regime it is sufficient to measure the fidelity with high-precision at early times, and then fit the resulting exponential decay rate. 

This gives another axis for fidelity extrapolation which may be particularly useful for digital quantum devices implementing random unitary circuits, which typically use a form of system size scaling to estimate the fidelity in the high-entanglement regime~\cite{Morvan2023PhaseSampling}. Such an approach should work for fidelity estimation of systems with largely arbitrary size and dimension, but does require the exact functional form of the fidelity decay be assumed \textit{a priori}, and thus would not apply for systems with more non-trivial noise sources like those affecting our present experiment (or potentially also some digital quantum computers~\cite{Arute2019QuantumProcessor}).

\subsubsection{Scalability of Monte Carlo inference for non-Markovian noise sources}
In Ext. Data Fig.~\ref{EFig:learnable}d, we investigated the minimum bond dimension of the classical (approximate) simulation, $\chi_\text{extrap}$, required for the Monte Carlo inference procedure to reach a reliable, saturated prediction for the fidelity at $N=60$ at the latest experimental time (14.5 cycles). There. we found $\chi_\text{extrap}$ saturated for a bond dimension around $\chi\sim1500$. However, an important question concerns the applicability at the Monte Carlo inference technique at larger system sizes, i.e. the scaling of $\chi_\text{extrap}$ with $N$. If the required bond dimension scales poorly with system size then it becomes difficult to trust the Monte Carlo inference results at larger system sizes, assuming there is a cap in the maximum bond dimension usable by the classical simulation. 

\begin{figure*}[h!]
	\centering
    \vspace{3mm}
	\includegraphics[width=180mm]{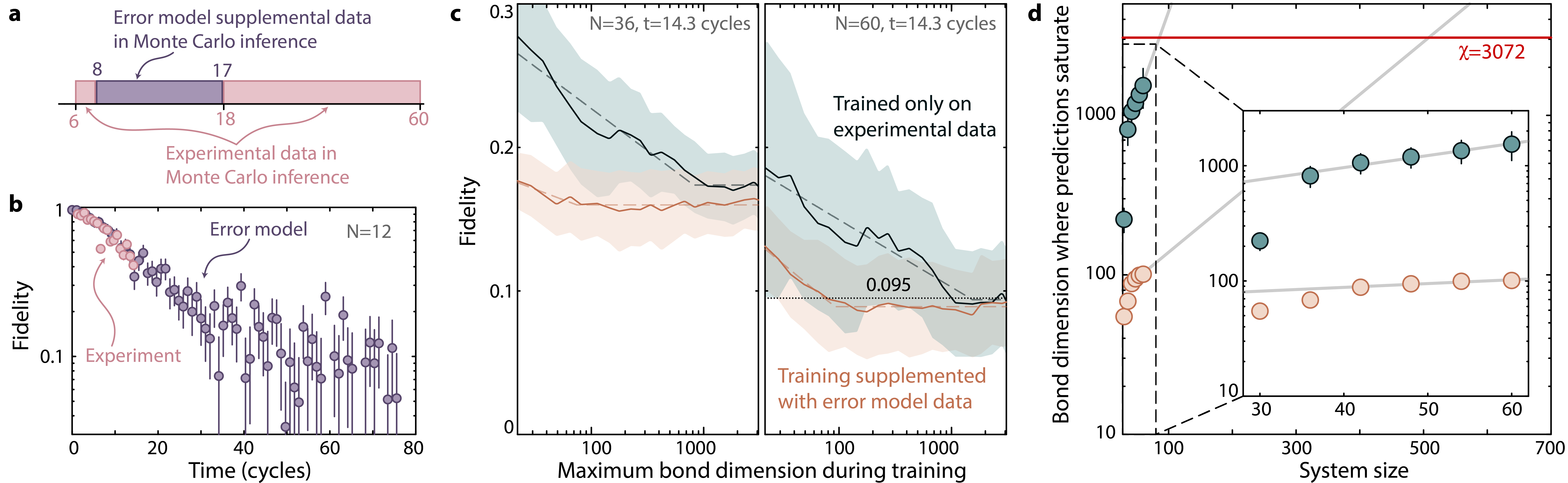}
	\caption{\textbf{Scalability of Monte Carlo inference. a.} In order to improve scalability of our Monte Carlo inference protocol, we supplement training with error model data evaluated out to late evolution times for system sizes from $N=8$ and $N=17$. \textbf{b.} Error model and experimental $F_d$ values are in good agreement at early times where we have data (see also Ext Data. Fig.~\ref{EFig:fd_vs_f}). Error bars in the late time regime are due to typicality error which reduces with increasing system size (see Ext. Data Fig.~\ref{EFig:sample_complexity}). \textbf{c.} We plot the predicted fidelity for $N=36$ and $N=60$ at the latest experimental time for two possible sets of training data: 1) considering only the experimental data (from $N=6$ to $N=60$ up to 15 cycles), and 2) supplementing with error model data out to late times (75 cycles). In both cases we vary the maximum bond dimension used during training, and then fit to find the bond dimension where predictions saturate, $\chi_\text{extrap}$. Fits are proportional to $\log(\chi)$ and then constant after $\chi_\text{extrap}$ (dashed lines). \textbf{d.} $\chi_\text{extrap}$ as a function of system size for the predicted fidelity to saturate at the latest experimental time. The $\chi_\text{extrap}$ is improved by roughly an order of magnitude by supplementing with error model data. Conservatively extrapolating with $\chi_\text{extrap}\propto\exp(N)$ (grey lines) and setting a maximum $\chi=3072$ shows maximum system sizes of ${\sim}500$ and ${\sim}90$ with and without the error model supplement. 
        } 
	\vspace{-0.0cm}
	\label{EFig:montecarloscale}
\end{figure*}

We consider two versions of the training procedure (Ext. Data Fig.~\ref{EFig:montecarloscale}). In the first, we perform Monte Carlo inference using just the experimental data, as done previously. In the second approach we replace the training data from $N=8$ to 17 with data obtained from error model simulations up to 75 evolution cycles (sampled every 1 cycle). $F_d$ values are calculated for error model data for $\chi$ ranging from 1 to 90. We replace with error model results, rather than simply augmenting, so that the training data does not have multiple outputs defined for the same set of inputs. 

We have extensively checked the validity of our error model both in terms of fidelity and KL divergence (also see Refs.~\cite{Choi2023PreparingChaos,Scholl2023ErasureSimulator} and Ext. Data Fig.~\ref{EFig:fd_vs_f}), and using error model allows us to generate training data for $F_d$ as a function of bond dimension in the low-$N$, long-time regime where the non-exponential fidelity decay becomes more clearly visible. We emphasize that access to an error model is not a general requirement, but is instead simply a tool we are employing in this case as the original experimental dataset was planned and taken without accounting for the non-exponential fidelity decay, as it was a novel observation of our present analysis. Using the error model alone (with no experimental data) for inference is not expected to give good fidelity extrapolation at large system sizes given the large difference between the largest simulable error model size $(N\approx20)$ and the largest experimental size.

In both cases (purely experimental data, or supplemented with error model data), we plot $\chi_\text{extrap}$ as a function of system size. Here $\chi_\text{extrap}$ is found from a phenomenological fit of the predicted fidelity, $F_p$, as a function of the maximum bond dimension used in training, $\chi$,
\begin{align}
F_p=\begin{cases} 
      F + A \log(\chi) & \chi<\chi_\text{extrap} \\
      F & \chi\geq\chi_\text{extrap},
   \end{cases}
\end{align}
where $F$ and $A$ are free fit parameters (Ext. Data Fig.~\ref{EFig:montecarloscale}c). In general, we see that supplementing with error model data produces consistent fidelity predictions as our original, all-experimental inferences, but saturates far earlier. This builds additional confidence in both the error model and the Monte Carlo inference procedure.

We then study how $\chi_\text{extrap}$ scales as a function of $N$, and set a cap of $\chi\sim3072$, which represents a significant, but still achievable bond dimension. We choose a conservative phenomological fit model of 
\begin{align}
    \chi_\text{extrap}=A\exp(BN),
\end{align}
where $A$ and $B$ are fit constants.

We find the all-experimental inference exceeds this cap for $N\sim90$, but the supplemented dataset goes as high as $N\sim500$. We emphasize that this value represents a lower bound that could be improved either through use of larger classical resources, or possibly through an improved classical algorithm. This indicates that as long as care is taken to acquire experimental data in a diverse regime of system sizes and evolution times, we conservatively believe the Monte Carlo inference procedure we show here should provide scalable predictions for an order of magnitude more atoms than we currently study.

\subsubsection{Sample complexity}
To estimate error bars on the raw data, $\sigma_F$, we propagate from separately calculating the sampling error of the numerator and denominator of $F_d$ as 
\begin{align}
\sigma=\sqrt{\left(\mathbb{E}(\tilde{p}(z)^2)-\mathbb{E}(\tilde{p}(z))^2\right)/M},
\label{eq:error_bars}
\end{align}
where $\tilde{p}$ is the rescaled bitstring probability distribution obtained from MPS simulation, $M$ is the number of samples, and the expectation value is taken over the set of experimentally (numerically) sampled bitstrings for the numerator (denominator). Enough samples are taken numerically that the dominant overall sampling error is from the experiment. See Section B for more details.

\begin{figure}[ht!]
	\centering
	\includegraphics[width=181mm]{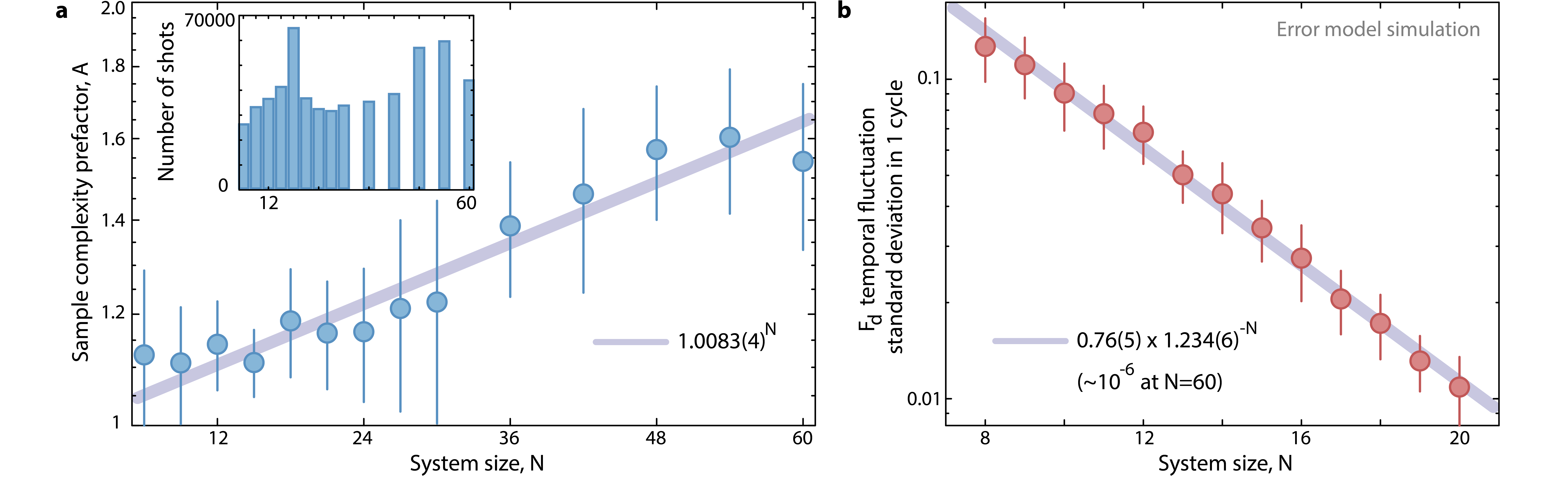}
	\caption{\textbf{Uncertainty sources of the benchmarking protocol.} There are two fundamental sources of noise in calculating $F_d$: statistical error, and systematic error arising from so-called typicality fluctuations~\cite{Choi2023PreparingChaos}. \textbf{a.} We first evaluate the statistical error behavior. At a fixed fidelity, the statistical error bars of our protocol are given~\cite{Mark2023BenchmarkingLett} by $A/\sqrt{M}$, where $M$ is the number of samples and $A$ is the sample complexity. At approximately a fixed fidelity of $F_d=0.5$ (see text), we find $A$ grows weakly with system size, indicating favorable scalability of our protocol from a statistical perspective. For this work, $M{\sim}40000$ for all system sizes, summing over all times (inset). \textbf{b.} Typicality error manifests as temporal fluctuations of the estimator $F_d$, with respect to the true fidelity $F$, due to benchmarking in a finite-sized Hilbert space. The standard deviation of these fluctuations scales~\cite{Choi2023PreparingChaos} as $1/\sqrt{D}$, where $D$ is the Hilbert space dimension. From error model simulations, we calculate the standard deviation of temporal fluctuations of $F_d$ in a 1-cycle period at late times, which we find decreases as $1.23^{-N}$ (or ${\sim}1.2/\sqrt{D}$), implying a negligible typicality error of ${\sim}10^{-6}$ at $N=60$.
        } 
	\vspace{2.0cm}
	\label{EFig:sample_complexity}
\end{figure}

We expect the statistical error of our benchmarking protocol~\cite{Choi2023PreparingChaos,Mark2023BenchmarkingLett} to scale as $A/\sqrt{M}$, where $A$ is the \textit{sample complexity} which scales exponentially in system size with a base which depends on the fidelity~\cite{Mark2023BenchmarkingLett}. We measure $A$ at an approximately fixed fidelity of $F_d=0.5$ by performing an average of all the experimentally measured error bars from Eq.~(\ref{eq:error_bars}), weighted by their inverse distance from $F_d=0.5$, such that
\begin{align}
A(N) = \frac{\sum_t \sigma_F(N,t)\sqrt{M(N,t)}w(N,t)}{\sum_t w(N,t)},
\end{align}
with $w(N,t)=1/|F_d(N,t)-0.5|$. With this approach, we find $A=1.0083(4)^N$ (Ext. Data Fig.~\ref{EFig:sample_complexity}a).

\clearpage
\newpage
\subsection{Extrapolating pure state entanglement growth}
We target the generation of high entanglement entropy states through effectively infinite-temperature quench dynamics. In order to estimate the mixed state negativity, we first must estimate the pure state entanglement, even for system sizes for which we can no longer exactly simulate the dynamics. To do this, we extrapolate the scaling behavior from small system sizes. Specifically, we fit entanglement growth, $S$, with the following functional form as a function of time and system size:
\begin{align}
S = S_0 + m_1 t + \frac{(m_2-m_1)(t-\tent-\sigma)}{1+\exp(-\frac{t-\tent}{\sigma})}.
\end{align}

To respect the physical constraints of the entanglement growth, we fix $m_1$ -- the early time linear entanglement growth -- to be system size independent. The Rydberg Hamiltonian induces fast entanglement growth over the first ${\sim}0.5$ cycles of evolution which makes fitting the linear growth erroneous without also including a system-size independent offset, $S_0$. We then expect $\tent$, which we identify with the approximate entanglement saturation time, to scale linearly with $N$, which we empirically find is true also for $\sigma$ (a smoothness parameter). We further empirically find that $m_2$ (the shallow entanglement growth slope after $\tent$) scales approximately quadratically with $N$. 

\begin{figure*}[h!]
	\centering
 \vspace{5mm}
	\includegraphics[width=86mm]{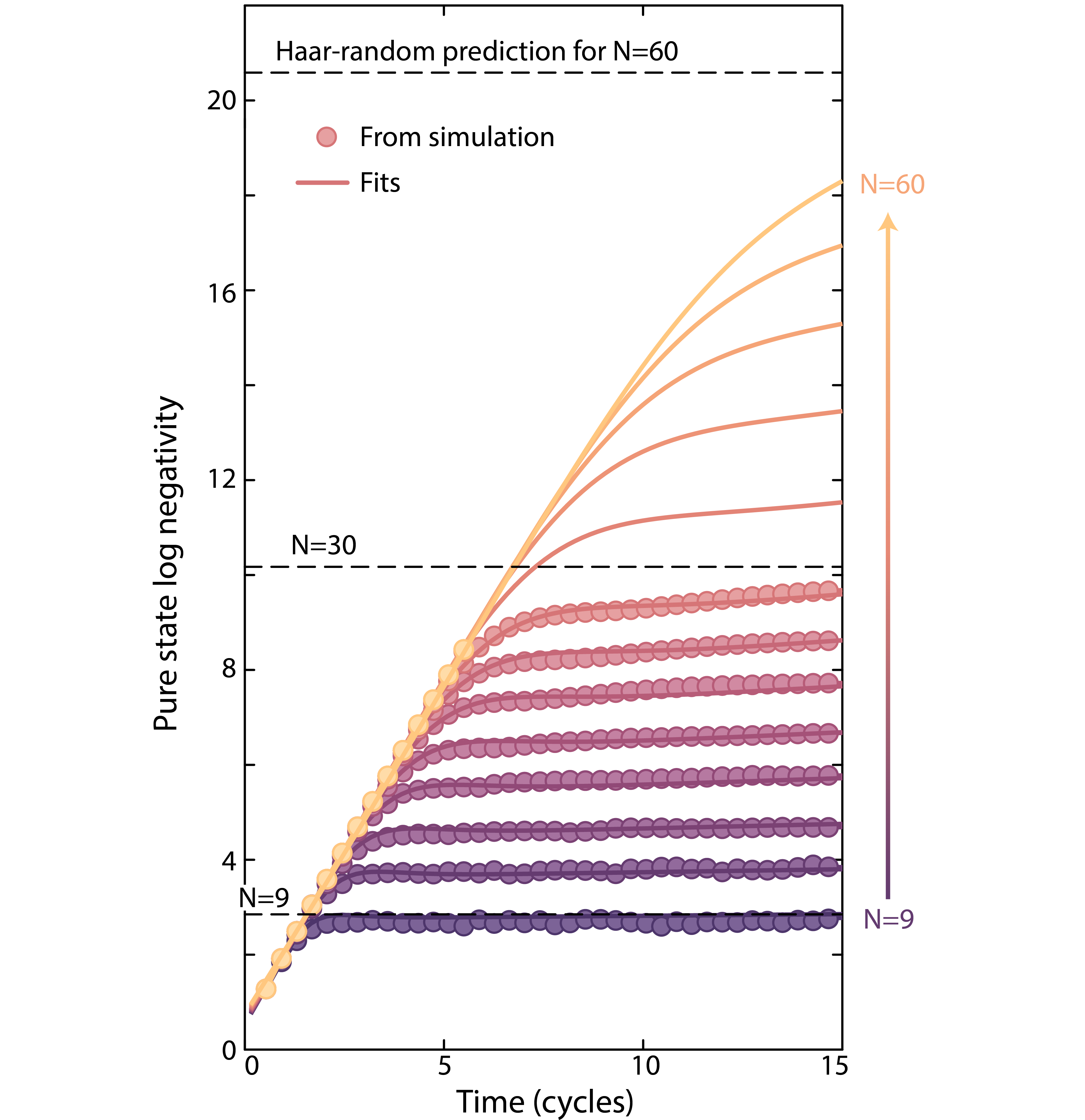}
	\caption{\textbf{Pure state entanglement growth dynamics for the Rydberg Hamiltonian.} We directly calculate the entanglement (here log negativity) for system sizes up to $N=60$, and fit the resulting data (see text) to extrapolate the entanglement dynamics for large $N$ at late times. We also compare against the prediction for Haar random states at various system sizes (black dashed lines), and find that as system size increases we gradually undershoot this expectation more, which we attribute potentially to imperfect parameter selection making the dynamics effectively less than infinite temperature at large $N$.
        } 
	\vspace{-0.0cm}
	\label{EFig:entanglement_growth}
\end{figure*}

We show the results of these fits in Ext. Data Fig.~\ref{EFig:entanglement_growth}, where markers are measured directly from simulation, and lines are fits. Though here we show log negativity, the von Neumann entropy scales and acts in the same way. 

It is important to note that as system size increases, we gradually reach relatively less entangled states at $\tent$ as compared to the expectation for a Haar random state (dashed lines in Ext. Data Fig.~\ref{EFig:entanglement_growth}). Given the Rydberg blockade constraint, we define the Haar random prediction as $S_\mathrm{Haar}\approx\log_2(D_0)-0.47$, where $D_0$ is the blockaded Hilbert space dimension of the half-cut chain, and the numerical offset of -0.47 is a type of Page correction~\cite{Page1993AverageSubsystem}, which is actually~\cite{Bhosale2012EntanglementStatistics,Datta2010NegativityStates} $\approx\log_2(\frac{64}{9\pi^2})$, as explained further below.

\newpage
\clearpage
\subsection{Bounds on the mixed state entanglement proxy}
In this work, we have introduced the entanglement-proxy, $\mathcal{E}_P$, which estimates the mixed state entanglement, $\mathcal{E}_N(\hat{\rho})$, in terms of the pure state entanglement, $\mathcal{E}_N(|\psi\rangle)$, and the fidelity, $F \equiv \bra{\psi}\hat{\rho}\ket{\psi}$ (see Eq.~\eqref{eqn:ln_lower_bound} of the main text).

\begin{figure}[b!]
	\centering
	\includegraphics[width=181mm]{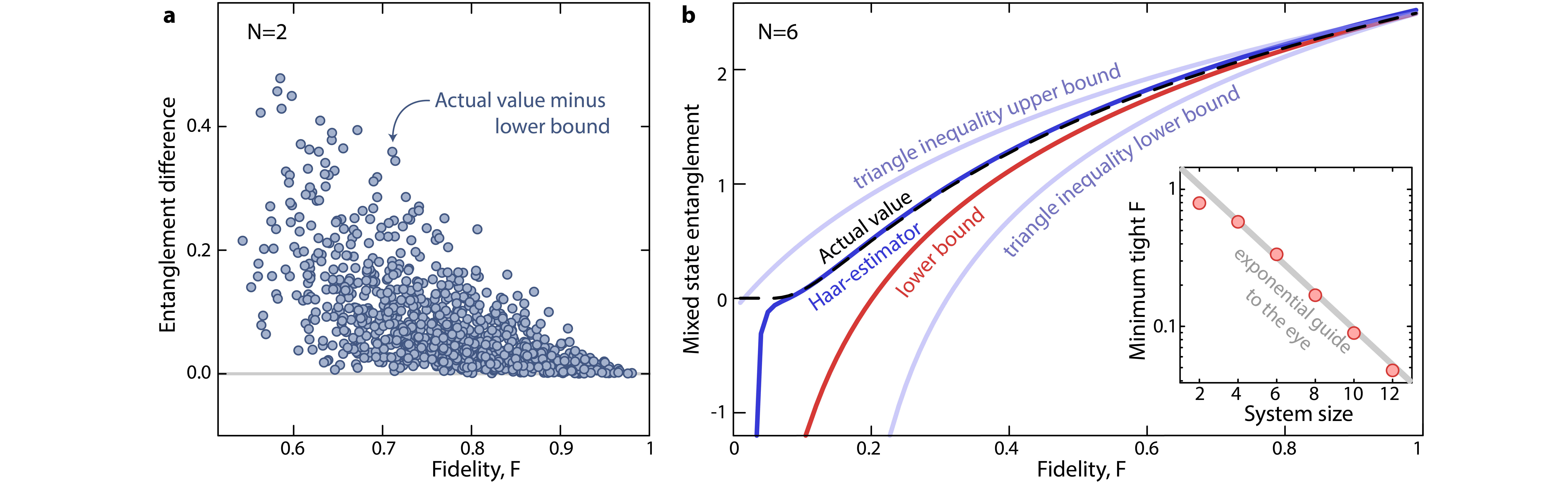}
	\caption{\textbf{Estimation of log negativity from fidelity. a.} We demonstrate the validity of the lower bound Eq.~\eqref{eqn:ln_lower_bound} of the main text, by plotting the difference $\mathcal{E}_N(\hat{\rho}) - \mathcal{E}_P$, with $\ket{\psi}$ the pure state with highest fidelity to $\hat{\rho}$. Here, we generate 1000 random two-qubit mixed states as uniform incoherent mixtures of two Haar random states. The difference is always above zero, indicating the validity of the lower bound. \textbf{b.} Dependence between log negativity $\mathcal{E}_N(\hat{\rho})$ and fidelity $F$ for globally depolarized Haar random states, Eq.~\eqref{eq:Haar_random_states}. For small values of $F$, the negativity $\mathcal{E}_N(\hat{\rho})$ (black dashed) deviates from the lower bound $\mathcal{E}_P$ (red), illustrated here for the half-chain negativity of six-qubit depolarized Haar random states. Remarkably, for such states, $\mathcal{E}_N(\hat{\rho})$ is uniquely determined by the fidelity $F$ and Hilbert space dimension $D$. We find an analytic series expression for this dependence, Eq.~\eqref{eq:Haar_negativity_estimator}, with first-order correction illustrated in blue. We also illustrate general lower and upper bounds for the log negativity, Eq.~\eqref{eq:triangle_ineq_bounds}, in light blue. Inset: The lower bound (red) improves exponentially as a function of system size, quantified here as the minimum fidelity for which the lower bound is $90\%$ of the actual value.
        } 
	\vspace{-0.0cm}
	\label{EFig:negativity_bounds}
\end{figure}

Explicitly, consider a system bipartitioned into two subsystems, $A$ and $B$, and described by density matrix, $\hat{\rho}_{AB}$. The log negativity~\cite{Vidal2002ComputableEntanglement} is given by 
\begin{align*}
\mathcal{E}_N=\log_2||\hat{\rho}_{AB}^{T_A}||_1,
\end{align*}
where $\hat{\rho}_{AB}^{T_A}$ is the partially transposed density matrix, and $||\cdot||_1$ is the trace norm. More plainly, the log negativity measures the log absolute sum of all eigenvalues of $\hat{\rho}_{AB}^{T_A}$. If $\hat{\rho}_{AB}$ is separable, the log negativity is zero, but for entangled states it generically serves as an upper bound to the distillable entanglement of the system~\cite{Plenio2005LogarithmicConvex}, and is an entanglement monotone~\cite{Plenio2005LogarithmicConvex}. For pure states the log negativity is equivalent to the R\'enyi-1/2 entropy.
\\

\noindent The following discussion is generally split into three subsections:

In the first, we study lower bounds and estimators of the mixed state entanglement. We prove the validity of our entanglement-proxy, $\mathcal{E}_P$, for the case when the target pure state, $|\psi\rangle$ is an eigenstate of the mixed state, $\hat{\rho}$, as for instance is the case if $|\psi\rangle$ is the highest fidelity pure state to $\hat{\rho}$. We further improve our result for the case where $|\psi\rangle$ is a Haar-random state. Finally, we prove more general unconditional upper and lower bounds to argue that our entanglement-proxy becomes exponentially tighter as the system size increases.

Next, we discuss possible violations of the lower bound given by $\mathcal{E}_P$, if we allow $|\psi\rangle$ to no longer be an eigenstate of $\hat{\rho}$. We consider two physically realistic possibilities: 1) globally correlated coherent errors, and 2) incoherent local errors. For each case we provide numeric or analytic evidence showing that at worst our entanglement-proxy lower bound is potentially violated by an $\mathcal{O}(1)$ amount.

Finally, we make a specific connection between the entanglement content of an erroneous mixed state, and of a truncated MPS state. We show that at equal fidelity, the truncated MPS is less entangled than the depolarized Haar random state, which explains an observed discrepancy between the experimental mixed state entanglement and $\chi^*$ found in the main text.

As a brief aside, we note an important point, arising from the fact that in the main text we have used either $F_d$ as the fidelity for experiment, or the linear cross-entropy, $F_\mathrm{XEB}$, as the fidelity for the literature examples. However, $F_d$ and $F_\mathrm{XEB}$ are both lowered by measurement error, which strictly speaking does not affect the mixed state entanglement. However, both $F_d$ and $F_\mathrm{XEB}$ are expected to decrease due to measurement errors, and so this only serves to make our lower bound less tight. In principle the mixed state entanglement-proxy could be measurement-corrected to account for this effect, but we choose to include measurement errors as a more general quality-factor of the experiments.

\subsubsection{Negativity lower bounds and estimators}
The validity of our entanglement-proxy has been shown in the past~\cite{Lee2003Convex-roofSystems} for the case of \textit{isotropic states}; here we extend this result and validate our entanglement-proxy when the target pure state, $|\psi\rangle$, is an eigenstate of the mixed state, $\hat{\rho}$ (Theorem~\ref{thm:simple_bound}). We then improve our result for the case of depolarized Haar-random pure states (Lemma~\ref{cor:Haar_random}), and then show more general lower bounds which allow us to argue our entanglement-proxy should tighten exponentially with system size (Lemma~\ref{cor:Triangle_ineq}). Numerical support of these analytical claims is shown in Ext. Data Fig.~\ref{EFig:negativity_bounds}.

We begin with the case of $|\psi\rangle$ being an eigenstate of $\hat{\rho}$
\begin{theorem}
\label{thm:simple_bound}
For any mixed state $\hat{\rho}$ and pure state $\ket{\psi}$ which is an eigenstate of $\hat{\rho}$, the logarithmic negativities $\mathcal{E}_N(\hat{\rho})$ and $\mathcal{E}_N(|\psi\rangle)$ are related by
\begin{equation}
    \mathcal{E}_N(\hat{\rho}) \geq \mathcal{E}_P \equiv \mathcal{E}_N(\ket{\psi}) + \log_2(F)~, \label{eq:negativity_lower_bound_app}
\end{equation}
where $F \equiv \bra{\psi}\hat{\rho}\ket{\psi}$ is their fidelity.
\end{theorem}

\vspace{5mm}
\begin{proof}
The proof of this Theorem follows in three steps. It is equivalent to derive the bound
\begin{equation}
    \Vert \hat{\rho}^{T_A} \Vert_1 \geq F \Vert \ketbra{\psi}^{T_A}\Vert_1~.
\label{eq:negativity_bound_proof}
\end{equation}
First, we express the state 
\begin{equation}
\hat{\rho} = F \ketbra{\psi} + (1-F) \hat{\rho}_\perp
\end{equation}
for some positive semi-definite state $\hat{\rho}_\perp$, which is guaranteed because $\ket{\psi}$ is an eigenstate of $\hat{\rho}$.

Next, we use the variational definition of the trace-norm as a maximum over projectors $\hat{P}$, i.e.~operators with eigenvalues either 0 or 1. 
\begin{equation}
    \Vert \hat{\rho}^{T_A} \Vert_1 = 2 \max_{P} \text{tr}(\hat{P} \hat{\rho}^{T_A}) -1~.
\end{equation}
This maximum is attained when $\hat{P}$ is the projector onto the positive-eigenvalue subspace of $\hat{\rho}^{T_A}$. This variational definition gives the lower bound:
\begin{equation}
    \Vert \hat{\rho}^{T_A} \Vert_1 \geq F \Vert \ketbra{\psi}^{T_A} \Vert_1 + (1-F) (2 \text{tr}(\hat{P}_\psi \hat{\rho}_\perp^{T_A}) -1)~,
    \label{eq:negativity_bound_step}
\end{equation}
where $\hat{P}_\psi$ is the projector onto the positive-eigenvalue subspace of $\ketbra{\psi}^{T_A}$.

Finally, we utilize an explicit construction of $\hat{P}_\psi$ to show that $\text{tr}(\hat{P}_\psi \hat{\rho}_\perp^{T_A}) \geq 1/2$, giving the desired bound Eq.~\eqref{eq:negativity_bound_proof}. This step is the most involved, which we show below.

The eigenvalues $\lambda$ and eigenvectors $\ket{v}$ of $\ketbra{\psi}^{T_A}$ can be expressed in terms of the Schmidt values $s_j$ and Schmidt basis $\ket{a_j}, \ket{b_j}$ of $\ket{\psi}$, with respect to the same bipartition~\cite{Lee2003Convex-roofSystems}. With $\ket{\psi} = \sum_j s_j \ket{a_j}\ket{b_j}$, the eigenvalues $\lambda$ are $s_j^2$ or $\pm s_i s_j$ (for $i\neq j$), and their corresponding eigenvectors are respectively $\ket{v_{jj}} \equiv \ket{a_j}\ket{b_j}$ and $\ket{v_{ij}^\pm} \equiv \frac{1}{2}\left(\ket{a_i}\ket{b_j} \pm\ket{a_j}\ket{b_i}\right)$. Then $\hat{P}_\psi$ has explicit expression:
\begin{equation}
    \hat{P}_\psi = \sum_j \ketbra{v_{jj}} + \sum_{i<j} \ketbra{v^+_{ij}}~.
\end{equation}
Next, we use the relation $\text{tr}(\hat{M}^{T_A}\hat{N}^{T_A}) = \text{tr}(\hat{M} \hat{N})$, valid for all operators $\hat{M}$ and $\hat{N}$ defined on $AB$. Thus $\text{tr}(\hat{P}_\psi \hat{\rho}_\perp^{T_A}) = \text{tr}(\hat{P}_\psi^{T_A}\hat{\rho}_\perp)$. $\hat{P}_\psi^{T_A}$ can be expressed as 
\begin{equation}
    \hat{P}_\psi^{T_A} = \frac{1}{2} \left(I+ \ketbra{\Phi}\right)~, 
\end{equation}
where $\ket{\Phi} \equiv \sum_j \ket{a_j}\ket{b_j}$ is the \textit{unnormalized} maximally entangled state in the Schmidt bases $\{|a_j\rangle\}$ and $\{|b_j\rangle\}$. Since $\hat{\rho}_\perp$ is a positive semidefinite state, we have 
\begin{equation}
\text{tr}(\hat{P}_\psi^{T_A}\hat{\rho}_\perp) = \frac{1}{2} \text{tr}(\hat{\rho}_\perp) +  \frac{1}{2} \bra{\Phi} \hat{\rho}_\perp \ket{\Phi} \geq \frac{1}{2}~.
\end{equation}
This proves the desired bound Eq.~\eqref{eq:negativity_bound_proof}.
\end{proof}
\vspace{5mm}

This Theorem validates the entanglement proxy in the simplest case. We next move to improve it, first specifically in the case where $|\psi\rangle$ is taken to be a Haar random state, $|\psi_\mathrm{Haar}\rangle$.

\begin{lemma}
\label{cor:Haar_random}
For depolarized Haar random states, i.e.~states of the form
\begin{equation}
    \hat{\rho} = F' \ketbra{\psi_\mathrm{Haar}} + (1-F')\frac{I}{D}~, \label{eq:Haar_random_states}
\end{equation}
where $F' = F - (1-F)/(D-1)$ (chosen so the fidelity $\bra{\psi_\mathrm{Haar}} \hat{\rho} \ket{\psi_\mathrm{Haar}}$ equals $F$), the mixed state entanglement has the form
\begin{equation}
 \mathcal{E}_N(\hat{\rho}) \approx \mathcal{E}_N(\ket{\psi}) + \log_2(\mathfrak{f}(F))~,\label{eq:haar_random_neg_pred}
\end{equation}
where the more general function $\mathfrak{f}$ can be expressed to low order as
\begin{align}
    \mathfrak{f}(F) = F'\label{eq:Haar_negativity_estimator}+\frac{9}{64} (F' D_A)^{-1} \left(2\ln(F' D_A)+8 \ln 2 - 1\right)+ \mathcal{O}((F'D_A)^{-3})~,
\end{align}
where $D_A$ is the Hilbert space dimension of subsystem $A$, here assumed to be for a half-chain bipartitions.

\end{lemma}

\vspace{5mm}
\begin{proof}

Here, we restrict our discussion to equal bipartitions, although our results can be generalized to any bipartition.

First, with Haar random pure states, with high probability we have~\cite{Bhosale2012EntanglementStatistics,Datta2010NegativityStates}
\begin{equation}
    \mathcal{E}_N(\ket{\psi}) = \frac{N}{2} + \log_2\frac{64}{9 \pi^2} \approx \frac{N}{2} -0.472
\end{equation}

This expression follows from random matrix theory (RMT), which states that the Schmidt values $s_j$ of a Haar random state follow the \textit{quarter-circle law}, while $s_j^2$ follow the \textit{Marchenko-Pastur distribution}.

This enables us to compute the distribution of eigenvalues $\lambda$ of $\ketbra{\psi}^{T_A}$. The eigenvalues of the form $\lambda = s_j^2$ contribute a constant amount to $\Vert \hat{\rho}^{T_A}\Vert_1$. The dominant (exponentially larger) contribution to the log negativity comes from the eigenvalues of the form $\lambda = \pm s_i s_j$ which follow the distribution~\cite{Znidaric2007DetectingWitness}
\begin{equation}
    P(\tilde{\lambda}) = \frac{2}{\pi^2} \left[\left(1+\frac{\tilde{\lambda}^2}{16}\right)K\left(1-\frac{\tilde{\lambda}^2}{16}\right) - 2 E\left(1-\frac{\tilde{\lambda}^2}{16}\right)\right] ~, 
\end{equation}
where $\tilde{\lambda} \equiv D_A \lambda$, $ P(\tilde{\lambda})$ is supported in $\tilde{\lambda} \in [-4,4]$, and $K$ and $E$ are complete elliptic integrals.

The negativity of the depolarized state $\hat{\rho}$ can be calculated using this distribution: the eigenvalues $\lambda'$ of $\hat{\rho}^{T_A}$ are given by $\lambda' = F' \lambda + (1-F')/D$. Therefore, the trace norm can be calculated from $P(\tilde{\lambda})$ as 
\begin{align}
    & \Vert \hat{\rho}^{T_A}\Vert_1 \approx  2 F' D_A \int_0^\infty \tilde{\lambda} P(\tilde{\lambda}) d \tilde{\lambda} + 2 F' D_A \int_0^{(F'D_A)^{-1}} \left[(F'D_A)^{-1} - \tilde{\lambda} \right] P(\tilde{\lambda}) d\tilde{\lambda}
\end{align}
The first term evaluates to $\frac{64}{9 \pi^2} F' D_A$, while the second term can be systematically computed by using the expansion $P(\tilde{\lambda}) \approx - \frac{2}{\pi^2} \ln \vert\tilde{\lambda}\vert + \frac{8 \ln 2 - 4}{\pi^2}$, valid for small $\tilde{\lambda}$, giving Eq.~\eqref{eq:Haar_negativity_estimator}.
\end{proof}
\vspace{5mm}
Importantly, if more terms are added in the expansion of Eq.~\eqref{eq:Haar_negativity_estimator} Lemma~\ref{cor:Haar_random} is an exact estimator of the mixed state entanglement. Further, if $D_A$ is large, as is the case for large system sizes, then the original simple proxy is restored as $\mathfrak{f}\sim F'\sim F$. We show the efficacy of this estimator, with $\mathfrak{f}(F)$ evaluated to first order, in Ext. Data Fig.~\ref{EFig:negativity_bounds}b.

Finally we prove an even more general set of upper and lower bounds. While not amenable to be computed experimentally, they allow us to argue our entanglement-proxy likely becomes exponentially tight as the system size increases.
\begin{lemma}
\label{cor:Triangle_ineq}
For any state pure state $|\psi\rangle$ and mixed state $\hat{\rho}$ with decomposition
\begin{equation}
\hat{\rho} = F \ketbra{\psi} + (1-F) \hat{\rho}_\perp~,
\end{equation}
where $\hat{\rho}_\perp$ is the (not necessarily positive semi-definite) remainder of the noisy state, we can bound the mixed state log negativity by 
\begin{align}
&\mathcal{E}_N(\hat{\rho}) \geq \log_2 \left(F 2^{\mathcal{E}_N(|\psi\rangle)} - (1-F) \Vert \hat{\rho}^{T_A}_\perp \Vert_1\right) \label{eq:triangle_ineq_bounds}\\
&\mathcal{E}_N(\hat{\rho}) \leq \log_2 \left( F 2^{\mathcal{E}_N(|\psi\rangle)} + (1-F) \Vert \hat{\rho}^{T_A}_\perp \Vert_1\right)~,\nonumber
\end{align}
where $F$ is a lower bound for the fidelity $\bra{\psi}\hat{\rho}\ket{\psi}$.
\end{lemma}

\vspace{5mm}
\begin{proof}
    Unlike above, $\hat{\rho}_\perp$ does not necessarily commute with $\ketbra{\psi}$.

    To bound the negativity for such a state, we simply apply the triangle inequality on the trace norm $\Vert \cdot \Vert_1$, yielding Eqs.~\eqref{eq:triangle_ineq_bounds}.
\end{proof}
\vspace{5mm}

This bound is particularly useful in settings such as ours. Here, we can expect the noisy state to be weakly entangled, i.e.~$\Vert \hat{\rho}^{T_A}_\perp \Vert_1 = \mathcal{O}(1)$, while the pure state is highly entangled: $\Vert \ketbra{\psi}^{T_A} \Vert_1 = \mathcal{O}(\exp(N))$, where $N$ is the system size. Then the above bounds imply that our entanglement-proxy is in fact exponentially close to the true mixed state negativity: $\mathcal{E}_N(\hat{\rho}) = \mathcal{E}_P + \mathcal{O}(\exp(-N))$, as observed in the inset of Ext.~Data Fig.~\ref{EFig:negativity_bounds}b.

\subsubsection{Robustness of our bounds to realistic experimental errors}
In the above section, we showed several proofs which let us argue our entanglement-proxy, $\mathcal{E}_P$ is a tight lower bound for the mixed state entanglement under specific assumptions. Here, we relax these assumptions, but show that under realistic noise sources, maximum possible violations of our bound are at most $\mathcal{O}(1)$.

\vspace{5mm}
\noindent\textbf{Error model simulations--}
To start, we reiterate and emphasize the results of Fig.~\ref{Fig4}a of the main text. There, we showed numerically that the entanglement-proxy appears to be a genuine lower bound of the mixed state entanglement either for a random unitary circuit (RUC) undergoing incoherent local noise, or for our experiment undergoing the full set of error sources (Ext. Data Fig.~\ref{EFig:error_model}). 

Now, our aim is to provide further analytical and numerical evidence for the robustness of the entanglement-proxy under: 1) global coherent errors, or 2) incoherent local errors.

\vspace{5mm}
\noindent\textbf{Global coherent errors--} In this section, we consider potential violations of our bound due to global coherent errors. To study this, we first prove a Theorem for a more general bound of the mixed state entanglement, based on studying the norm of the commutator between the mixed state $\hat{\rho}$ and pure state $\ket{\psi}$. This bound is equivalent to our original entanglement-proxy up to a correction term, which we then show analytically is small if the normalized entanglement of the target state is large. Finally, we present numerical support of this discussion.

\begin{theorem}[Commutator based bound of the mixed state entanglement]
\label{thm:commutator_negativity_bound}
For any mixed state $\hat{\rho}$ and pure state $\ket{\psi}$ whose commutator has Frobenius norm $\mathcal{C} \equiv \Vert \left[\hat{\rho},|\psi\rangle\langle\psi|\right] \Vert_F$, the logarithmic negativities $\mathcal{E}_N(\hat{\rho})$ and $\mathcal{E}_N(|\psi\rangle)$ are related by
\begin{equation}
\mathcal{E}_N(\hat{\rho}) \geq \mathcal{E}_N(|\psi\rangle)+\log_2(F)+ 2\log_2(1 - (\mathcal{C}/F)\sqrt{(1-\alpha)/(2\alpha)})~.
\label{eq:commutator_negativity_bound}
\end{equation}
where $F \equiv \bra{\psi}\hat{\rho}\ket{\psi}$ is the fidelity, $\alpha \equiv 2^{\mathcal{E}_N(|\psi\rangle)}/D_A \in [0,1]$ is a normalized entanglement of $\ket{\psi}$, and $D_A$ is the Hilbert space dimension of the subsystem $A$ (assumed to be smaller than its complement).
\end{theorem}

\vspace{5mm}
\begin{proof}
    
We write $\hat{\rho}$ in a suitable basis as: 
\begin{equation}
\hat{\rho} = \begin{pmatrix} F & B^\dagger \\ B & \hat{\rho}’ \\ \end{pmatrix}~, \label{eq:matrix_eq}
\end{equation}
where the first row/column denotes the basis vector $\ket{\psi}$, and the second denotes the subspace perpendicular to $\ket{\psi}$. With $P_\perp$ the projector onto this subspace, the matrix $\hat{\rho}’ \equiv P_\perp \hat{\rho} P_\perp$ is positive semi-definite. The off-diagonal element $B \equiv P_\perp \hat{\rho} \ket{\psi}$ is a $(d-1)\times 1$ vector, which we take to be an unnormalized vector $\ket{B}$ in the full Hilbert space (with first entry equal to zero). 

Using the variational definition of the trace-norm used in the proof of Theorem~\ref{thm:simple_bound}, we write
\begin{align}
\Vert \hat{\rho}^{T_A} \Vert_1 &\geq \text{tr}(\hat{P}_\psi \hat{\rho}^{T_A}) = \langle \Phi|\hat{\rho}|\Phi\rangle \\
&= F \abs{\langle\psi|\Phi\rangle}^2 + \bra{\Phi}\hat{\rho}’ \ket{\Phi} + \langle \Phi | \psi \rangle \langle B | \Phi \rangle + \langle \Phi | B \rangle \langle \psi | \Phi \rangle,\label{eq:expansion_of_rho}
\end{align}
where $\ket{\Phi}$ is the \textit{unnormalized} maximally entangled state in the full Hilbert space, in the Schmidt bases $\{|a_j\rangle\}$ and $\{|b_j\rangle\}$ of $\ket{\psi}$. We then bound the terms of this expression.

First, the Frobenius norm of the commutator $\mathcal{C}=\Vert [ \hat{\rho}, |\psi\rangle\langle \psi|]\Vert_F$ has the expression
\begin{equation}
   \mathcal{C}^2 = 2\langle \psi|\hat{\rho}^2 |\psi\rangle - 2\langle \psi|\hat{\rho} |\psi\rangle^2 = 2\langle B | B \rangle~.
\end{equation}
Therefore, $\langle B|B\rangle\leq\mathcal{C}^2/2$. We then use the \textit{Schur complement} condition for positive semi-definite matrices. With our decomposition, the fact that $\hat{\rho}$ is positive semi-definite implies that the Schur complement $\hat{\rho}’ - F^{-1} |B\rangle \langle B|$ is positive semi-definite. We conclude that 
\begin{equation}
\langle \Phi | B \rangle \langle B |\Phi \rangle \leq F \langle \Phi | \hat{\rho}’ |\Phi \rangle~,
\label{eq:Schur_complement}
\end{equation}
Using the fact that, $|B\rangle\perp|\psi\rangle$ and that $|\Phi\rangle$ is an unnormalized state with norm $D_A$ and overlap  $|\langle\Phi|\psi\rangle|^2\equiv\alpha D_A$, then accounting for the norms of $|\Phi\rangle$ and $|B\rangle$, the maximum possible value of $|\langle B|\Phi\rangle|$ is $|\langle B|\Phi\rangle|\leq\sqrt{(1-\alpha)D_A}\mathcal{C}/\sqrt{2}$. Eq.~\eqref{eq:Schur_complement} also implies that $\langle \Phi | \hat{\rho}’ |\Phi \rangle\geq(1-\alpha)D_A\mathcal{C}^2/(2F)$.

Combining these ingredients with Eq.~\eqref{eq:expansion_of_rho}, we then have
\begin{align}
\Vert \hat{\rho}^{T_A} \Vert_1 \geq\langle\Phi|\hat{\rho}|\Phi\rangle&\geq F\alpha D_A+(1-\alpha)D_A\mathcal{C}^2/(2F)-2\sqrt{\alpha D_A^2(1-\alpha)\mathcal{C}^2/2}\\
&=\left(\sqrt{F\alpha}-\sqrt{(1-\alpha)\mathcal{C}^2/(2F)}\right)^2D_A\\
&=F\alpha D_A\left(1-(\mathcal{C}/F)\sqrt{(1-\alpha)/(2\alpha)}\right)^2
\end{align}

Taking the logarithm of both sides, we arrive at Theorem~\ref{thm:commutator_negativity_bound}. 
\end{proof}
\vspace{5mm}

With this Theorem~\ref{thm:commutator_negativity_bound} in hand, to validate our entanglement proxy we need only bound the size of the quantity $\mathcal{C}/F$, and show it is $\mathcal{O}(1)$. Here, we show this is the case under the physically realistic scenario of the system undergoing coherent global errors. which are a dominant error source in our experiment (Ext. Data Fig.~\ref{EFig:error_model}). First, we prove a useful Lemma rewriting the coefficient to the correction term, $\mathcal{C}/F$, in terms of only $F$.

\begin{lemma}
    For normally distributed global coherent quasistatic parameter errors, the coefficient to the correction term in Theorem~\ref{thm:commutator_negativity_bound}, $\mathcal{C}/F$, can be written entirely in terms of the fidelity as

    \begin{align}
        \frac{\mathcal{C}}{F\sqrt{2}}=\sqrt{\frac{2}{\sqrt{3+2F^2-F^4}}-1}~.\label{eq:commutator_result}
    \end{align}
    
    \label{lemma:c_over_F}
\end{lemma}

\vspace{5mm}
\begin{proof}
To begin, consider evolution under a fixed parameter value, which results in the pure state $\ket{\Psi(t,\theta)} \equiv e^{-i(\hat{H}_0+\theta \hat{V})t} \ket{\Psi_0}$. In Theorem~\ref{thm:gaussian_fidelity}, we found a Gaussian fidelity dependence between states corresponding to any two parameter values $\theta_1$ and $\theta_2$:
\begin{equation}
    F(\theta_1,\theta_2) \equiv \abs{\langle\Psi(t,\theta_1)|\Psi(t,\theta_2)\rangle}^2 = \exp\left(-N \lambda t^2 \frac{(\theta_1-\theta_2)^2}{2} \right)~,~\label{eq:gaussian_fidelity}
\end{equation}
For some constant $\lambda$ [Eq.~\eqref{eq:Fidelity_Gaussian}]. Assuming a normal distribution $P(\theta)$ of parameter values $\theta$ with mean $\theta_0$ and standard deviation $\sigma$, we obtain the mixed state $\hat{\rho}(t) = \int d\theta P(\theta) |\Psi(t,\theta)\rangle \langle \Psi(t,\theta)|$. In Eq.~\eqref{eq:nonexp_decay}, we computed the fidelity of $\rho(t)$ to the pure state $\ket{\Psi(t,\theta_0)}$ and found it to be $F = 1/\sqrt{\Lambda+1}$, where $\Lambda \equiv N t^2 \lambda \sigma^2$.

The Frobenius norm of the commutator, $\mathcal{C}$, then has expression
\begin{align}
    \mathcal{C}^2&=\Vert \left[\hat{\rho}(t),|\Psi(t,\theta_0)\rangle\langle\Psi(t,\theta_0)|\right] \Vert^2_F\\
    &= 2 \langle \Psi(t,\theta_0)|\hat{\rho}(t)^2 |\Psi(t,\theta_0)\rangle - 2 \langle \Psi(t,\theta_0)|\hat{\rho}(t) |\Psi(t,\theta_0)\rangle^2\\
    &= 2\int d \theta_1 P(\theta_1) \int d\theta_2 P(\theta_2) \langle \Psi(t,\theta_0)|\Psi(t,\theta_1)\rangle \langle \Psi(t,\theta_1)|\Psi(t,\theta_2)\rangle \langle \Psi(t,\theta_2)|\Psi(t,\theta_0)\rangle - 2 F^2\\
    &= \frac{4}{\sqrt{3 \Lambda^2 + 8 \Lambda + 4}}  - \frac{2}{\Lambda+1}~.
\end{align}
In this computation, the unknown phases of the inner products $\langle \Psi(t,\theta_i)|\Psi(t,\theta_j)\rangle$ cancel, allowing the use of Theorem~\ref{thm:gaussian_fidelity}. Rearranging, it is straightforward to then arrive at Eq.~\eqref{eq:commutator_result}.
\end{proof}
\vspace{5mm}

We can then combine Theorem~\ref{thm:commutator_negativity_bound} and Lemma~\ref{lemma:c_over_F}, to arrive at the simplified result
\begin{align}
\mathcal{E}_N(\hat{\rho}) &\geq \mathcal{E}_N(|\psi\rangle) + \log_2(F) + 2\log_2\left(1- \sqrt{\left(\frac{2}{\sqrt{3+2F^2-F^4}}-1\right) \frac{1-\alpha}{\alpha}} \right)\label{eq:rydberg_negativity_bound}\\
&\approx\begin{cases} 
      \mathcal{E}_N(|\psi\rangle) + \log_2(F) & F\approx1 \\
      \mathcal{E}_N(|\psi\rangle)+\log_2(F) + 2 \log_2\left(1-0.39 \sqrt{\frac{1-\alpha}{\alpha}}\right) & F\ll1\label{eq:rydberg_negativity_bound_approx}
   \end{cases}
\end{align}
which shows that our entanglement-proxy is robust in the presence of quasistatic parameter fluctuations, up to a $\mathcal{O}(1)$ deviation which decreases with the normalized entanglement $\alpha\equiv 2^{\mathcal{E}_N(|\psi\rangle)}/D_A$. 

\begin{figure}[b!]
    \centering
    \vspace{5mm}
\includegraphics[width=181mm]{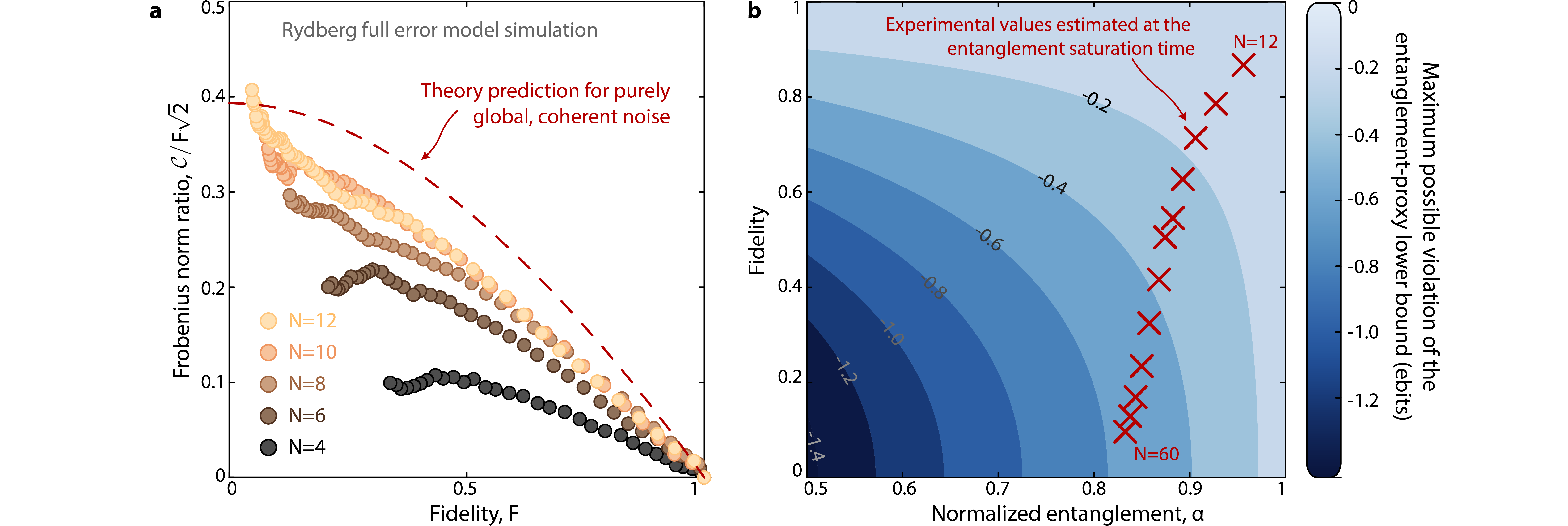}
    \caption{\textbf{Bounding the entanglement-proxy in the presence of errors. a.} We consider full error model simulations of the Rydberg dynamics studied in the main text, i.e. simultaneously under both global and local errors which each can be either Markovian or non-Markovian. We study the ratio of the commutator norm to the fidelity versus fidelity, truncating to only show data for which $F>1/D_A$, where $D_A$ is the half chain Hilbert space dimension. The data appears to converge as a function of system size, approaching the theory prediction obtained for the case of purely global, coherent Hamiltonian fluctuations, see Eq.~\eqref{eq:rydberg_negativity_bound}. \textbf{b.} Here we plot the expected maximum violation of our entanglement-proxy lower bound as a function of the fidelity and the normalized entanglement; more specifically, we plot the correction term from Eq.~\eqref{eq:rydberg_negativity_bound}. We find that in the reasonable parameter regime shown, the violation is always $\lesssim1$ ebit. From analysis shown in Ext. Data Fig.~\ref{EFig:entanglement_growth}, we plot the expected experimental values from $N=12$ to 60, and find the maximum violation appears to saturate around ${\approx}$0.6 ebits. Together these results imply that our entanglement-proxy is robust up to $\mathcal{O}(1)$ corrections in the face of realistic noise.}
    \label{Efig:errors_of_negativity_rydberg}
\end{figure}

More concretely, let us assume a normalized entanglement of $\alpha=0.8$ (as is approximately the expectation at $N=60$ and $t=14.3$ cycles, see Ext. Data Fig.~\ref{EFig:entanglement_growth}). In that case, Eq.~\eqref{eq:rydberg_negativity_bound_approx} reduces to
\begin{align}
\mathcal{E}_N(\hat{\rho}) &\gtrapprox\mathcal{E}_P(\hat{\rho}) -0.6.
\end{align}
In other words, we expect the maximum possible violation of our bound is at most around half an ebit of entanglement. We emphasize that this does not mean the bound is in fact violated by this amount - only that it \textit{potentially could be}.

In order to validate that the coefficient to the correction term is small, in Ext. Data Fig.~\ref{Efig:errors_of_negativity_rydberg}a we show $\mathcal{C}/(F\sqrt{2})$ as a function of $F$ for various system sizes, calculated with our full Rydberg error model (which includes both Markovian and non-Markovian terms, and both global and local errors). Despite the error model including more noise terms than just global parameter fluctuations, we see curves as a function of $N$ increase but appear to converge near the prediction of Eq.~(\ref{eq:commutator_result}). Then, assuming the prediction of Eq.~(\ref{eq:commutator_result}), we show the maximum possible violation of the lower bound provided by $\mathcal{E}_P$ (i.e. the correction term in Eq.~\eqref{eq:rydberg_negativity_bound}) as a function of normalized entanglement and fidelity. We further include estimated experimental values up to $N=60$ (Ext. Data Fig.~\ref{Efig:errors_of_negativity_rydberg}b). We find the experimental values appears to saturate around a maximum possible violation of ${\approx}0.6$ ebits.

\begin{figure}[hbt!]
	\centering
    \vspace{0.2cm}\includegraphics[width=110mm]{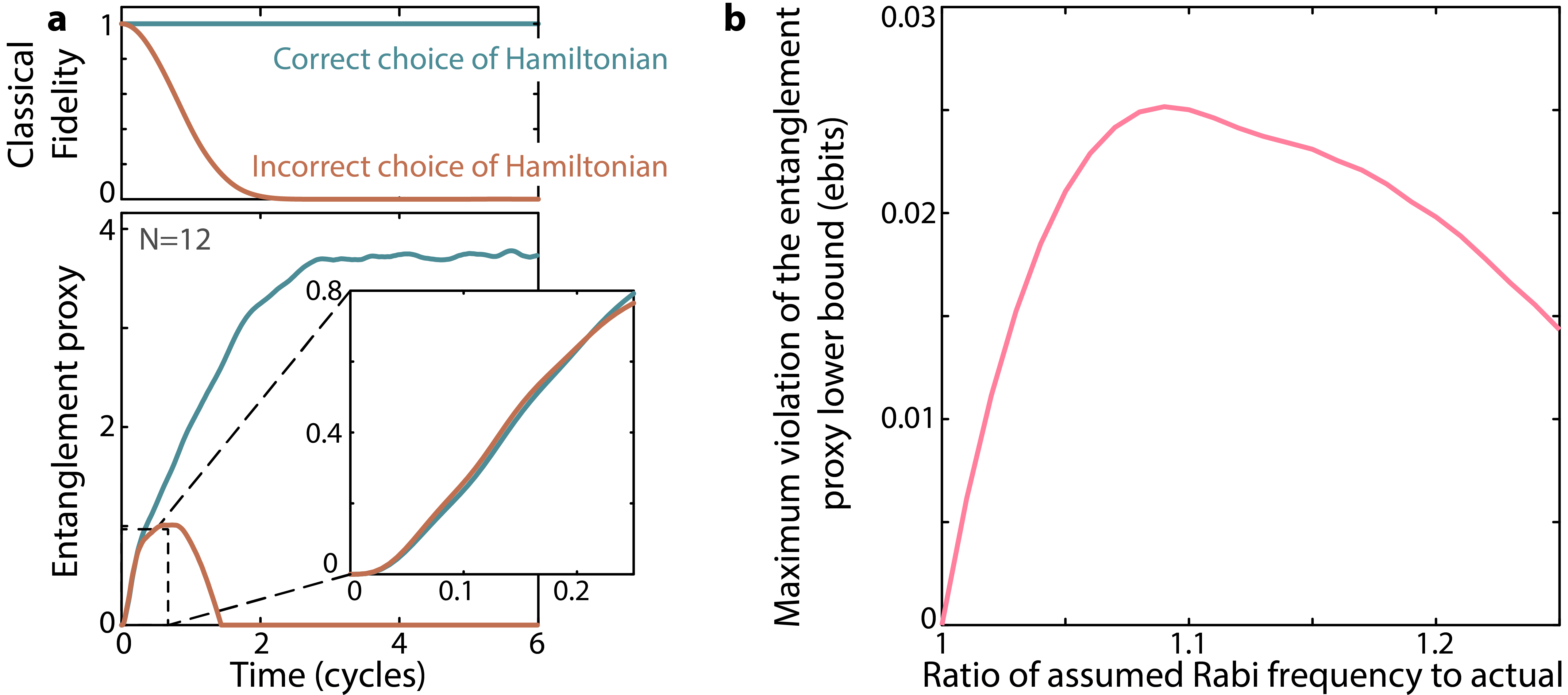}
	\caption{\textbf{Sensitivity to incorrect pure state. a.} We study a possible error source in our entanglement proxy, that the pure state negativity could be assumed incorrectly, for instance if the wrong Hamiltonian parameters are chosen in simulation. We consider error-free quench evolution with $\Omega/2\pi=6.9$ MHz, leading to normal linear growth and then saturation of pure state negativity (which is equal to the entanglement proxy, as $F=1$). If we calculated the entanglement proxy assuming the Rabi frequency was instead $\Omega/2\pi=7.2$ MHz, then at very early times the calculated entanglement proxy can actual slightly exceed the true entanglement, showcasing an adversarial example where our entanglement proxy fails to be a lower bound (inset). However, loss of fidelity due to the Hamiltonian parameter mismatch will eventually dominate (top), restoring the validity of the lower bound for large system sizes or late evolution times. \textbf{b.} Even for a wide range of possible Rabi frequency miscalibrations (represented by the ratio of the miscalibrated Rabi frequency to the actual Rabi frequency), the maximum possible violation of the proxy lower bound is only $\sim0.02$ ebits.
        } 
	\vspace{0.2cm}
    \label{EFig:bad_target_state}
\end{figure}

\vspace{5mm}
\noindent\textbf{Systematic miscalibration errors--}
Next, we comment on the case where the target state $|\psi\rangle$ is set incorrectly, for instance because of a miscalibration of the Hamiltonian parameters. While this type of error can lead to a violation of our lower bound, the violation is small and the period for which it occurs is ephemeral, because the incorrect choice of $|\psi\rangle$ will lead to a fidelity decay which is stronger than any typical increase in pure state entanglement. To showcase this, we consider quench dynamics resulting in linear growth of negativity followed by saturation (Ext. Data Fig.~\ref{EFig:bad_target_state}). We study the two entanglement regimes (saturated or growing linearly) separately.

First, we consider the case where the entanglement has saturated. We imagine the true evolution was performed with a Rabi frequency $\Omega_0$, resulting in a state $|\psi_0\rangle$ at time $t$. However, we consider the situation where the parameters in the Hamiltonian have been mischosen, for instance with Rabi frequency $\Omega_1 = \Omega_0+\omega$. From Theorem~\ref{thm:gaussian_fidelity} we know the resulting fidelity overlap between the two states is given by
\begin{align}
F=|\langle\psi_0|\psi_1\rangle|^2=2^{-N\lambda\omega^2t^2/2}.
\end{align}
Because we are studying the entanglement saturated regime, the pure state negativities of $|\psi_0\rangle$ and $|\psi_1\rangle$ are approximately equal, but the mixed state negativity will now be calculated as
\begin{align}
\mathcal{E}_P(|\psi_1\rangle)&=\mathcal{E}_N(|\psi_1\rangle) +\log_2(F)\\
&\approx\mathcal{E}_P(|\psi_0\rangle)-N\lambda\omega^2t^2/2.
\end{align}
Therefore, any change in the assumed Rabi frequency will lead to a quadratic loss in the mixed state entanglement proxy, and the proxy will remain a lower bound.

The situation is largely the same if we now work in the linear growth regime, where we assume the pure state entanglement grows as $\mathcal{E}(|\psi_0\rangle)=\alpha\Omega_0 t$, and $\mathcal{E}(|\psi_1\rangle)=\alpha\Omega_1 t$, where $\alpha$ is the entanglement velocity related to Lieb-Robinson bounds. In this case, the entanglement proxy can be written as 
\begin{align}
\mathcal{E}_P(|\psi_1\rangle)&=\mathcal{E}_N(|\psi_1\rangle) +\log_2(F)\\
&=\alpha\Omega_1 t -N\lambda\omega^2t^2/2\\
&=\mathcal{E}_P(|\psi_0\rangle) + \alpha\omega t -N\lambda\omega^2t^2/2\\
&= \mathcal{E}_P(|\psi_0\rangle) + \omega t(\alpha-N\lambda\omega t/2).
\end{align}

Thus, as long as $N\lambda\omega t/2>\alpha$, the erroneous choice of target state will still result in a lower overall mixed state entanglement proxy, thus maintaining the lower bound. This will then become increasingly tight for larger system sizes and later evolution times. In Ext. Data Fig.~\ref{EFig:bad_target_state} we show simulations demonstrating this, indicating that Hamiltonian parameter miscalibration should not substantially affect the veracity of our mixed state entanglement lower bound past very early times.

\vspace{5mm}
\noindent\textbf{Incoherent local errors--}
Next, we turn to prove our entanglement proxy is a valid lower bound for the case of random unitary circuits (RUCs) undergoing realistic noise, namely local depolarization error. To do so, we will introduce a new theorem, which bounds the violation in our entanglement proxy by a value related to the fidelity overlap with eigenstates of $\hat{\rho}$, and then will show that this correction term is small.

We proceed first by proving a more general Lemma which will assist us; the proof has similarities to the proof of Theorem~\ref{thm:commutator_negativity_bound}, but we will present it again here for completeness.

\begin{lemma}
\label{lemma:general_bound}
    For any mixed state $\hat{\rho}$ and pure state $\ket{\psi}$
\begin{equation}
 \mathcal{E}_N(\hat{\rho}) \geq \mathcal{E}_N(|\psi\rangle)+\log_2(F)+2\log_2(1 - \sqrt{(1-F)(1-\alpha)/(F\alpha)})~, \label{eq:unconditional_bound}
\end{equation}
where $F \equiv \bra{\psi}\hat{\rho}\ket{\psi}$ is their fidelity, $\alpha \equiv 2^{\mathcal{E}_N(|\psi\rangle)}/D_A \in [0,1]$ is a normalized entanglement of $\ket{\psi}$, and $D_A$ is the Hilbert space dimension of the subsystem $A$ (assumed to be smaller than its complement).
\end{lemma}


\vspace{5mm}
\begin{proof}
In deriving Theorem~\ref{thm:simple_bound}, we utilized the fact that when $\ket{\psi}$ is an eigenstate of $\hat{\rho}$, the remainder state $\hat{\rho}_\perp$ is positive semi-definite, which allowed us to bound $\bra{\Phi}\hat{\rho}_\perp \ket{\Phi} \geq 0$. To show Lemma~\ref{lemma:general_bound}, we use constraints that arise from the fact that $\hat{\rho}$ is positive semidefinite. We write $\hat{\rho}$ in a suitable basis as: 
\begin{equation}
\hat{\rho} = \begin{pmatrix} F & B^\dagger \\ B & \hat{\rho}’ \\ \end{pmatrix}~, \label{eq:matrix_eq}
\end{equation}
where the first row/column denotes the basis vector $\ket{\psi}$, and the second denotes the subspace perpendicular to $\ket{\psi}$. With $P_\perp$ the projector onto this subspace, the matrix $\hat{\rho}’ \equiv P_\perp \hat{\rho} P_\perp$ is positive semi-definite. The off-diagonal element $B \equiv P_\perp \hat{\rho} \ket{\psi}$ is a $(d-1)\times 1$ vector, which we take to be an unnormalized vector $\ket{B}$ in the full Hilbert space (with first entry equal to zero). 

Using Eq.~\eqref{eq:negativity_bound_step}, the negativity can be obtained from 
\begin{equation}
\Vert \hat{\rho}^{T_A} \Vert_1 \geq \langle \Phi|\hat{\rho}|\Phi\rangle = F \abs{\langle\psi|\Phi\rangle}^2 + \bra{\Phi}\hat{\rho}’ \ket{\Phi} + \langle \Phi | \psi \rangle \langle B | \Phi \rangle + \langle \Phi | B \rangle \langle \psi | \Phi \rangle~.
\end{equation}
Only the last two terms above may be negative. We bound their magnitude using the \textit{Schur complement} condition for positive semi-definite matrices. With our decomposition, the fact that $\rho$ is positive semi-definite implies that the Schur complement $\hat{\rho}’ - F^{-1} |B\rangle \langle B|$ is positive semi-definite. We conclude that 
\begin{equation}
\langle \Phi | B \rangle \langle B |\Phi \rangle \leq F \langle \Phi | \hat{\rho}’ |\Phi \rangle~,
\end{equation}
which allows us to further conclude that
\begin{equation}
\langle \Phi |\hat{\rho} |\Phi \rangle \geq (\sqrt{F} \abs{\langle \Phi |\psi \rangle} - \sqrt{\langle \Phi | \hat{\rho}’ |\Phi \rangle})^2~.
\end{equation}
We then use the fact that $\ket{\Phi}/\sqrt{D_A}$ is a normalized state. Since $\abs{\langle\Phi |\psi\rangle}^2 = \alpha D_A = \Vert |\psi\rangle \langle \psi |^{T_A} \Vert_1$, $\text{tr}\hat{\rho}’ = 1-F$ and $\bra{\psi} \hat{\rho}’ \ket{\psi} = 0$, we conclude that
\begin{equation}
\bra{\Phi}\hat{\rho}’ \ket{\Phi} \leq (1-F)(1-\alpha)D_A~.
\end{equation}  
Then when $F \geq 1-\alpha$, we obtain the claimed Eq.~\eqref{eq:unconditional_bound}.
\end{proof}
\vspace{5mm}

We then simply generalize Lemma~\ref{lemma:general_bound} to the case where $|\psi\rangle$ has fidelity overlap with a particular eigenstate of $\hat{\rho}$
\begin{theorem}[Eigenstate-fidelity based bound]
\label{thm:fidelity_negativity_bound}
For any mixed state $\hat{\rho}$ and pure state $\ket{\psi}$ which has a fidelity $f \equiv |\langle \psi|\lambda\rangle|^2$ with an eigenstate $|\lambda\rangle$ of $\hat{\rho}$, the logarithmic negativities $\mathcal{E}_N(\hat{\rho})$ and $\mathcal{E}_N(|\psi\rangle)$ are related by
    \begin{equation}
\mathcal{E}_N(\hat{\rho}) \geq \mathcal{E}_N(|\psi\rangle) +  \log_2(F) + 2\log_2(1-\sqrt{(1-f)(1-\alpha)/(f\alpha)})~,
\label{eq:robust_negativity_bound}
\end{equation}
where $F \equiv \bra{\psi}\hat{\rho}\ket{\psi}$ is the fidelity, $\alpha \equiv 2^{\mathcal{E}_N(|\psi\rangle)}/D_A \in [0,1]$ is a normalized entanglement of $\ket{\psi}$, and $D_A$ is the Hilbert space dimension of the subsystem $A$ (assumed to be smaller than its complement).
\end{theorem}

\vspace{5mm}
\begin{proof} 
We first write $\hat{\rho} = F_t |v\rangle\langle v| + (1-F_t) \hat{\rho}_\perp$, where $|v\rangle$ is an eigenstate of $\hat{\rho}$ with eigenvalue $F_t$, and $\hat{\rho}_\perp$ is the state projected onto the orthogonal subspace to $\ket{v}$. We write the state $\ket{\psi}$ that we benchmark against $\hat{\rho}$ in terms of $\ket{v}$ as $\ket{\psi} = \sqrt{f} \ket{v} + \sqrt{1-f} \ket{\psi_\perp}$. Then writing $\hat{\rho}$ in the same basis as Eq.~\eqref{eq:matrix_eq} gives
\begin{equation}
\hat{\rho} = \begin{pmatrix} F_t f & F_t \sqrt{f(1-f)} |\psi\rangle \langle \psi_\perp| \\  F_t \sqrt{f(1-f)} |\psi_\perp \rangle \langle \psi|~~  & \hat{\rho}_\perp + F_t (1-f) |\psi_\perp\rangle\langle\psi_\perp| \\ \end{pmatrix}~, \label{eq:matrix_eq2}
\end{equation}
As with the above, the only negative contribution to $\langle \Phi |\hat{\rho} |\Phi \rangle $ arises from the off-diagonal terms of Eq.~\eqref{eq:matrix_eq2}. This allows us to improve our bound to the claimed Theorem ~\ref{thm:fidelity_negativity_bound}. 
\end{proof}
\vspace{5mm}

We now show that under incoherent local errors, the correction term presented in Theorem~\ref{thm:fidelity_negativity_bound} is small. We study the case of random unitary circuits (RUCs) in the presence of local depolarization in order to treat the problem analytically and make connection to the digital quantum circuits we compare against in the main text. 

We study the fidelity and purity and bound the fidelity $f\equiv |\langle \psi(t)|v\rangle|^2$ between the ideal evolved state $|\psi(t)\rangle$ with the largest eigenstate $|v\rangle$ of the noisy evolved state $\hat{\rho}(t)$, by computing both the purity and fidelity of $|\psi(t)\rangle$ and $\hat{\rho}(t)$. The relationship between these quantities in fact bounds the fidelity of $\ket{\psi(t)}$ to the largest eigenstate of $\hat{\rho}$,
allowing us to apply Theorem~\ref{thm:fidelity_negativity_bound} to evaluate the robustness of our negativity proxy.

Physically, our result means that the state $\hat{\rho}$ is a mixture of a particular pure state $\ket{v_1}$ with large population $F=\lambda_1$ and many other orthogonal states with smaller populations. For a circuit with spacetime volume $V = N t$, the largest eigenvalue decays exponentially $\lambda_1 \approx (1-p)^V$, yet remains larger than all other eigenvalues. The other states correspond to trajectories with errors. The average number of errors is $k=pV $, and there are roughly $\frac{V!}{k!(V-k)!}$ different spacetime locations of $k$ errors. When the number of trajectories are much less than the Hilbert space dimension $D$, we expect the resultant wavefunctions resultant to be orthogonal to one another. This is true if the $k$ errors are sparsely spread over the system in spacetime (such that their separations are sufficiently large) and the effects of errors are scrambled quickly. When $p$ is sufficiently small and when the quantum dynamics is sufficiently chaotic, these assumptions are expected to hold.

\begin{figure}
    \centering
\includegraphics[width=181mm]{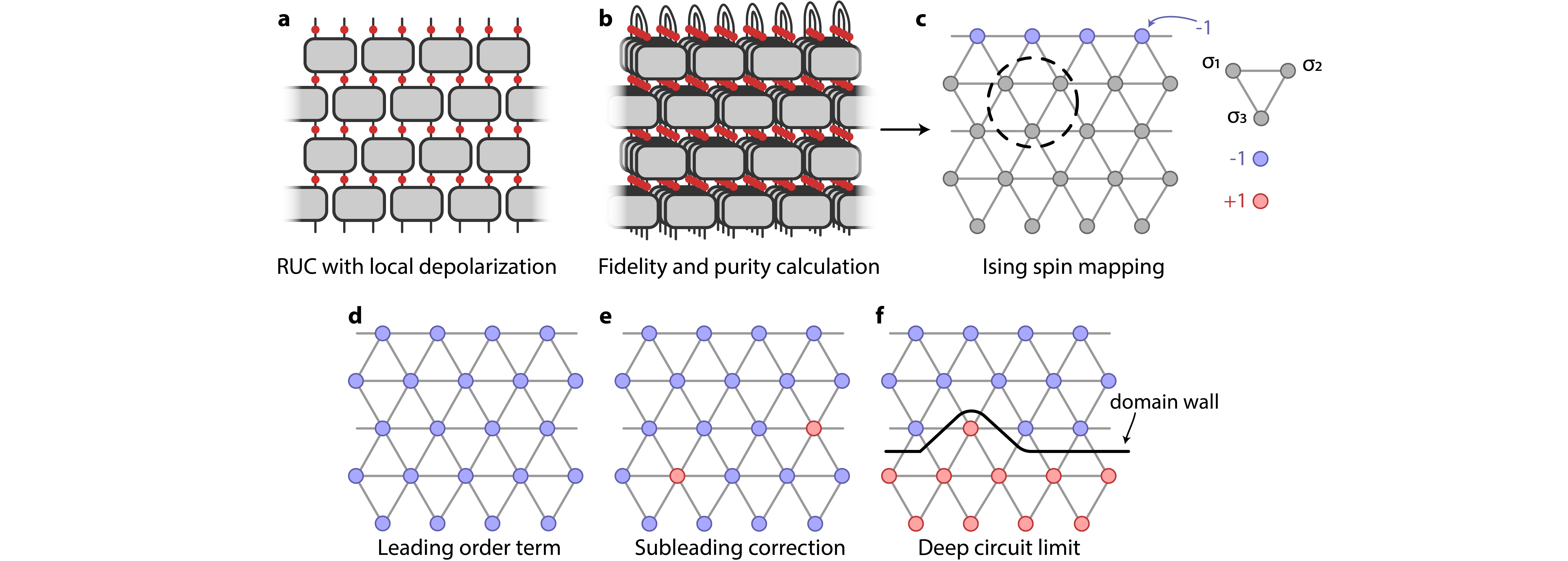}
    \caption{\textbf{Purity and fidelity in random unitary circuits (RUCs) with local depolarization. a.} We consider a brickwork RUC, where local depolarization with strength $p$ is applied after every layer and at every site (red dots). \textbf{b.} Typical values of purity or fidelity can be computed by random circuit averages. In both quantities, swap boundary conditions are applied on four copies of the circuit, representing two copies of the states. For purity, dephasing is applied to both copies (depicted here), while for fidelity, dephasing is only applied to one copy. \textbf{c.} After averaging over random circuits, the purity and fidelity can be expressed as the partition functions of effective statistical mechanical Ising models, with three-spin Boltzmann weights on every downward-facing triangle (listed in Table~\ref{tab:stat_mech}), the top boundary fixed to be `-' (blue circles) and the bottom boundary free. \textbf{d.} The dominant Ising spin configuration for both quantities is where all Ising spins are in the `-' (swap) state (blue circles). \textbf{e.} Subleading corrections arise from configurations with sparse, isolated ``bubbles" of `+' Ising spins (red circles). \textbf{f.} For sufficiently deep circuits, the dominant configurations are those with a boundary domain of `-' spins and a bulk domain of `+' spins. However, this is only relevant in a regime where the fidelity is trivially small. }
    \label{fig:RUC_purity_fidelity}
\end{figure}

We proceed to justify this intuition by averaging over random circuits. At every time step, we apply random unitary two-qubit gates in a brickwork circuit geometry, and subject it to local depolarization, which locally depolarizes the state $\rho$ at every site with probability $p$, equivalent to applying the local channel $\rho \mapsto (1-p) \rho + p (I_j/d)\otimes \text{tr}_j(\rho)$ independently to every site $j$, where $d=2$ is the local Hilbert space dimension (Ext. Data Fig.~\ref{fig:RUC_purity_fidelity}a). For simplicity, we also take the initial state $\ket{\psi_0}$ to be the product state $|0\rangle^{\otimes N}$. After averaging over RUCs, the purity $P\equiv \text{tr}[\rho(t)^2]$ and fidelity $F\equiv \langle \psi(t)|\rho(t)|\psi(t)\rangle$ can be expressed as the partition functions of two different Ising ferromagnets, with the Ising spins $\sigma = \pm 1$ respectively denoting the local identity and swap permutations (Ext. Data Fig.~\ref{fig:RUC_purity_fidelity}b). 

 The derivation of the mapping between RUCs and classical spin models is described in detail in many works including Refs.~\cite{Fisher2023RandomCircuits,Bao2020TheoryMeasurements}.
Here we simply quote the result:
\begin{align}
    P &= \sum_{\{ \sigma \} } \prod_\triangledown W_P(\sigma_1, \sigma_2, \sigma_3; p)~, \label{eq:partition_P_sum}\\
    F &= \sum_{\{ \sigma \} } \prod_\triangledown W_F(\sigma_1, \sigma_2, \sigma_3; p), \label{eq:partition_F_sum}
\end{align}
where the summation is over all possible Ising configurations of classical spin variables $\sigma =  \pm1$ and the product is over all downward-facing triangles in the resulting triangular lattice (Ext. Data Fig.~\ref{fig:RUC_purity_fidelity}c).
$W_{P/F}(\sigma_1, \sigma_2, \sigma_3;p)$ act as local ``Boltzmann weights," parameterized by the error probability $p$.
Their values are enumerated in Ext. Data Table~\ref{tab:stat_mech}. The Ising spins at the bottom are determined by the initial condition and are effectively unconstrained (open boundary condition), while the purity and fidelity expressions result in an additional layer of Ising spins at the top, which are pinned to be the swap permutation `$-$'.

\begin{table}[]
    \centering
\begin{tabular}{ c  c  c }
\toprule
 $\sigma_1, \sigma_2, \sigma_2$ & $W_P(\sigma_1,\sigma_2,\sigma_3;p)$ & $W_F(\sigma_1,\sigma_2,\sigma_3;p)$ \\
 \midrule
 $+,+,+$ & $1$ & 1 \\  
 $+,+,-$ & $0$ & 0  \\ 
 $+,-,+$ & $\frac{d}{d^2+1} + \frac{2d}{d^4-1} p + \frac{4d}{3(d^4-1)} p^2$ & $\frac{d}{d^2+1} + \frac{d}{d^4-1} p$  \\ 
 $+,-,-$ & $\frac{d}{d^2+1} + \frac{-2d^3}{d^4-1}p + \frac{4d^3}{3(d^4-1)} p^2$ & $\frac{d}{d^2+1} + \frac{-d^3}{d^4-1}p$ \\
 $-,+,+$ & $\frac{d}{d^2+1} + \frac{2d}{d^4-1} p + \frac{4d}{3(d^4-1)} p^2$ & $\frac{d}{d^2+1} + \frac{d}{d^4-1} p$ \\  
 $-,+,-$ & $\frac{d}{d^2+1} + \frac{-2d^3}{d^4-1}p + \frac{4d^3}{3(d^4-1)} p^2$ & $\frac{d}{d^2+1} + \frac{-d^3}{d^4-1}p$  \\ 
 $-,-,+$ & $\frac{4d^2}{d^4-1} p + \frac{-20d^2}{3(d^4-1)}p^2 + \frac{16d^2}{3(d^4-1)} p^3 + \frac{-16d^2}{9(d^4-1)}p^4$ 
 & $\frac{2d^2}{d^4-1} p + \frac{-d^2}{d^4-1}p^2$  \\ 
 $-,-,-$ & $ 1 + \frac{-4d^4}{d^4-1} p + \frac{20d^4}{3(d^4-1)}p^2 + \frac{-16d^4}{3(d^4-1)} p^3 + \frac{16d^4}{9(d^4-1)} p^4 $ 
 & $1 + \frac{-2d^4}{d^4-1} p + \frac{d^4}{d^4-1}p^2$ \\
 \bottomrule
\end{tabular}
    \caption{Ising Boltzmann weights of local downward-facing triangle configurations for the effective statistical mechanical models. Their partition functions equal the purity $P$ [Eq.~\eqref{eq:partition_P_sum}] and fidelity $F$ [Eq.~\eqref{eq:partition_F_sum}] of random unitary circuits (RUCs) with local depolarization. The geometric arrangement of $\sigma_1,\sigma_2,\sigma_3$ is indicated in Fig.~\ref{fig:RUC_purity_fidelity}c. In this work, we consider RUCs of qubits, with local Hilbert space dimension $d=2$, and only consider the $\mathcal{O}(1)$ and $\mathcal{O}(p)$ terms. In this table, however, we state the full expressions in terms of general $d$ for future reference.}
    \label{tab:stat_mech}
\end{table}

When $p=0$, notice that $W_{P/F}(-,-,+) = 0$.
Combined with the fact that all top boundary spins have $\sigma = -$, the only non-vanishing contribution to $P$ or $F$ is the one where all $\sigma = -$ (Ext. Data Fig.~\ref{fig:RUC_purity_fidelity}d).
In this case, $P = \prod_{\triangledown} W_P(-,-,-) = 1$ and similarly $F = 1$. This is expected since in the absence of errors, $\rho$ is a pure state with perfect fidelity.

Now consider small, nonzero $p$. $P$ and $F$ have contributions from configurations where some of the spin variables are flipped to the `+' state, with weight $\mathcal{O}(p)$ (Ext. Data Fig.~\ref{fig:RUC_purity_fidelity}e). We call such spin flips ``bubbles," because such bubbles are penalized from growing by an effective line-tension term, which weights larger bubbles by a factor of $(2/5)^l$, where $l$ is the length of the boundary of the bubble. These small bubbles will effectively have renormalized weights, but will remain $\mathcal{O}(p)$ and whose precise values are not consequential to our conclusions. 
Therefore, in this regime, the number of such spin flips will be rare and sparse. To this end, we approximate $P$ and $F$ by simply summing over the contributions from sparse, disjoint bubbles. However, this approximation breaks down at larger values of $p$, discussed below. Within this approximation,
\begin{align}
    P &\approx  \left( 1- \frac{4d^2}{d^4-1} p + \mathcal{O}(p^2)\right)^V\exp(A_P V)\equiv \exp(-B_P V) ~,\\
    F &\approx \left(1-\frac{2d^2}{d^4-1}p + \mathcal{O}(p^2)\right)^V 
     \exp(A_F V) \equiv \exp(-B_F V)~,
\end{align}
where $V$ is the spacetime volume (number of Ising spins) and $A_F$ and $A_P$ are the renormalized relative weights of a single bubble. 

We are interested in the ratio $P/F^2\approx e^{(2B_F - B_P) V} \geq 1$. When $p \ll 1$,
\begin{align}
    2B_F-B_P &= -2A_F + A_P 
    - 2 \ln  \left( 1- \frac{2d^2}{d^4-1} p + \mathcal{O}(p^2)\right)
    + \ln \left( 1- \frac{4d^2}{d^4-1} p + \mathcal{O}(p^2) \right)   \nonumber\\
    &= A_P - 2A_F + \mathcal{O}(p^2).
\end{align}
Because each bubble already has $\mathcal{O}(p)$ weight, the leading order contribution to $A_P- 2A_F$ is determined by that of $W_P(-,-,+)$ and $W_F(-,-,+)$, which have ratio precisely 2. Therefore, we conclude that all $\mathcal{O}(p)$ terms vanish and
\begin{align}
    P/F^2 = e^{C p^2 V} \label{eq:purity_fidelity_ratio},
\end{align}
for some $\mathcal{O}(1)$ constant $C$.

\begin{figure}[t!]
    \centering
\includegraphics[width=181mm]{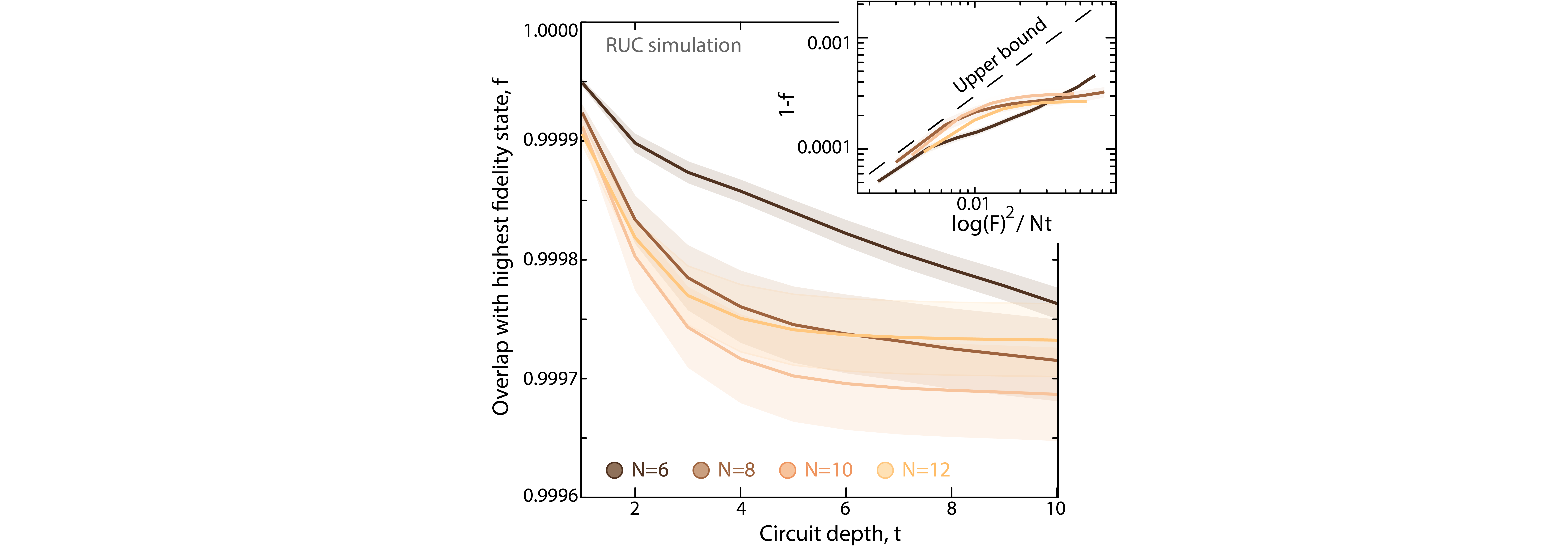}
    \caption{\textbf{Bounding the entanglement-proxy in the presence of errors.} We consider the case of incoherent local errors, specifically for random unitary circuits undergoing local depolarization. The overlap, $f$, between the target pure state and the highest fidelity state to $\hat{\rho}$ remains near unity. Further, $1-f$ is linearly bounded by the quantity $\log(F)^2/Nt$ with a small proportionality constant $0.03$ (inset), consistent with our RUC analysis leading up to Eq.~\eqref{eq:one_minus_f_bound}, and giving the overall bound Eq.~\eqref{eq:ruc_negavitity_bound}.}
    \label{Efig:errors_of_negativity}
\end{figure}

In order to translate this into a bound on the eigenstate-fidelity $f$, we denote the fidelities between $|\psi\rangle$ and the eigenstates $|v_i\rangle$ of $\hat{\rho}$ as $\mu_i \equiv |\langle v_i | \psi \rangle |^2$, with $f\equiv \mu_1$. With $\lambda_i$ the eigenvalues of $\hat{\rho}$, we have $F = \sum_i \lambda_i \mu_i$ and $P = \sum_i \lambda_i^2$.  The Cauchy-Schwarz inequality implies that 
\begin{align}
\mu_1^2 + (1-\mu_1)^2 \geq \sum_i \mu_i^2 \geq F^2/P \Rightarrow 1-\mu_1 \leq \frac{P/F^2-1}{2P/F^2}~.
\end{align}
Using Eq.~\eqref{eq:purity_fidelity_ratio}, we obtain
\begin{align}
1-f \equiv 1- \mu_1 \leq (1-\exp(-Cp^2 V))/2 = \mathcal{O}(p^2V)\label{eq:one_minus_f_bound}.
\end{align} 

Using Theorem~\ref{thm:fidelity_negativity_bound}, we obtain 
\begin{equation}
    \mathcal{E}_N(\hat{\rho}) \gtrapprox \mathcal{E}_N(|\psi\rangle) + \log_2(F) + 2\log_2\left(1- \sqrt{\frac{1-\alpha}{\alpha}} C_2 \frac{\ln(F)^2}{V}\right)~\label{eq:ruc_negavitity_bound},
\end{equation}
for another $\mathcal{O}(1)$ constant $C_2$, indicating that the entanglement-proxy $E_P$ is valid up to a small correction which is negligible as long as $F \gg \exp(-\sqrt{V}) = \exp(-\sqrt{N t})\sim \exp(-N)$, i.e.~$F$ is not trivially small. We give numerical support for this in Ext. Data Fig.~\ref{Efig:errors_of_negativity}.

Finally, we remark that additional configurations may come to dominate the partition function sums Eqs.~\eqref{eq:partition_P_sum},~\eqref{eq:partition_F_sum} when the circuit is sufficiently deep. This is easiest to understand in the statistical mechanical picture. For both the purity and fidelity, the bulk has a small preference for the `+' state, because depolarization breaks the Ising symmetry and favors the identity element. Meanwhile, the boundary condition (associated with evaluating purity or fidelity) pins the boundary to the `-' state, acting as a strong boundary field. When the circuit is shallow, the boundary term dominates, giving rise to the global `-' domain assumed above. However, when the circuit is sufficiently deep, it becomes energetically favorable for a large domain wall to form, dividing the spins into a `-' domain near the boundary, and a `+' domain in the bulk (Ext. Data Fig.~\ref{fig:RUC_purity_fidelity}f), at the expense of a domain wall with free-energy scaling with the spatial size. In our above calculations, we neglect this phenomenon because it is only relevant when the fidelity is trivially small, $\mathcal{O}(D^{-1})$, where $D$ is the total Hilbert space dimension. In this regime, the purity and fidelity decouple from the bulk dynamics and assume this constant value which arises from the boundary Ising `-' domain.

\subsubsection{Negativity of truncated MPS states}
Finally, we investigate an interesting observation in the main text, that at $N=60$ the entanglement-proxy found for the experiment in Fig.~\ref{Fig4} is ${\sim}3$ ebits higher than $\log_2(\chi^*)$ in Fig.~\ref{Fig5}. In other words, the fidelity-equivalent MPS simulation is less entangled than the experiment. Here we study the fidelity and entanglement (negativity) of truncated pure state MPS representations to explain this behavior.

For a theoretical treatment, we consider a simplified setup where we start with an ideal pure state and its equal bipartite Schmidt decomposition $\ket{\psi}=\sum_{j=1}^{D_A} s_j \ket{a_{j}}\ket{b_{j}}$ where $D_A$ is the Hilbert space dimension of the half system and the Schmidt values $s_j$ are assumed to be in descending order. Then the Schmidt rank is truncated to a maximum bond dimension $\chi\le D_A$. Denote the truncated state and its Schmidt decomposition as $\ket{\psi_{\oldchi}} = \sum_{j=1}^{\oldchi} s'_{j}\ket{a_{j}}\ket{b_{j}}$, with $s'_j = s_j/\sqrt{\sum_{k=1}^{\oldchi} s_k^2}$ (normalization condition). It is straightforward to see that
\begin{equation}
    F = \abs{\langle \psi_{\oldchi}|\psi\rangle}^2 = \sum_{j=1}^\oldchi s_j^2.
\end{equation}
To derive the entanglement negativity, the eigenvalues $\lambda$ of the partial transposed state $\ket{\psi}\bra{\psi}^{T_A}$ is given by $s_j^2$ or $\pm s_is_j$ as mentioned above. This gives
\begin{equation}
||\ket{\psi_\oldchi}\bra{\psi_\oldchi}^{T_A}||_1 = \frac{\left(\sum_{j=1}^\oldchi s_j\right)^2}{\sum_{j=1}^{\oldchi} s_j^2},
\end{equation}
and therefore
\begin{equation}
    \mathcal{E}(|\psi_\oldchi\rangle) = \log_2 ||\ket{\psi_\oldchi}\bra{\psi_\oldchi}^{T_A}||_1 =  2 \log_2 \sum_{j=1}^\oldchi s_j - \log_2 \sum_{j=1}^{\oldchi} s_j^2.
\end{equation}
For generic states with unknown $s_j$, it is hard to draw any meaningful relationship between fidelity and entanglement negativity. Therefore, we consider two cases: maximally entangled states and Haar random states.
\\

\noindent\textbf{Maximally entangled states---}For a maximally entangled state, the Schmidt values are given by $s_j=1/\sqrt{D_A}$. It follows directly that
\begin{align}
    F &= \frac{\chi}{D_A}; \\
    \mathcal{E}(|\psi_\oldchi\rangle) &= \log_2 \chi = \log_2 D_A + \log_2 F,
\end{align}
with $\log_2 D_A$ the log negativity of the ideal maximally entangled pure state. Thus, the log negativity for a fidelity-equivalent truncated maximally entangled state is equal to that of an isotropic mixed state.
\\

\noindent\textbf{Haar random states---}The case for Haar random states is more interesting, as such states arise from random unitary circuit evolution, and are closer to the output of our Rydberg experiment as well. As mentioned previously, the Schmidt values of Haar random states follow the \emph{quarter-circle law} as
\begin{equation}
    P(\tilde s) = \frac{1}{\pi}\sqrt{4-\tilde{s}^2},
\end{equation}
where $\tilde{s}\equiv \sqrt{D_A} s$. Setting a maximum Schmidt rank can be translated to setting a minimum cutoff $s_\mathrm{min}$ (and the corresponding $0\le\tilde{s}_\mathrm{min}\le 2$) of the Schmidt values (keeping only Schmidt values above $s_\mathrm{min}$), where the relation between $\chi$ and $\tilde{s}_\mathrm{min}$ is given by
\begin{equation}
    \chi = D_A \int_{\tilde{s}_\mathrm{min}}^{2} P(\tilde s) d\tilde{s} = D_A\left[\frac{4}{\pi}\sin^{-1}\left(\frac{\sqrt{2 - \tilde{s}_\mathrm{min}}}{2}\right)-\frac{1}{2\pi} \tilde{s}_\mathrm{min} \sqrt{4-\tilde{s}_\mathrm{min}^2}\right].
\end{equation}
Depending on the cutoff, 
\begin{align}
    {\sum_{j=1}^{\oldchi(\tilde{s}_\mathrm{min})} s_j} &= \sqrt{D_A} \int_{\tilde{s}_\mathrm{min}}^{2} \tilde{s} P(\tilde s) d\tilde{s} = \frac{\sqrt{D_A}}{3\pi}\left(4 - \tilde{s}_\mathrm{min}^2\right)^{\frac{3}{2}}; \\
    {\sum_{j=1}^{\oldchi(\tilde{s}_\mathrm{min})} s_j^2} &= \int_{\tilde{s}_\mathrm{min}}^{2} \tilde{s}^2 P(\tilde s) d\tilde{s} = \frac{4}{\pi}\sin^{-1}\left(\frac{\sqrt{2 - \tilde{s}_\mathrm{min}}}{2}\right) + \frac{1}{4\pi} \tilde{s}_\mathrm{min} (2-\tilde{s}_\mathrm{min}^2) \sqrt{4-\tilde{s}_\mathrm{min}^2}.
\end{align}
Therefore, the fidelity and log negativity can be written in parametric forms as
\begin{align}
    F(\tilde{s}_\mathrm{min}) &= \frac{4}{\pi}\sin^{-1}\left(\frac{\sqrt{2 - \tilde{s}_\mathrm{min}}}{2}\right) + \frac{1}{4\pi} \tilde{s}_\mathrm{min} (2-\tilde{s}_\mathrm{min}^2) \sqrt{4-\tilde{s}_\mathrm{min}^2}; \\
    \mathcal{E}(|\psi_{\oldchi(\tilde{s}_\mathrm{min})}\rangle) &= \log_2 D_A + \log_2 \frac{(4-\tilde{s}_\mathrm{min}^2)^3}{9\pi^2} - \log_2 F(\tilde{s}_\mathrm{min}).
\end{align}
It is easy to verify that when $\tilde{s}_\mathrm{min}=0$, the equations reduce to the case of ideal Haar random states, with $\chi=D_A$, $F=1$, and $\mathcal{E}=\log_2 D_A + \log_2 64/9\pi^2$, where $64/9\pi^2 \approx 0.72$ is the Page correction~\cite{Page1993AverageSubsystem}.

Importantly, we can then compare this prediction against the prediction for a depolarized Haar random state, where we use the first-order approximation from Eq.~\eqref{eq:Haar_negativity_estimator}, in other words taking $D_A\rightarrow\infty$. At a given fidelity for each, the difference in estimated entanglement, $\delta$, is given as

\begin{align}
\delta&=\mathcal{E}_N(\hat{\rho}_\text{Haar})-\mathcal{E}_N(|\psi_{\oldchi(\tilde{s}_\mathrm{min})}\rangle)\\
&=3\log_2\left(\frac{4}{4-\tilde{s}_\mathrm{min}^2}\right) +2\log_2(F)\label{eq:entanglement_difference_parametric}\\
&\approx\left(2+\log_2\left(\frac{64}{9\pi^2}\right)\right)(1-F)^{2/3} \approx \frac{3}{2}(1-F)^{2/3}
\end{align}

\begin{figure}[t!]
	\centering
    \vspace{0.0cm}\includegraphics[width=49mm]{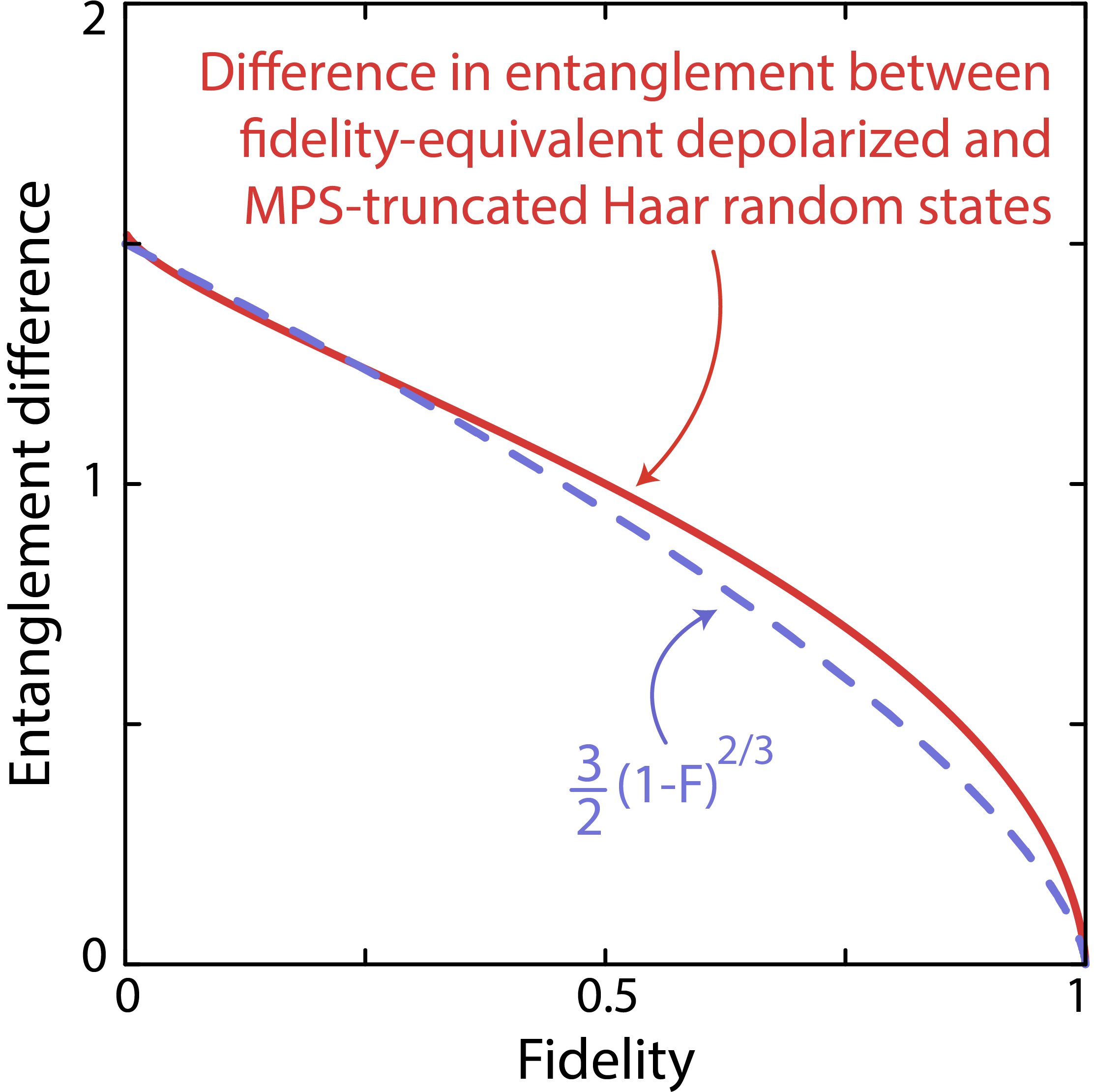}
	\caption{\textbf{Entanglement difference between truncated and depolarized Haar random states.} At fixed fidelity, MPS truncation and depolarization affect the entanglement content of a pure state differently. For a Haar random state, we can solve this relation analytically, finding the depolarized state has a higher entanglement (log negativity) content than the MPS-truncated stated at equal fidelity.
        } 
	\vspace{0.5cm}
    \label{EFig:truncation_fidelity}
\end{figure}

In the last line we have performed series expansions of both $F$ and $\delta$ in terms of $\tilde{s}_\mathrm{min}$ to arrive at a non-parametric form. In Ext. Data Fig.~\ref{EFig:truncation_fidelity} we plot $\delta$ (as well as the non-parametric approximation) as a function of $F$, which is always positive, indicating that for all $F$, the MPS representation of a truncated Haar random state is less entangled (by $\mathcal{O}(1)$ ebits) than a equivalent-fidelity depolarized Haar random mixed state. Given that the experiment does not produce a perfect Haar random state, as well as the fact that a real MPS simulation performs many truncations between potentially correlated timesteps, these results are in good agreement with the experimentally found discrepancy of ${\sim}3$ ebits.

\clearpage
\newpage
\subsection{Mixed state entanglement estimation for quasi-adiabatic state preparation}
The mixed state entanglement proxy we have introduced (Eq. 2 of the main text) is a general lower bound in generic situations for any target pure state of interest (see Section F). We have demonstrated its applicability in quantum quench dynamics, and we anticipate its direct applicability in a much broader range of quantum simulation experiments. To exemplify this, we consider in this section the estimation of mixed state entanglement during quasi-adiabatic state preparation and at quantum critical points.

To estimate the mixed state entanglement with our proxy, we require knowledge of the entanglement of the pure target state and the fidelity of experimental state to the pure target state. At quantum critical points, the scaling of entanglement is well studied~\cite{Eisert2010Colloquium:Entropy}, and the pure state entanglement can be extrapolated from exact simulations at small system size. 

\begin{figure}[b!]
	\centering
    \vspace{0.0cm}\includegraphics[width=180mm]{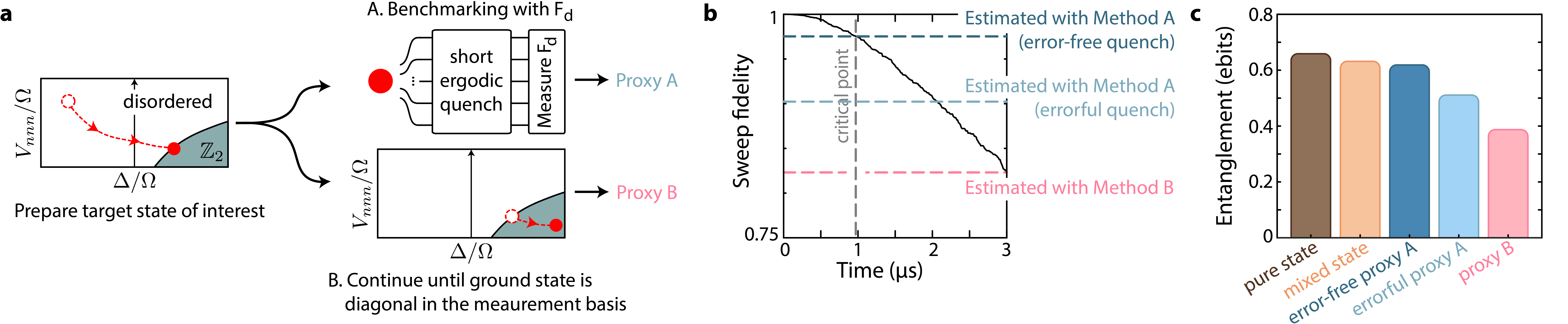}
	\caption{\textbf{Estimating mixed state entanglement of adiabatic state preparatiom. a.} The mixed state entanglement proxy presented in the main text can be applied to the case of adiabatic state preparation. For instance, a quasi-adiabatic sweep may be performed to prepare the critical state at a quantum phase transition. To estimate the fidelity, there are two possible methods: A) use the $F_d$ formula presented here (and as further discussed in this context of \textit{target state benchmarking} in Ref.~\cite{Choi2023PreparingChaos}), or B) continue the sweep into a classical phase for which the ground state fidelity can be directly measured. \textbf{b, c.} We use our error model to simulate a sweep to the quantum phase transition between the disordered and $\mathbb{Z}_2$ phases of the ground state Rydberg phase diagram in 1D. For method A, after reaching the critical point we performed a disordered quench for 4 cycles and calculate $F_d$ to estimate the fidelity. For method B, we use the $\mathbb{Z}_2$ population at the end of the entire sweep to estimate the fidelity. While both options work, for the particular sweep simulated here, method A yields a tighter lower bound on the actual mixed state negativity at the critical point, particularly if the measurement quench is error-free.
        } 
	\vspace{0.3cm}
    \label{EFig:sweep_entanglement}
\end{figure}

It remains then to estimate the fidelity, for which we identify two possible methods (Ext. Data Fig.~\ref{EFig:sweep_entanglement}a):

Method A) For particular states of interest, $F_d$ may not work directly. However, $F_d$ can still be applied by following the quasi-adiabatic sweep preparation with a short, ergodic quench~\cite{Choi2023PreparingChaos,Mark2023BenchmarkingLett}. Importantly, $F_d$ is accurate for short quenches, meaning there is little fidelity loss in this process, and that tensor network methods can be used to generate exact classical simulation references even for large system sizes where critical states have area law entanglement with potentially logarithmic corrections~\cite{Eisert2010Colloquium:Entropy}. Further, the performance of this method could be improved by calibrating the fidelity loss due to the quench via measurements at multiple times.

Method B) Continue the adiabatic sweep into a phase with a ground state for which the fidelity is easily measured (such as a product state). The fidelity of preparing the critical state can then be approximated as the fidelity at the end of the sweep.

We demonstrate both fidelity estimation methods using our standard error model (with no preparation or measurement errors, see Section J) to simulate a quasi-adiabatic sweep to prepare the ground state of the quantum phase transition between the disordered and classical $\mathbb{Z}_2$ ordered phases in the 1D Rydberg phase diagram (Ext. Data Fig.~\ref{EFig:sweep_entanglement}b) with $N=11$ atoms. The sweep in question has a tangent detuning profile with maximum detuning range of $\pm2\pi\times40$ MHz, and Rabi frequency of $2\pi\times5$ MHz. The sweep duration is 3 $\mu$s.

For Method A, we simulate the preparation of the target state at the critical point. We then use the error model to simulate a quench for 4 cycles with a time-independent Hamiltonian with $\Omega=2\pi\times6.9$ MHz, and with $2\pi\times\pm2$ MHz random on-site disorder drawn from a uniform distribution. Such a technique in a similar parameter regime has been well studied theoretically in our previous works~\cite{Choi2023PreparingChaos,Mark2023BenchmarkingLett}. For Method B, we measure the population of the classical $\mathbb{Z}_2$ ordered state ($|10101\ldots0101\rangle$) at the end of the entire quench. Similar techniques can also be used for estimating fidelities of two-dimensional arrays~\cite{Scholl2021QuantumAtoms,Ebadi2021QuantumSimulator}. For both methods the mixed state entanglement proxy remains a lower bound to the actual mixed state negativity at the quantum phase transition, with Method A providing a tighter bound. For Method A we also provide the value assuming the measurement quench is error-free (or where the fidelity loss due to the quench is calibrated away), which closely estimates the actual mixed state negativity (Ext. Data Fig.~\ref{EFig:sweep_entanglement}c).

\newpage
\subsection{Simulation methods}
\subsubsection{Description of the Caltech computing cluster}
This work extensively utilizes the Caltech Resnick High Performance Computing (HPC) Center, running CentOS Linux 7. In our work, each simulation is assigned to a single node, using 16 cores of Intel Cascade Lake CPU (at 2.2 or 2.4 GHz), with a total node memory of 386 GB (at 2933 MT/s). We note that many classical simulation times are reported as core-time, which we simplify as the wall time multiplied by the number of cores. 

\begin{figure}[h!]
	\centering
    \vspace{5mm}
    \includegraphics[width=89mm]{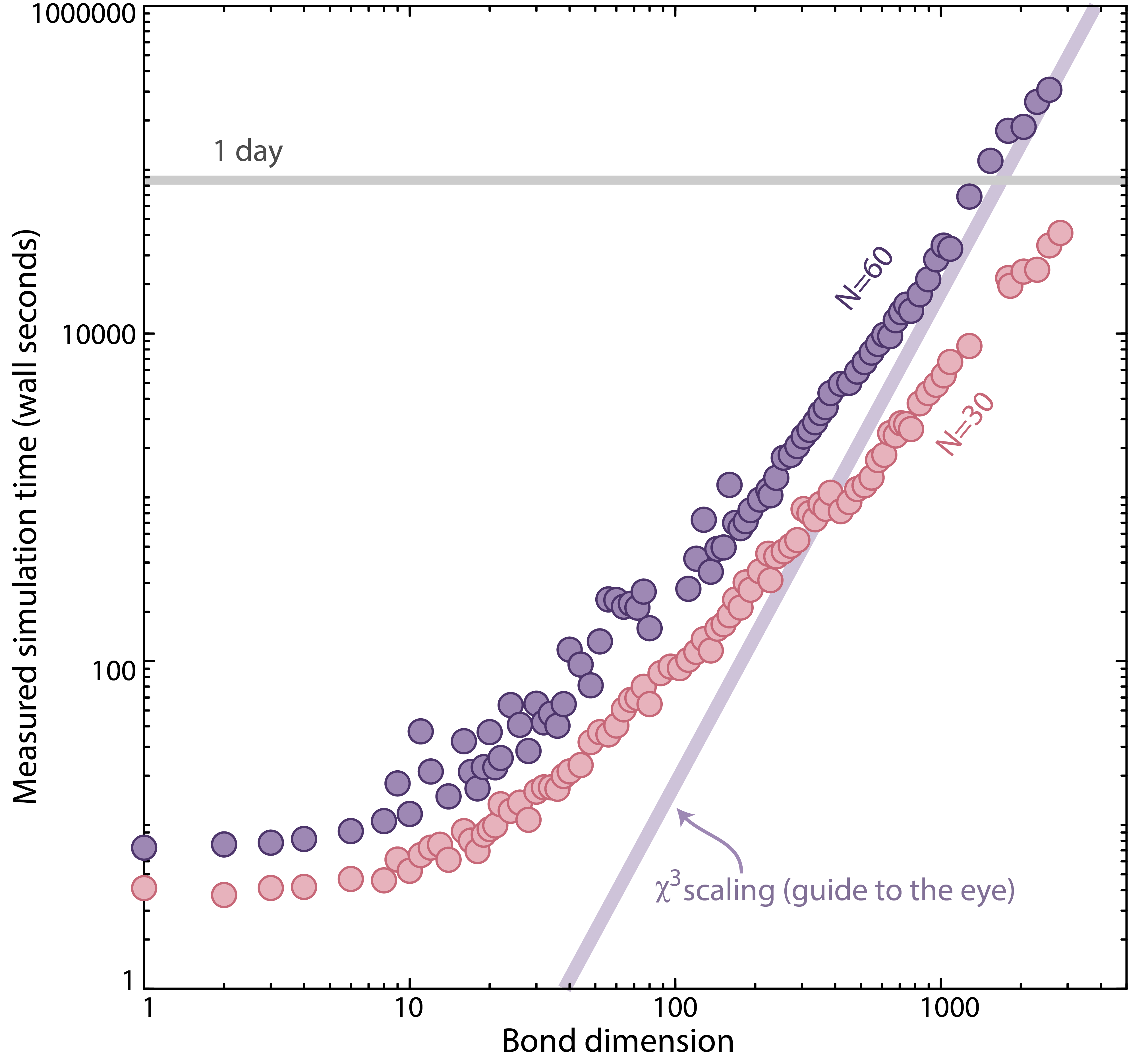}
	\caption{\textbf{Simulation time as a function of bond dimension.} After an initial overhead plateau, the Lightcone-MPS simulation time (measured in raw wall time on the Caltech central cluster) to reach the last experimental time increases as a function of $N$, approaching the expected asymptotic $\chi^3$ scaling.
        } 
	\vspace{0.0cm}
	\label{EFig:simulation_time}
\end{figure}

In terms of raw wall time, the largest simulation we performed (for $N=60$ and $\chi=3072$) took ${\sim}8.3$ days (${\sim}11.1$ days if including the additional time points past the last experiment time in order to calculate $p_\mathrm{avg}$). Representative times to simulate up to the last experimental time point are shown as a function of bond dimension in Ext. Data Fig.~\ref{EFig:simulation_time}; extrapolating the expected $\mathcal{O}(\chi^3)$ scaling we predict simulation for $N=60$ and $\chi^*=3400$ would entail a simulation time of $11.3$ wall-days (180 core-days, as reported in the main text). In total, we generated roughly ${\sim}$100 TB of data in order to create and store the MPS tensors at various bond dimensions used in this work.

\subsubsection{Exact simulation with a Schrieffer-Wolff transformation} We first consider the possibility of exact brute-force simulation of our system. The current state-of-the-art in exact RUC simulation is 45 qubits~\cite{Haner20170.5Circuit}, while the state-of-the-art for continuous Hamiltonian time evolution is 38 qubits~\cite{Hauru2021SimulationEvolution}. While clearly neither of these approaches would work natively for the largest experimental system sizes of $N=60$, it is possible they could become applicable through a Schrieffer-Wolff transformation by rewriting the Rydberg Hamiltonian only in the blockaded subspace, applying appropriate level shifts to mimic the effect of higher blockade sectors, as described in the supplement of Ref.~\cite{Bluvstein2021ControllingArrays}. Because of the Rydberg blockade constraint, the effective system size is given by $\tilde{N}=\log_2(\text{Fib}(N+2))\approx0.7N$, where $\text{Fib}$ is the Fibonacci function. For the 60-atom system, $\tilde{N}{\sim}$42 qubits, so such brute-force techniques could allow for potentially pseudo-exact simulation.

\begin{figure*}[t!]
	\centering
	\vspace{0.0cm}\includegraphics[width=183mm]{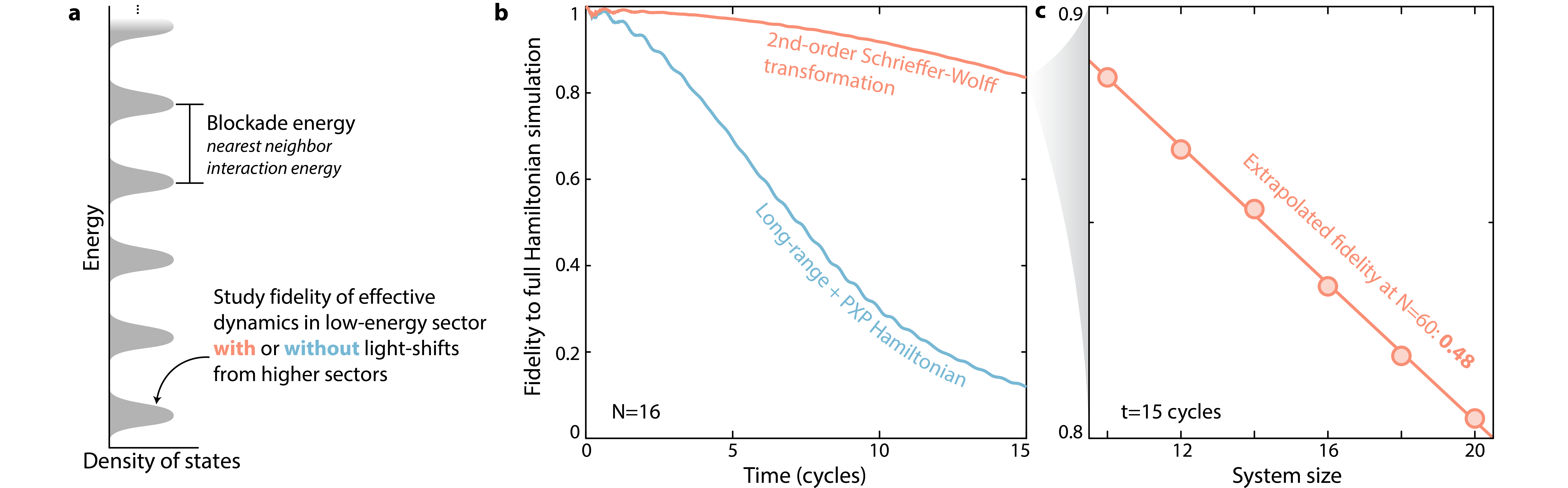}
	\caption{\textbf{Using an effective Hamiltonian simulation. a.} The Rydberg blockade constraint stratifies the energy levels of the Rydberg Hamiltonian, enabling approximating the dynamics in the full Hilbert space by an effective set of dynamics in the lowest energy Hilbert space (relative size of energy sectors are not to scale). \textbf{b.} While a lowest-order approximation (blue) quickly causes the fidelity of the reduced-Hilbert-space simulation to drop compared to simulation in the full space, adding the second-order term from a Schrieffer-Wolff Hamiltonian transformation~\cite{Bluvstein2021ControllingArrays} (orange) improves the fidelity. \textbf{c.} Even with the second-order correction, the simulation fidelity decreases as a function of system size. It is possible that numerically solving for higher order terms in the Schrieffer-Wolff transformation will improve this fidelity, but it likely can never achieve unity given the finite blockade violation probability (Ext. Data Fig.~\ref{EFig:blockaded}d).
        } 
	\vspace{0.5cm}
	\label{EFig:levelshifts}
\end{figure*}

However, here we discount such approaches for multiple reasons. First they both rely on specialized hardware -- a supercomputing system~\cite{Haner20170.5Circuit} with 100s of TB of RAM, or tensor processing units~\cite{Hauru2021SimulationEvolution} which are not widely available. Further, even with this hardware it is not clear that our system would be easily simulable,  because of difficulties associated with treating the blockaded subspace. In particular, applying local gate operations becomes more complicated to implement efficiently, because the blockaded subspace is not factorizable into the product of local bases. 

Finally, applying this transformation is still an approximation to the true dynamics, for which it is not clear if the resulting fidelities will actually be competitive at these large system sizes and late evolution times. In Ext. Data Fig.~\ref{EFig:levelshifts} we test the accuracy of the Schrieffer-Wolff transformation at second order for system sizes up to $N=20$ ($\tilde{N}\sim14$), where we see the fidelity falls by the latest experimental time. Importantly, the fidelity decay increases with system size. A simple linear extrapolation implies a simulation accuracy of ${\sim}0.48$ for $N=60$. Further, employing this approach will always have some fundamental inaccuracy as the blockade constraint is imperfect, and we expect the ideal target state to always have some finite population in the blockade-violating sectors, a symptom which grows stronger with increasing system size and evolution time (Ext. Data Fig.~\ref{EFig:blockaded}d). However, we do not discount the possibility that by adding successively higher-order terms in the Schrieffer-Wolff transformation, the accuracy of this method could be at least improved. 

We strongly emphasize that storing even a single copy of the wavefunction at $\tilde{N}{\sim}42$ still requires ${\sim}65$ TB of memory, and that $\tilde{N}{\sim}42$ exceeds the current state-of-the-art for Hamiltonian simulation~\cite{Hauru2021SimulationEvolution}, $N=38$. Still, we welcome collaboration on studying the boundaries of nearly-exact classical simulation with the most advanced hardware.

\subsubsection{Standard Trotterization-based TEBD-MPS}
A standard approach for simulating large one-dimensional (1D) quantum systems is to use a matrix product state (MPS) representation~\cite{Vidal2003EfficientComputations}. In this representation, the quantum state is decomposed into a tensor network in the form of a 1D chain through repeated Schmidt decompositions. The key insight of such schemes is that if particles are weakly entangled at any bipartition of the system, then the on-site tensors will be low rank, and so can be efficiently \textit{truncated} to a smaller size while preserving most of the information. The size to which the tensor is truncated is user-controllable, and is known as the \textit{bond dimension}, $\chi$. Thus, larger bond dimensions equate to truncating less information when converting a state into its MPS representation. For a more general review of MPS, see Ref.~\cite{Bridgeman2017Hand-wavingNetworks}.

An important aspect of the tensor decomposition of the MPS is that unitary evolution can be efficiently applied to just part of the tensor network. This enables the standard time-evolving block decimation (TEBD) algorithm~\cite{Vidal2004EfficientSystems}, in which Hamiltonian evolution is discretized (or \textit{Trotterized}) using the Suzuki--Trotter formula
\begin{equation}
    e^{-i (\sum_i \hat{H}_i) \delta} = \prod_i e^{-i \hat{H}_i \delta} + \mathcal{O}(\delta^2).
\end{equation}
For a many-body Hamiltonian with (geometrically) local interactions $\hat{H}_i$, each term is applied sequentially onto the MPS. With suitable implementation, Trotterization results in a second-order decomposition of the time evolution operator. 

\begin{figure*}[b!]
	\centering
	\vspace{0.0cm}\includegraphics[width=183mm]{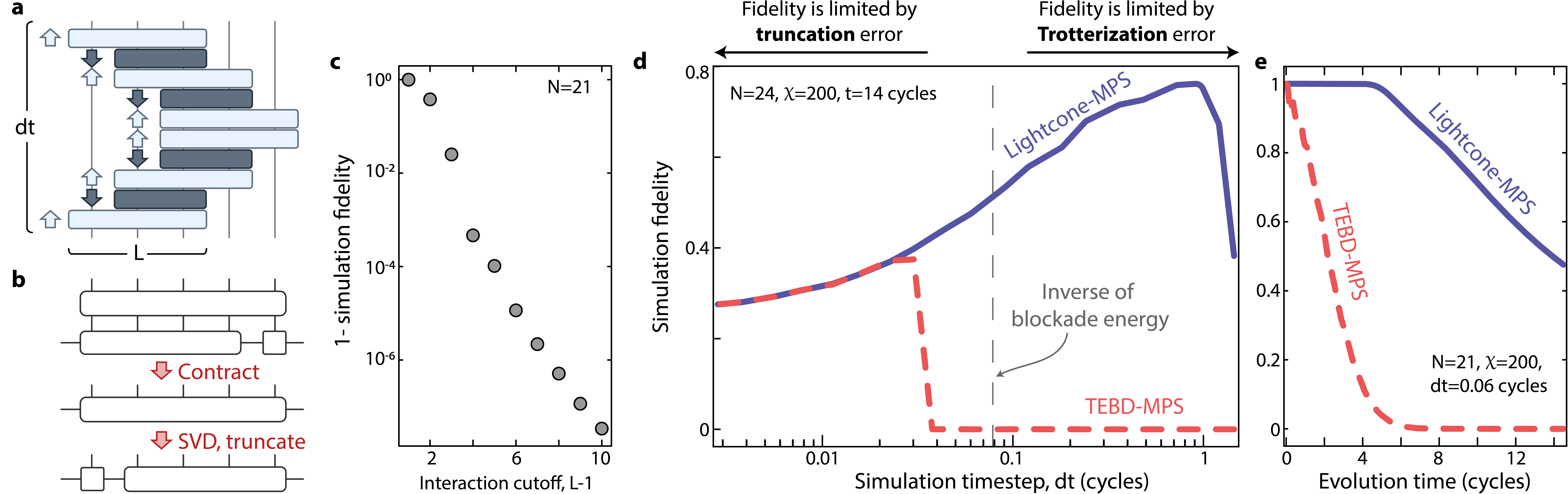}
	\caption{\textbf{Comparing Lightcone-MPS and TEBD algorithms. a.} Illustration of the Lightcone-MPS Hamiltonian decomposition over a single time step $dt$. The long range Rydberg interaction is truncated after a cutoff distance of $L$. \textbf{b.} Illustration of the MPS update procedure. \textbf{c.} We choose the Lightcone-MPS block size to be $L=6$, which we estimate as introducing an error of $<0.5\%$ at $N=60$. \textbf{d, e.} For MPS algorithms there exists an optimal simulation timestep; for too large a timestep, the Hamiltonian Trotterization introduces simulation errors. For too small a timestep, the MPS is contracted and truncated too often, leading to loss of accuracy. For the TEBD algorithm, the maximum timestep is set by the inverse blockade energy, but the Lightcone-MPS algorithm can be used accurately with much larger timesteps. This ultimately means that at equal bond dimension, $\chi$, the Lightcone-MPS algorithm can be optimized to higher fidelities than the TEBD. 
        } 
	\vspace{0.5cm}
	\label{EFig:tebd_vs_haah}
\end{figure*}

Thus, the basic algorithmic loop of the TEBD approach is: 1) perform Schmidt decompositions to write the state as a tensor network, 2) \textit{truncate} the tensor ranks, 3) \textit{Trotterize} the Hamiltonian, 4) and apply it to the relevant tensors, after which we repeat step 1. The two fundamental sources of error in this approach come from steps 2 (the truncation) and 3 (the Trotterization) (Ext. Data Fig.~\ref{EFig:tebd_vs_haah}).

The error of the Trotterization decomposition depends on the product of the commutator of the terms in the Hamiltonian and the time step $dt$ (here $dt$ is defined as a full cycle of application of local evolution from the leftmost site to the rightmost site and then back to the leftmost site). For the Rydberg system, the commutator is on the order of the nearest neighbor interaction $V_{nn}$. Therefore, even with the second-order Suzuki--Trotter decomposition, the perturbative expansion converges only for $V_{nn} \cdot dt \lesssim 1$. Due to the Rydberg blockade, the nearest neighbor interaction for Rydberg Hamiltonian is strong, putting a significant constraint on the maximum $dt$ possible for the simulation. In practice, with a second-order decomposition of the time-evolution operator, we found that a maximum of $dt\approx1/100$ Rabi cycle is necessary to have a $0.5\%$ Trotterization error. 

Importantly, however, for too small of a timestep, the TEBD algorithm will perform many more truncations, each time lowering the fidelity. This leads to a fundamental tension, where an optimal $dt$ must be chosen which balances these competing effects of truncation and Trotterization errors (Ext. Data Fig.~\ref{EFig:tebd_vs_haah}d). 

Moreover, a truncation of the long-range interaction must be chosen to efficiently use the TEBD algorithm. In practice, we found that keeping an interaction length of $5$ results in the desired error of less than $0.5\%$ (Ext. Data Fig.~\ref{EFig:tebd_vs_haah}c). This results in a local evolution involving at most $6$ atoms. 

We emphasize that we do not believe this performance could be strongly improved through the use of a Schrieffer-Wolff transformation as described above. This is because in performing the Schmidt decompositions, MPS-based methods already find the optimal working basis, which is nearly entirely within the blockaded Hilbert space (Ext. Data Fig.~\ref{EFig:blockaded}d). In fact, it is likely such a transformation will \textit{decrease} the MPS performance, because of the added complexity in local gate application due to the non-locally-factorizable structure of the blockaded subspace.

\subsubsection{The Lightcone-MPS algorithm}
One of the challenges of the Trotterization TEBD algorithm is the small $dt$ required to faithfully simulate the quantum dynamics. However, this difficulty can be circumvented by efficiently utilizing the lightcone dynamics of quantum systems, allowing for simulation with much larger $dt$, and thus higher fidelity as truncation errors are minimized. In particular, in Ref.~\cite{Haah0QuantumHamiltonians} it was shown that a global time evolution operator could be decomposed into small blocks of local evolution by combining forward and backward time evolution. In this proposed non-perturbative decomposition, the time step is only limited by the Lieb-Robinson velocity $v$ and the block size $L$ of the local time evolution, not the blockade interaction strength~\cite{Tran2019LocalityInteractions}.

Here, we propose a modified version of the Hamiltonian decomposition in Ref.~\cite{Haah0QuantumHamiltonians}, which is especially amenable to efficient integration with MPS state representations, and which we call Lightcone-MPS. This modification allows for a smaller block size $L$ with the same cutoff of long-range interactions. This algorithm consists of alternating forward-time and backward-time evolution, where the forward-time evolution has a block size $L$, and the backward evolution has a block size of $L-1$. Each block performs the time evolution for all Hamiltonian terms strictly inside the block. Crucially, in this implementation, $dt$ is no longer limited by blockade interactions at the edges of the local evolution blocks (Ext. Data Fig.~\ref{EFig:tebd_vs_haah}d).

Specifically, for the Rydberg Hamiltonian on $N$ atoms
\begin{align}
\hat{H}/h=\Omega\sum_{i=1}^N \hat{S}_i^x -\Delta\sum_{i=1}^N \hat{n}_i + \frac{C_6}{a^6} \sum_{i=1}^N\sum_{j=i+1}^N \frac{\hat{n}_i \hat{n}_j}{|i-j|^6},
\end{align}
we consider the Hamiltonian defined locally from $k$ to $k+L-1$ sites as
\begin{equation}
    \hat{H}_{k, k+L-1}/h =  \Omega\sum_{i=k}^{k+L-1} \hat{S}_i^x -\Delta\sum_{i=k}^{k+L-1} \hat{n}_i + \frac{C_6}{a^6} \sum_{i=k}^{k+L-1}\sum_{j=i+1}^{k+L-1} \frac{\hat{n}_i \hat{n}_j}{|i-j|^6},
\end{equation}
with the forward evolution operator on $L$ sites and backward evolution operator on $L-1$ sites defined as
\begin{align}
    \hat{U}_{k, k+L-1}^\mathrm{forward} &= \exp\left( -i \frac{\hat{H}_{k, k+L-1}}{h} \frac{dt}{2}\right); \\
    \hat{U}_{k, k+L-2}^\mathrm{backward} &= \exp\left( +i \frac{\hat{H}_{k, k+L-2}}{h} \frac{dt}{2}\right)
\end{align}

Then, the evolution within one timestep is decomposed (Ext. Data Fig.~\ref{EFig:tebd_vs_haah}a) as
\begin{equation}
    \exp\left(-i \frac{\hat{H}}{h} dt\right) \approx \hat{U}_{1, L}^\mathrm{forward} \prod_{i=2}^{N-L+1} \left(\hat{U}_{i, i+L-2}^\mathrm{backward} \hat{U}_{i, i+L-1}^\mathrm{forward}\right) \prod_{i=N-L+1}^{2} \left(\hat{U}_{i, i+L-1}^\mathrm{forward} \hat{U}_{i, i+L-2}^\mathrm{backward}\right) \hat{U}_{1, L}^\mathrm{forward} \label{eq:lightcone_decomposition}
\end{equation}

Here, the order of applying the time evolution operator is reversed in every other layer (i.e. left to right for odd layers and right to left for even layers), and each two layers are defined as one timestep. In practice, each forward evolution can be combined with the backward evolution (each pair of operators in Eq.~\eqref{eq:lightcone_decomposition}) into a single operator, reducing the simulation cost. 

The simulation cost can be further reduced by always keeping a tensor with $L-1$ atoms fused together, without performing addition decompositions of the MPS. (Ext. Data Fig.~\ref{EFig:tebd_vs_haah}b). More specifically, we can keep an MPS in the form
\begin{equation}
    \langle z_1, z_2, \dots, z_N | {\psi}_{MPS} \rangle = \sum_{\alpha_1, \alpha_2, \dots, \alpha_{N-1}}A_{z_1}^{\alpha_1} A_{z_2}^{\alpha_1\alpha_2}\cdots M_{z_j, \dots, z_{j+L-2}}^{\alpha_{j-1}\alpha_{j+L-2}} \cdots B_{z_N}^{\alpha_{N-1}},
\end{equation}
without decomposing $M_{z_j, \dots, z_{j+L-2}}^{\alpha_{j-1}\alpha_{j+L-2}}$ into $L-1$ tensors. Here, $z_i$ denotes the physical indices and $\alpha_i$ denotes the bond dimensions. In addition, we use $A$ to denote the tensors in the left canonical form, $B$ to denote tensors in the right canonical form, and $M$ to denote tensors in the center canonical form. Basically, all tensors to the left of $M_{z_j, \dots, z_{j+L-2}}^{\alpha_{j-1}\alpha_{j+L-2}}$ should be in the left canonical form and all the tensors to the right of $M_{z_j, \dots, z_{j+L-2}}^{\alpha_{j-1}\alpha_{j+L-2}}$ should be in the right canonical form. When applying the operator of length $L$ between site $j$ and site $j+L-1$, we can first combine $ M_{z_j, \dots, z_{j+L-2}}^{\alpha_{j-1}\alpha_{j+L-2}}$ with $ B_{z_{j+L-1}}^{\alpha_{j+L-2}\alpha_{j+L-1}}$ into a tensor of length $L$
\begin{equation}
    M_{z_j, \dots, z_{j+L-1}}^{\alpha_{j-1}\alpha_{j+L-1}} = \sum_{\alpha_{j+L-2}} M_{z_j, \dots, z_{j+L-2}}^{\alpha_{j-1}\alpha_{j+L-2}} B_{z_{j+L-1}}^{\alpha_{j+L-2}\alpha_{j+L-1}}.
\end{equation}
Then, we update the tensor by applying the operator of length $L$
\begin{equation}
    {M'}_{z_j, \dots, z_{j+L-1}}^{\alpha_{j-1}\alpha_{j+L-1}} = \sum_{z'_j, \dots, z'_{j+L-1}} U_{z_j, \dots, z_{j+L-1}; z'_j, \dots, z'_{j+L-1}}M_{z'_j, \dots, z'_{j+L-1}}^{\alpha_{j-1}\alpha_{j+L-1}},
\end{equation}
and decompose the updated tensor back into two tensors of length $1$ and $L-1$ 
\begin{equation}
    {M'}_{z_j, \dots, z_{j+L-1}}^{\alpha_{j-1}\alpha_{j+L-1}} \xrightarrow{SVD} {A'}_{z_j}^{\alpha_{j-1}\alpha_{j}} {M'}_{z_{j+1}, \dots, z_{j+L-1}}^{\alpha_{j}\alpha_{j+L-1}}.
\end{equation}
Here, the process of truncating the Schmidt coefficient and combining it into ${M'}_{z_{j+1}, \dots, z_{j+L-1}}^{\alpha_{j}\alpha_{j+L-1}}$ is omitted for simplicity. The same process can be carried on over the rest of the system, and reversed if the operators are applied in the reverse order.

The Lightcone-MPS algorithm allows for a maximum $dt{\sim} L/v$. Although computing the Lieb-Robinson velocity is difficult, based on empirical results, we find in practice that $dt\approx0.5$ Rabi cycle is sufficient for a $0.5\%$ decomposition error with a block size of $6$ (long-range interaction cutoff at $5$). Again, the $dt$ is defined as a full cycle of application of local evolution from the left-most site to the right-most site and then back to the left-most site. 

Crucially, there is still a tension on $dt$ from the competing effects of truncation and Trotterization errors, but for the Lightcone-MPS algorithm the Trotterization errors are minimized up to much larger $dt$. This means at equal bond dimension, if selecting the optimal $dt$, the Lightcone-MPS can reach higher fidelities than TEBD (Ext. Data Fig.~\ref{EFig:tebd_vs_haah}de). Equivalently, to reach the same fidelity necessitates the TEBD to use a larger bond dimension, as is visible in Fig.~\ref{Fig5} of the main text.

Additionally, even if a larger bond dimension is used for the TEBD (where truncation error is small), the optimal $dt$ in TEBD is empirically ${\sim}50\times$ smaller than that for the Lightcone-MPS, increasing the raw simulation time further. In practice, to benchmark the experiment, we need to choose $dt$ such that we can match the time step chosen in the experiment (around $0.77$ Rabi cycle). Therefore, we use half the experimental time step as the $dt$ for the Lightcone-MPS algorithm, and use $1/80$ the experimental time step as the $dt$ for the TEBD algorithm. This still results in a $1/40$ reduction in number of simulation steps thanks to the Lightcone-MPS algorithm.

\subsubsection{Time-dependent variational principle}
The Time-Dependent Variational Principle~\cite{Haegeman2011Time-DependentLattices} (TDVP) algorithm is another MPS time evolution algorithm. In this algorithm, the Schr\"odinger equation is projected onto the variational manifold of MPS, where an effective (nonlinear) partial differential equation (PDE) is obtained to update the MPS parameters. This method is better than TEBD in certain situations such as simulating emergent hydrodynamics, due to its ability to preserve all symmetries and conservation laws. In addition, TDVP can naturally handle long-range interactions. 

However, the TEBD algorithm (and the Lightcone-MPS algorithm) is expected to be better suited in our setup. First, the TEBD algorithm is optimal for minimizing the infidelity in each evolution time step. This is done by truncating the Schmidt coefficients in a way that the infidelity is minimized. In the TDVP algorithm, on the other hand, the infidelity is not necessarily minimized at the cost of preserving conserved quantities. Second, when combined with MPS representation, the TDVP implementation is nontrivial. This is due to the existence of the gauge degree of freedom in MPS representation. More specifically, the representation of MPS is not unique due to gauge transformation. This makes it difficult to utilize MPS-based TDVP by leveraging existing PDE solvers. Ref.~\cite{Haegeman2016UnifyingStates} presents a solution to this problem, which shows that TDVP can be implemented by using the canonical form of MPS, but it relies on evolving the system locally in small time steps, akin to the Trotterization of the Hamiltonian. In addition, it turns out that this solution is a minor modification of the original TEBD algorithm, where a backward time evolution is added. This backward evolution is already naturally incorporated in our Lightcone-MPS algorithm. 

Moreover, regardless of the implementation of the TDVP algorithm, the PDE needs to be solved numerically, with the maximum time step limited by the largest term of the Hamiltonian or the largest term in the PDE. Therefore, the TDVP algorithm has the same bottleneck as the TEBD algorithm ---the time step must be small compared to the blockade interaction strengths, limiting its utility. In our Lightcone-MPS algorithm, however, this limitation is lifted by utilizing extra physical properties such as Lieb-Robinson bound and our knowledge about the geometry of the system (Ext. Data Fig.~\ref{EFig:tebd_vs_haah}). Given this, we do not explicitly test a TDVP implementation in this work.

\subsubsection{Noisy MPO-based simulation}

Instead of simulating perfect quantum dynamics with limited fidelity, an alternative method of benchmarking with the experiment is to (almost) exactly simulate noisy quantum dynamics $\hat{\rho}(t)$ with similar fidelity. We explore this possibility using the recently developed matrix product operator (MPO) based simulation~\cite{Noh2020EfficientDimension}. In MPO simulation, each local tensor has an extended Hilbert space dimension of $4$ instead of $2$. The increased local dimension significantly increases the cost of simulation using the efficient decomposition of Ref.~\cite{Haah0QuantumHamiltonians}. Therefore, we fall back to simulating the Trotterized Hamiltonian dynamics and ignore the long-range interactions. In addition, we choose $dt$ to be half the Rabi cycle, and ignore Trotterization error, in order to consider this method in the most favorable light.

\begin{figure}[ht!]
	\centering
	\includegraphics[width=86mm]{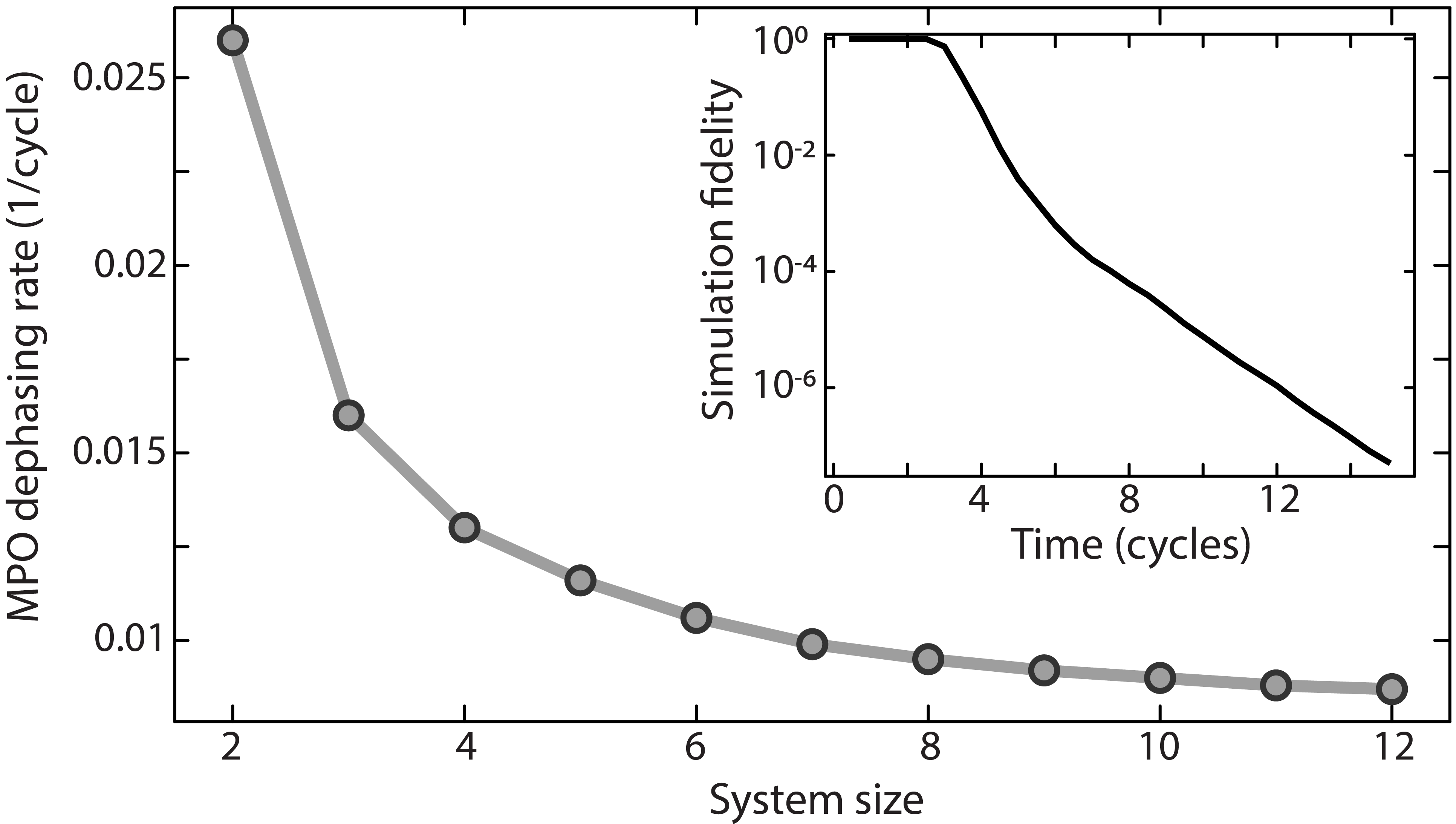}
	\caption{\textbf{Matrix product operator (MPO) simulation.} Effective dephasing rate per qubit to obtain noisy dynamics with a fidelity similar to the exponential decay fit of the experiment. Inset) Truncation accuracy of the MPO method with $\chi=1100$ at $N=60$. The effective dephasing rate is taken as the dephasing rate at $N=12$, which should serve as an upper bound of the effective dephasing rate at $N=60$.}
	\vspace{-0.0cm}
	\label{EFig:dephase_mpo}
\end{figure}

For simplicity, we consider local dephasing error as the only error source and find the effective dephasing rate $\gamma$ per qubit as a function of system size such that the fidelity of the simulation matches the fidelity of the experiment (Ext. Data Fig.~\ref{EFig:dephase_mpo}), extrapolated using the exponential decay fit. As shown in the figure, the effective dephasing rate decreases as the system size increases, which means the dephasing rate at $N=12$ overestimates the dephasing rates at larger system sizes. Since the MPO entanglement decreases as the decoherence rate increases, this leads to an underestimation of the MPO simulation complexity. To make the case more favorable for the MPO method, we thus use the dephasing rate at $N=12$ for the simulation at $N=60$. 

Assuming the early time exponential fit of the experimental fidelity, the Lightcone-MPS algorithm achieves similar fidelity with the experiment at $N=60$ using $\chi=913$ (Ext Data Fig.~\ref{EFig:chi_extrap}). This motivates us to test if the MPO method with a similar or larger bond dimension can achieve the same fidelity. We simulate the noisy quantum dynamics at $N=60$ using $\chi=1100$ and the dephasing rate upper bound, yielding the un-normalized simulation mixed state $\hat{\rho}_{trun}$, where we don't explicitly normalize $\hat{\rho}_{trun}$ at each time step. Here, $\mathrm{tr}(\hat{\rho}_{trun})$ serves as a measure of the simulation accuracy, where it equals 1 when the simulation is perfect and decreases as we truncate the density matrix. We measure $\mathrm{tr}(\hat{\rho}_{trun})$ as a function of time (Ext. Data Fig.~\ref{EFig:dephase_mpo}, inset) and find that it drops to ${\sim} 10^{-7}$. This shows that the MPO algorithm is far from an exact simulation of the noisy quantum dynamics. We note that the $\mathrm{tr}(\hat{\rho}_{trun})$ measures the accuracy of the algorithm compared to the noisy dynamics. It does not measure the fidelity of the simulation compared to the ideal dynamics, which can be even smaller. Thus, this study, in combination with the Trotterization errors not considered here, implies the MPO-based approach will achieve far lower fidelity than the experiment and the lightcone-MPS method for reasonable simulation parameters.

\subsubsection{Path integral formulation}
A hybrid path integral algorithm was introduced in Ref.~\cite{Arute2019QuantumProcessor} for quantum circuit simulation. In this algorithm, the system is cut into two patches. When no gate is applied between the two patches, each patch evolves independently under the Sch\"ordinger equation. When a gate is applied across the two patches, the gate is decomposed into sums of gates that apply independently on the two patches, where one path was sampled out of this decomposition. More formally, we write the gate as $\hat{U} = \sum_i \hat{V}_i \otimes \hat{W}_i$. For each path, we only choose a particular $\hat{V}_i\otimes \hat{W}_i$, so that the two patches stay in a product state. The wave function of each path can be then summed over to obtain the desired result. Suppose each gate across the cut can be minimally decomposed as $r$ product gates and there are $m$ gates across the cut in total, $r^m$ paths are required to fully describe the system.

\begin{figure*}[ht!]
	\centering
	\includegraphics[width=183mm]{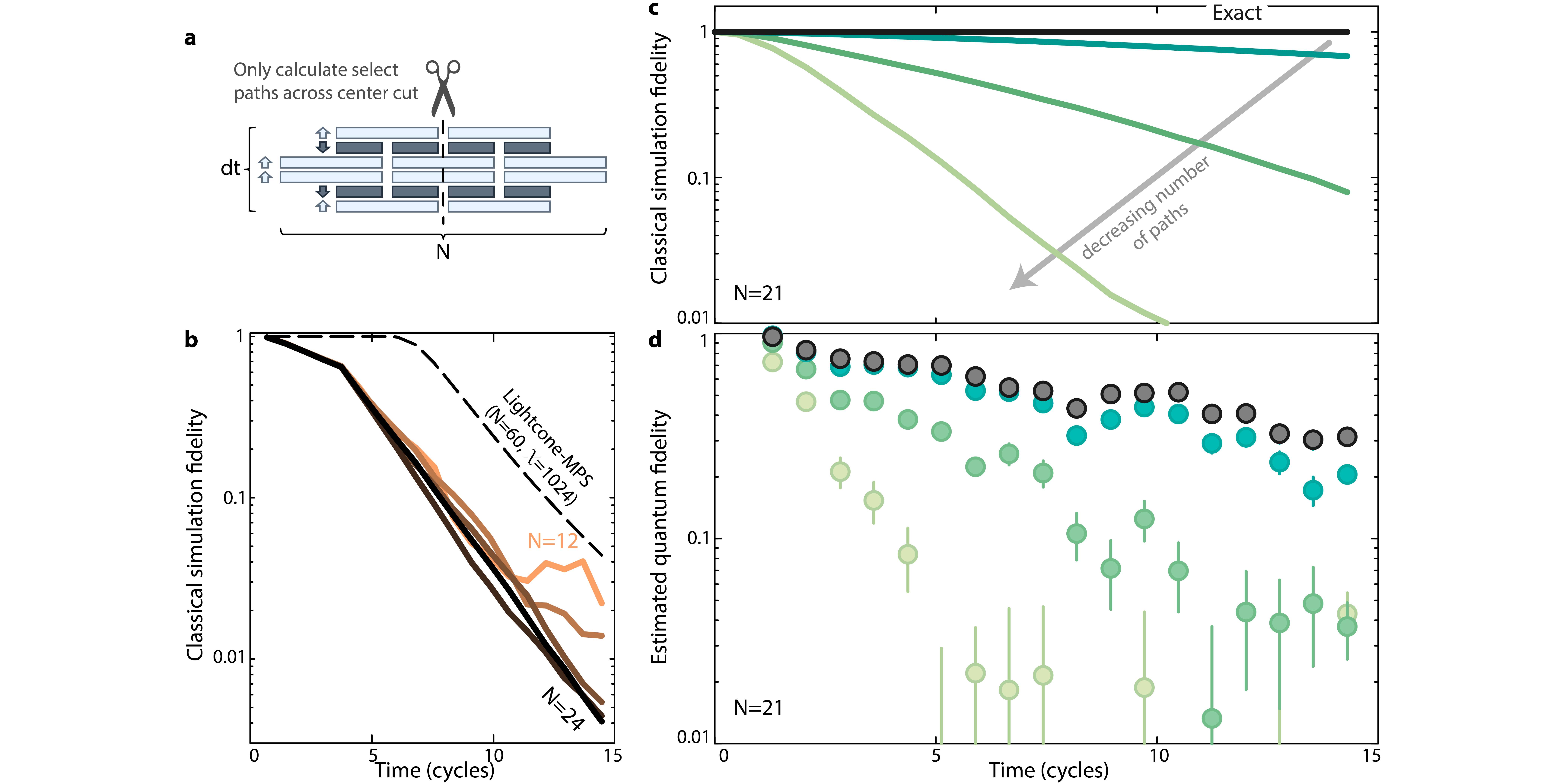}
	\caption{\textbf{A hybrid path integral simulation. a.} Illustration of the hybrid lightcone-path integral (Lightcone-PI) method. \textbf{b.} Simulation fidelity of the Lightcone-PI with $4\times 10^6$ paths, showing decreasing fidelity as a function of $N$. We compare against $F_\text{svd}$ for the Lightcone-MPS algorithm at $\chi=1024$ and $N=60$, which outperforms the L-PI method. \textbf{c, d.} At a fixed $N$, the infidelity in the Lightcone-PI method from using too few paths is visible when benchmarking directly against experimental data, as was the case for the Lightcone-MPS algorithm (Fig.~\ref{Fig2} of the main text).
        } 
	\vspace{-0.0cm}
	\label{EFig:path_integral}
\end{figure*}

In this work, we adapt the hybrid path integral algorithm to the long-range Hamiltonian dynamics. We first decompose~\cite{Haah0QuantumHamiltonians} the global time evolution operator into local gates with block size $L=10$. We further reverse the order of the gate application at every other layer so the gates can be fused between the two layers (Ext. Data Fig.~\ref{EFig:path_integral}). This reduces the number of gates that need to be decomposed at the cut. This allows us to have a minimum of 19 cuts throughout the simulation to match the experiment time steps. Similar to the original hybrid path integral algorithm, we cut the system into two halves. For gates applied across the two halves of the system, we use Schmidt decomposition to write the gate as $\hat{U} =  \sum_i \sigma_i \hat{V}_i \otimes \hat{W}_i$, where $\sigma_i$ can be then absorbed into either of the halves. However, a na\"ive decomposition results in a large interaction between the blockade and non-blockade sectors, which results in poor convergence. Therefore, we project the gate into the blockade sector before performing the decomposition.

We test this algorithm for up to 24 atoms where we truncate the rank of the gates across the cut to $r_i$ to mimic actually simulating the path integral with $\prod_i r_i$ paths. We find that the resulting fidelity mostly depends on the number of paths and only weakly depends on the system size, where the dependency becomes weaker as the system size increases. This is because, for all system sizes, we have the same number of gates across the cut, which we expect to be the only factor affecting the fidelity for large enough system sizes. Extrapolating to 60 atoms, we find that we need at least $4\times 10^6$ paths to obtain a fidelity of ${\sim}0.01$, where \textit{each path} requires an exact simulation of two 30-atom systems. Even using our Lightcone-MPS method, simulating each 30-atom system takes 60 core-hours (with $\chi=1536$, which is nearly exact). This adds up to a formidable amount of computational resources. In comparison,  a direct application of the Lightcone-MPS algorithm for 60 atoms achieves better fidelity with a bond dimension of $\chi=1024$ (Ext. Data Fig.~\ref{EFig:path_integral}b).

Interestingly, we can still benchmark the experimental fidelity using the hybrid path integral approach, which we see shows qualitatively the same behavior as the Lightcone-MPS algorithm in Fig.~\ref{Fig2} of the main text. That is, when the number of paths is too low, the classical simulation fidelity drops, and this drop is visible directly on the benchmarked experimental fidelity (Ext. Data Fig.~\ref{EFig:path_integral}cd). 

This raises the interesting potential prospect of using the experimental benchmarking protocol in reverse to determine the fidelity of the classical algorithm, which we leave to future work.

\subsubsection{Neural network based Hamiltonian simulation}
Neural networks have shown great potential in simulating quantum dynamics~\cite{Carleo2017SolvingNetworks,Gutierrez2022RealStates}. However, such protocols are still in an early age of development and their application to the Rydberg Hamiltonian with ergodic, long-time dynamics is still challenging. We considered two different neural network algorithms used in Refs~\cite{Carleo2019NetKet:Systems,Luo2023Gauge-invariantModels}, but neither results in a fidelity higher than the Lightcone-MPS algorithm, even for comparing both at system sizes as low as $N=12$. While it is possible that neural network based algorithms require additional fine-tuning and hyper-optimization, as it is not the main focus of this work, we leave them for future study. We also note that neural network algorithms are under active development, and we welcome readers to benchmark their algorithms on our data.

\newpage
\subsection{Estimating the MPS fidelity}
\label{seq:estimating_fsvd}
\subsubsection{Approximating the MPS fidelity with truncation errors}
\begin{figure}[ht!]
	\centering
 \vspace{5mm}
	\includegraphics[width=181mm]{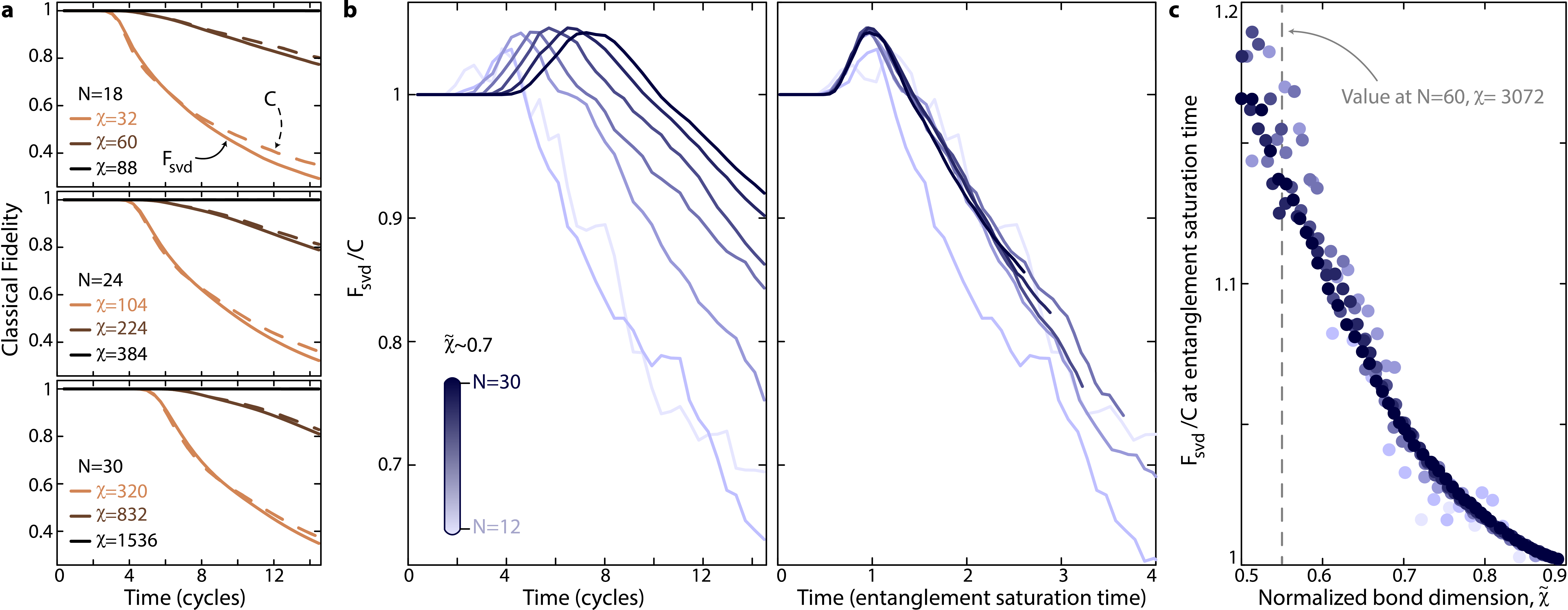}
	\caption{\textbf{Estimating the classical fidelity. a.}  MPS truncation accuracy, $F_\text{svd}$, and MPS fidelity, $C$, for various bond dimensions and system sizes. The MPS fidelity is measured against another MPS calculation with $\chi=3072$, which is almost exact for $N\le 30$. \textbf{b.} For a fixed log-normalized bond dimension, $\tilde{\chi}{\sim}0.7$, $F_\text{svd}$ first over-estimates $C$, before under-estimating it at late time (left). However, rescaling the time by the entanglement saturation time (which is linearly proportional to system size, see Fig.~\ref{Fig3}e of the main text) causes the data to collapse (right). This implies that before $\tent$, $F_\text{svd}$ will over-estimate $C$, so in the parameter regimes we study in this work, using $F_\text{svd}$ is the conservative choice when comparing quantum and classical fidelities. \textbf{c.} The ratio of $F_\text{svd}/C$ around the entanglement saturation time, showing the over-estimation increases with decreasing $\tilde{\chi}$. Given the data collapse, we expect that $F_\text{svd}$ overestimates by a factor of 1.15 at the entanglement saturation time for $N=60$. Taking this effect into account implies the actual $\chi^*(N{=}60)\approx4000$, higher than the value of 3400 quoted in the main text. However, we choose to present the lower value as a more conservative estimate. We note that this relationship between $F_\text{svd}$ and $C$ is only observed to hold for the case of the Lightcone-MPS algorithm, because the timestep is long enough that truncations are relatively independent. The agreement is not observed with a TEBD MPS implementation (which requires much finer timesteps, see text).
        } 
	\vspace{0.5cm}
	\label{EFig:f_trun}
\end{figure}

While MPS gives a controlled approximation of the exact wave function, an exact calculation of the MPS fidelity, $C$ -- defined as the overlap between the MPS state and the ideal target state -- is not possible over multiple time steps for system sizes greater than $N{\sim}30$. However, it is possible to estimate the MPS fidelity from the truncation fidelity
\begin{equation}
    F_\text{svd} = \prod_i \sum_{\alpha=1}^\oldchi s_{i, \alpha}^2,
\end{equation}
where $i$ runs over all steps involving Schmidt value truncations, and $s_{i, \alpha}$ are the Schmidt values at truncation step $i$. Here, we assume the wave function is normalized where $\sum_{\alpha=1}^\infty s_{i, \alpha}^2=1$. While this estimation is only approximate, it can be extremely accurate when successive truncations are independent~\cite{Zhou2020WhatComputers}. For $N\le 30$, a bond dimension of $\chi=3072$ almost exactly captures the dynamics, therefore, it is possible to compute the MPS fidelity, $C$, for various bond dimensions. In Ext. Data Fig.~\ref{EFig:f_trun}a, we plot $F_\text{svd}$ versus the actual MPS fidelity, $C$, to show the general agreement when using the Lightcone-MPS algorithm. This is expected due to the large $dt$ of the algorithm, which makes successive truncations approximately independent. 

\begin{figure}[ht!]
	\centering
	\includegraphics[width=89mm]{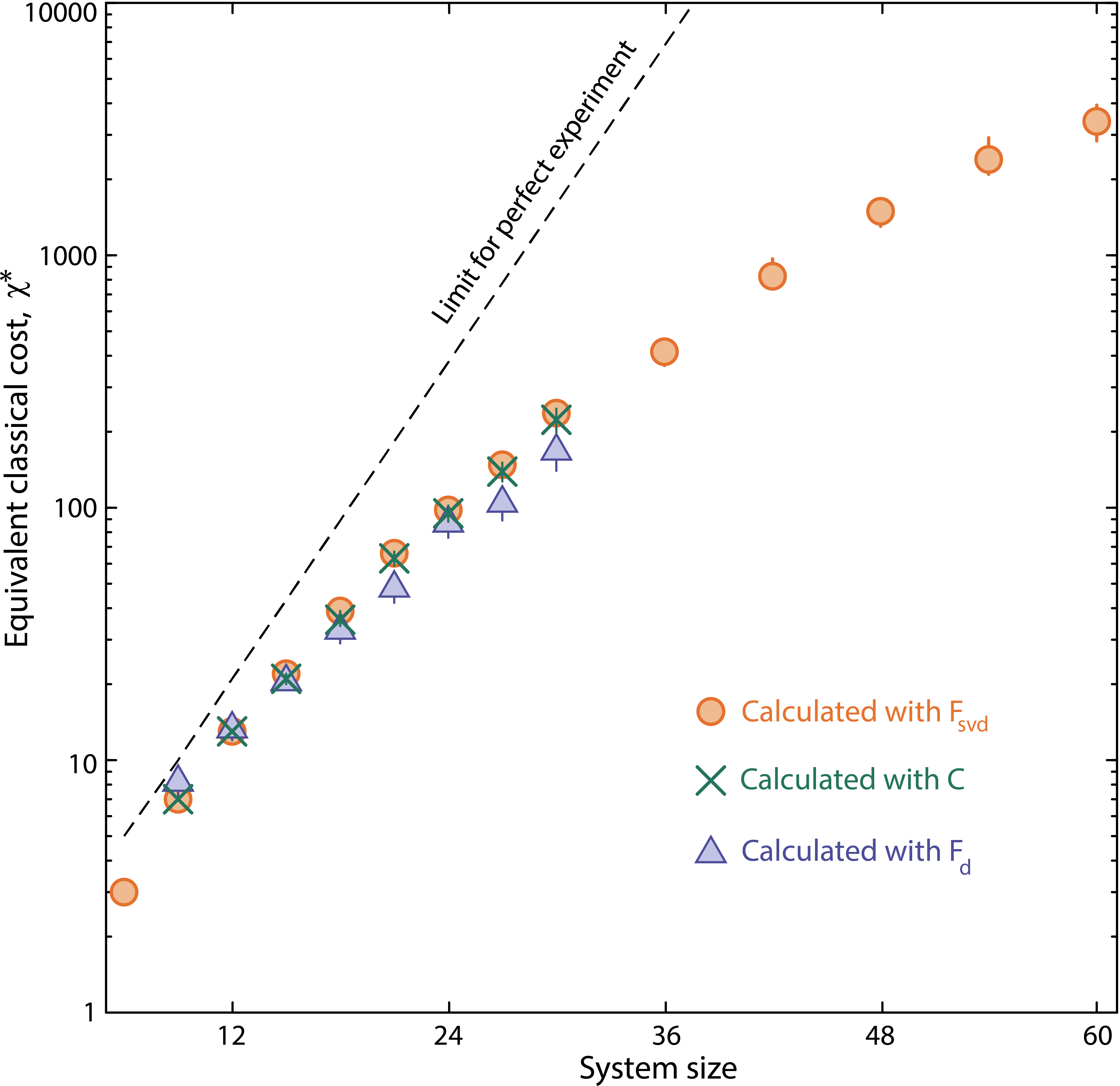}
    \caption{\textbf{Estimating $\chi^*$ from $F_d$ and true classical fidelity.} For system sizes larger than $N>30$, we are not easily able to calculate the exact classical MPS fidelity (as it is impractical to simulate the exact reference state), so in the main text we used $F_\text{svd}$ to compare against the experimental quantum fidelity. However, for $N\leq30$, we can directly compare against the classical MPS fidelity, $C$, which we find gives consistent values of $\chi^*$ in the applicable range. Finally, for these small $N$, we can directly calculate $F_d$ between an approximate MPS simulation and an exact simulation. We find the corresponding $\chi^*$ is also consistent with the values based solely on fidelity.
        } 
	\vspace{-0.0cm}
	\label{EFig:chi_star_sampling}
\end{figure}

In addition, we notice that the agreement becomes better when the system size increases, indicating that $F_\text{svd}$ can be a good estimator of the MPS fidelity for large systems (Ext. Data Fig.~\ref{EFig:f_trun}b). If anything, it appears that in the parameter regime of evolution times shorter than the entanglement saturation time, $F_\text{svd}$ slightly over-estimates $C$, making it a conservative quantity to study (Ext. Data Fig.~\ref{EFig:f_trun}bc). Further, we emphasize that for $N<30$, where we have access to both $C$ and $F_\text{svd}$, the values of $\chi^*$ are consistent (Ext. Data Fig.~\ref{EFig:chi_star_sampling}).

\subsubsection{Extrapolating the experimental $\chi^*$ value}
\begin{figure*}[ht!]
	\centering
	\includegraphics[width=181mm]{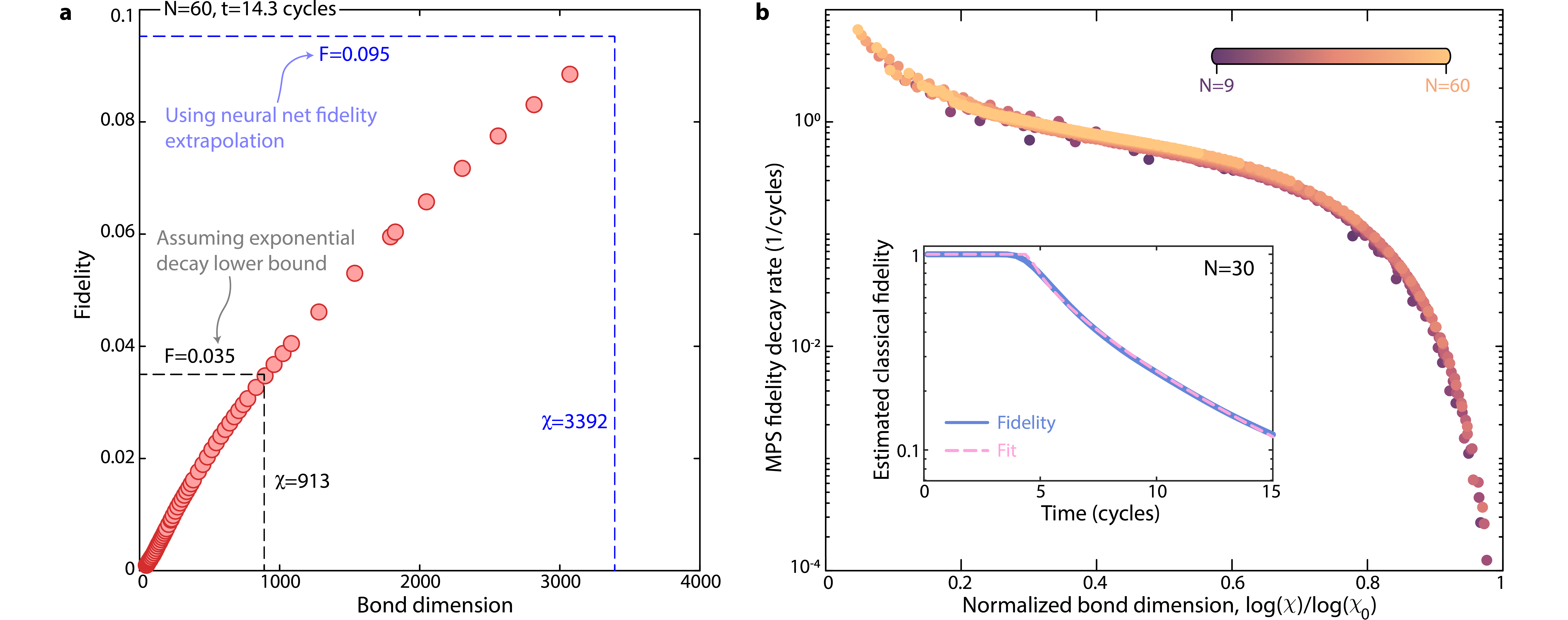}
    \caption{\textbf{Extrapolating $\chi^*$. a.} MPS fidelity (as quantified by the MPS truncation, $F_\mathrm{svd}$) for bond dimensions up to $\chi=3072$, the largest directly simulated. None of the MPS fidelities (markers) exceed the experimental value (blue dashed lines), but given the functional smoothness in this regime, we use a linear extrapolation of the last few data points to predict the $\chi^*$ for the experimental fidelity, yielding $\chi^*=3392\approx3400$. \textbf{b.} For estimating the $\chi$ required to reach a broader range of fidelities (as for instance is used in Fig.~\ref{Fig5}b of the main text), we fit the Lightcone-MPS fidelity as a stretched exponential decay past the exact simulation time (inset). We observe a universal collapse of the decay rate as a function of the log-normalized bond dimension, for all system sizes and bond dimension up to $N=60$ and $\chi=3072$.
        } 
	\vspace{-0.0cm}
	\label{EFig:chi_extrap}
\end{figure*}

We only directly perform Lightcone-MPS simulations up to $\chi=3072$. At this highest $\chi$, the MPS fidelity for $N=60$ at the latest experimental time is 0.088, as quantified by $F_\mathrm{svd}$. Given that this is lower than the experimental fidelity of 0.095(11), we need to extrapolate in order to find the predicted value of $\chi^*$. The fidelities are very close, so we make a simple linear extrapolation approximation, from which we find $\chi^*=3392\approx3400$ as quoted in the main text (Ext. Data Fig.~\ref{EFig:chi_extrap}a).

\subsubsection{Fitting and extrapolating the MPS fidelity}
In the previous subsection, we used a linear extrapolation of MPS fidelity as a function of bond dimension in order to predict $\chi^*$ beyond the regime we directly simulated. However, in Fig.~\ref{Fig5}b of the main text, we compare the experimental and MPS fidelity over a much larger hypothetical fidelity range, where the linear approximation breaks down (see for instance Ext. Data Fig.~\ref{EFig:chi_exact}ab). 

In order to make the comparisons in Fig.~\ref{Fig5}b, we perform fits of the MPS fidelity, approximated by $F_\text{svd}$. For RUCs, $F_\text{svd}$ is essentially exponentially decaying past $\tex$, as the truncations are independent~\cite{Zhou2020WhatComputers}. In our case this behavior is not perfect, and so we fit the MPS fidelity as a stretched exponential decay after $\tex$.

We are able to fit the exponential parameters, and in particular observe an apparent universality in the decay rate as a function of the log-normalized bond dimension (Ext. Data Fig.~\ref{EFig:chi_extrap}b), an interesting phenomena which we leave to future study. We ultimately use these fits to predict the MPS fidelity at arbitrary bond dimension to compare against experiment; we find the fits generally agree to within a few percent with the true fidelities.

To generate Fig.~\ref{Fig5}b, we assume an effective per-atom error rate, $\mathcal{F}$, and find the minimum $\chi^*$ such that the fitted MPS fidelity is always greater than $F=\mathcal{F}^{N t}$. The experimental $\mathcal{F}$ is determined as the value required for exponential decay of the experiment to match the fidelity value of 0.095(11) for the experiment at $t=14.3$ cycles and $N=60$. We then convert $\chi^*$ into runtime using the known conversion of $\chi=3072\rightarrow136$ core-days, measured empirically on the Caltech cluster, and assuming the runtime scales as $\mathcal{O}(\chi^3)$ (Ext. Data Fig.~\ref{EFig:simulation_time}).

\newpage
\clearpage
\subsection{Error model simulations}
\subsubsection{Random unitary circuits}
The dynamics of the one-dimensional random unitary circuit (RUC) shown in Fig.~\ref{Fig4}a are simulated using randomly sampled two-qubit SU(4) unitary gates from the Haar measure. Specifically, the time evolution of an $N$-qubit system starting from the initial state $|0\rangle^{\otimes N}$ can be described as follows:
\begin{align}
|\psi(t) \rangle = \hat{\mathcal{U}}_t  \hat{\mathcal{U}}_{t-1} \cdots \hat{\mathcal{U}}_{2}  \hat{\mathcal{U}}_1 |0\rangle^{\otimes N}, 
\end{align}
where $\hat{\mathcal{U}}_\text{odd} = \{ \hat{\mathcal{U}}_1, \hat{\mathcal{U}}_3, \cdots \}$ and $\hat{\mathcal{U}}_\text{even} = \{ \hat{\mathcal{U}}_2, \hat{\mathcal{U}}_4, \cdots \}$ are the odd- and even-time unitaries composed of local two-qubit unitaries as
\begin{align}
 \hat{\mathcal{U}}_\text{odd} = \prod_{i=1}^{N/2} \hat{U}_{2i-1,2 i}, \quad \hat{\mathcal{U}}_\text{even} &= \prod_{i=1}^{N/2} \hat{U}_{2i,2i+1}
\end{align} 
with open boundary condition, and $t$ is the circuit depth and $\hat{U}_{\mu,\nu}$ is the randomly-sampled SU(4) gate acting on two qubits at site $\mu$ and $\nu$. Note that at each circuit depth $t$, we randomly sample two-qubit random unitaries to generate a many-body unitary $\hat{\mathcal{U}}$, which leads to chaotic dynamics. To emulate noisy quantum dynamics, we utilize a stochastic evolution method~\cite{Molmer1993} incorporating local errors represented by the Pauli operators $\hat{S}^{x,y,z}$. These local errors are stochastically applied to individual qubits at a single-qubit error rate of $p_\text{err} = 0.007$ per circuit layer. By repeating the simulations of the noisy dynamics more than ${\sim}$2000 times with a fixed $p_\text{err}$ for 10 different circuit realizations, we obtain good approximations of the density matrices, $\hat\rho_\text{RUC}$, from which we extract the logarithmic negativity as a measure of entanglement in the mixed state.

\subsubsection{The Rydberg Hamiltonian}
The chaotic dynamics of a one-dimensional Rydberg-atom array presented in this study can be described by time evolution governed by the Rydberg Hamiltonian given in Eq.~(\ref{eq:RydbergHam}) of the Methods. To simulate the open quantum dynamics of the Rydberg atom array, we employ an \textit{ab initio} error model that incorporates realistic error sources observed in our experiment. For detailed information about the error model, please refer to Refs.\cite{Choi2023PreparingChaos,Scholl2023ErasureSimulator}. Similar to the case of the RUC, we simulate the noisy quantum evolution using the stochastic wavefunction method~\cite{Molmer1993} to obtain the density matrix. This allows us to calculate the logarithmic negativity (Fig.~\ref{Fig4}a of the main text). We generally find excellent agreement between our error model and the experimental fidelities across a range of system sizes (Ext. Data Fig.~\ref{EFig:fd_vs_f}), and note that this is the same error model employed in our recent study of record high-fidelity two-qubit Bell state generation\cite{Scholl2023ErasureSimulator}.

\begin{figure*}[ht!]
	\centering
	\includegraphics[width=183mm]{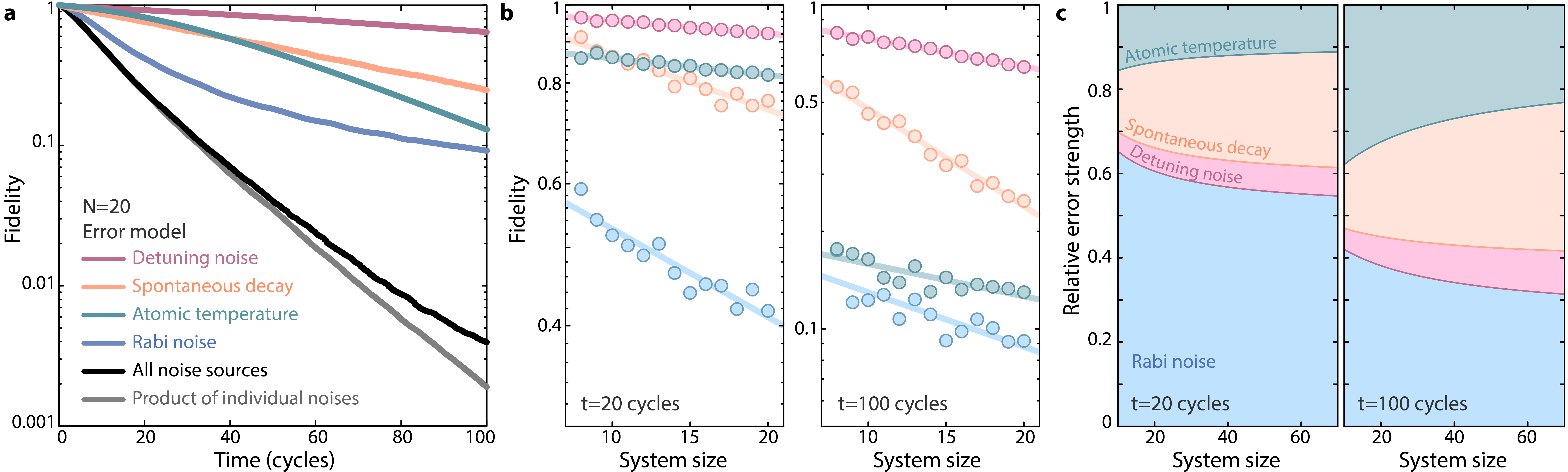}
	\caption{\textbf{Impact of different error sources on fidelity. a.} Simulated fidelity decay for a variety of choices of noise sources affecting our Rydberg quantum simulator. \textbf{b.} Fidelities for error sources in \textbf{a} at short (left) and long (right) times. Solid lines are exponential fits as a function of system size. \textbf{c.} We use the fits in \textbf{b} to estimate the relative contribution of different error sources for increasing system size. Here relative error contribution is defined as $\log(F_i)/\log(F)$, where the subscript, $i$, indexes the error source in question, and $F=\Pi_i F_i$, the product of all individual errors. Note that collective Rydberg decay processes are not accounted for here, and may further lower the fidelity measured for spontaneous decay noise at late times.
        } 
	\vspace{0.0cm}
	\label{EFig:error_model}
\end{figure*}

Of present interest is understanding the main limitations to our experimental fidelity, to best improve the system in the future. To this end, we perform error budget simulations, grouping error sources generally into four main categories: noise on the detuning, noise on the Rabi frequency, spontaneous decay, and atomic temperature effects (Ext. Data Fig.~\ref{EFig:error_model}). We generally find that Rabi frequency noise (in particular, as arising from shot-to-shot laser intensity fluctuations) is a dominant noise source across all times. We extrapolate from error model simulations at small system sizes to larger systems, and find that likely spontaneous decay can become of similar strength for large systems (Ext. Data Fig.~\ref{EFig:error_model}). We note that in these simulations, collective Rydberg decay effects are not accounted for~\cite{Nill2022Many-BodyEnsembles}, but we do not yet observe evidence of such effects in the $t<15$ cycles time window in which we experimentally operate.

\clearpage

\bibliography{references.bib}
\bibliographystyle{adamref}

\FloatBarrier